\IfFileExists{venue_preamble.tex}{\def\VenueMode{1}
  \input{venue_preamble}
}{\documentclass[11pt]{article}
  \usepackage[margin=1in]{geometry}
}

\usepackage[T1]{fontenc}
\usepackage{microtype}
\usepackage{amsmath}
\usepackage{amssymb}
\usepackage{amsthm}
\usepackage{aliascnt}
\usepackage{mathtools}
\usepackage{graphicx}
\usepackage{booktabs}
\usepackage{array}
\usepackage{xcolor}
\usepackage{tikz}
\usetikzlibrary{arrows.meta,decorations.pathreplacing,decorations.markings,positioning}
\definecolor{DSPBlue}{HTML}{0072B2}
\definecolor{DSPOrange}{HTML}{D55E00}
\definecolor{DSPPurple}{HTML}{7057A3}
\tikzset{
  dsp path/.style={line width=1.15pt,line cap=round},
  dsp arrow/.style={-{Stealth[length=4pt]}},
  dsp mid arrow/.style={postaction={decorate},decoration={markings,mark=at position 0.58 with {\arrow{Stealth[length=8pt,width=6pt]}}}},
  dsp label/.style={font=\normalsize,inner sep=2pt},
  dsp vertex/.style={circle,fill=white,draw,line width=0.8pt,inner sep=1.8pt},
  dsp cut/.style={rectangle,fill=white,draw=DSPPurple,line width=1pt,inner sep=2pt}
}
\usepackage{hyperref}
\usepackage[nameinlink,capitalize,noabbrev]{cleveref}

\ifdefined\VenueMode
\else
  \hypersetup{
    colorlinks=true,
    citecolor=blue,
    linkcolor=blue,
    urlcolor=blue
  }
\fi

\graphicspath{{figures/}}
\allowdisplaybreaks

\DeclareMathOperator{\dist}{dist}
\DeclareMathOperator{\spn}{span}
\DeclareMathOperator{\len}{len}
\DeclareMathOperator{\wit}{wit}
\DeclareMathOperator{\mem}{mem}
\DeclareMathOperator{\stored}{stored}
\DeclareMathOperator{\held}{held}
\newcommand{\Term}{\mathcal{T}}
\newcommand{\R}{\mathbb{R}}
\newcommand{\Sweep}{\mathcal{S}}
\newcommand{\reach}[1]{\le_{#1}}

\theoremstyle{plain}
\newtheorem{theorem}{Theorem}[section]
\newaliascnt{lemma}{theorem}
\newtheorem{lemma}[lemma]{Lemma}
\aliascntresetthe{lemma}
\newaliascnt{proposition}{theorem}

\aliascntresetthe{proposition}
\newaliascnt{corollary}{theorem}

\aliascntresetthe{corollary}
\newaliascnt{question}{theorem}
\newtheorem{question}[question]{Question}
\aliascntresetthe{question}

\theoremstyle{definition}
\newaliascnt{definition}{theorem}
\newtheorem{definition}[definition]{Definition}
\aliascntresetthe{definition}

\theoremstyle{remark}
\newaliascnt{remark}{theorem}

\aliascntresetthe{remark}

\newcommand{\PaperTitle}{Directed \(k\)-Disjoint Shortest Paths in Polynomial Time}
\newcommand{\PaperPdfTitle}{Directed k-Disjoint Shortest Paths in Polynomial Time}
\newcommand{\AuthorCao}{Yang Cao}
\newcommand{\AuthorQi}{Haoran Qi}
\newcommand{\AuthorWang}{Hanzhi Wang}
\newcommand{\PaperAuthors}{\AuthorCao\thanks{\texttt{yangcao@psu.edu}. Pennsylvania State University.} \and \AuthorQi\thanks{Jilin University.} \and \AuthorWang\thanks{Tsinghua University.}}
\newcommand{\PaperPdfAuthors}{\AuthorCao, \AuthorQi, \AuthorWang}
\newcommand{\PaperDate}{}

\newif\ifPaperPrintReferences
\PaperPrintReferencestrue

\title{\PaperTitle}
\author{\PaperAuthors}
\date{\PaperDate}
\hypersetup{pdftitle={\PaperPdfTitle},pdfauthor={\PaperPdfAuthors}}

\providecommand{\PaperMakeTitle}{\maketitle}
\providecommand{\PaperPrintBibliography}{\ifPaperPrintReferences
    \bibliographystyle{alpha}
    \bibliography{references}
  \fi}
\providecommand{\PaperEndMatter}{}

\begin{document}

\PaperMakeTitle

\begin{abstract}
The \(k\)-Disjoint Shortest Paths problem asks for pairwise vertex-disjoint shortest paths connecting \(k\) prescribed terminal pairs in a graph with positive edge weights. Eilam-Tzoreff [DAM 1998] asked whether it is polynomial-time solvable for every fixed \(k\). Lochet [SODA 2021] answered this for undirected graphs, while for directed graphs it has remained open for every \(k\ge3\). We give a polynomial-time algorithm for directed graphs and every fixed \(k\), resolving this open problem for all graphs.

\end{abstract}

\clearpage

\section{Introduction}
\label{sec:intro}

In the \(k\)-Disjoint Shortest Paths problem (\(k\)-DSP) we are given a graph with positive edge weights and \(k\) terminal pairs \((s_1,t_1),\dots,(s_k,t_k)\), and we ask for pairwise vertex-disjoint paths \(P_1,\dots,P_k\) such that every \(P_i\) is a shortest \(s_i\)-\(t_i\) path. It combines the disjoint paths problem with the requirement, as in routing schemes, that every demand be sent along a shortest route. We write \(n\) and \(m\) for the numbers of vertices and arcs. Weights are real numbers, and we count additions and comparisons of weights at unit cost.

Eilam-Tzoreff introduced the problem, solved two pairs in undirected graphs, and asked whether \(k\)-DSP can be solved in polynomial time for every fixed \(k\)~\cite{et98}. For undirected graphs the answer is yes: Lochet gave the first polynomial algorithm for every fixed \(k\)~\cite{loc21}, and Bentert, Nichterlein, Renken and Zschoche a faster geometric one~\cite{bnr+21}. Directed graphs behave differently. Without the shortest-path requirement, already two disjoint paths are NP-hard to find in a digraph~\cite{fhw80}, and with zero weights allowed, so is directed \(2\)-DSP~\cite{bk17}. With positive weights, Bérczi and Kobayashi showed that directed \(2\)-DSP can be solved in polynomial time~\cite{bk17}, and faster algorithms for two pairs followed~\cite{cks26,cqw26}. For three or more pairs, neither a polynomial algorithm nor a hardness result was known. Akmal and Wein note that it is open whether even directed \(3\)-DSP can be solved in polynomial time~\cite{aw22}, and Choudhary, Kumar and Saggi list it as an open problem~\cite[Section~6, Open Problem~3]{cks26}. When \(k\) is part of the input, the problem is NP-hard~\cite{et98}, so the question concerns a constant number of pairs.

\subsection{Results}

Our main result resolves the open problem of Eilam-Tzoreff for directed graphs, and thereby for all graphs.

\begin{theorem}[Informal version of \cref{thm:general-formal}]\label{thm:general}
For every fixed \(k\), there is a deterministic polynomial-time algorithm for directed \(k\)-DSP with positive real weights. It outputs a solution if one exists, and its running time is \(n^{g(k)+O(1)}\) for a function \(g\) of \(k\) alone.
\end{theorem}

The theorem also covers undirected graphs, the special case of symmetric digraphs. An exponent that does not depend on \(k\) is not to be expected: Chitnis, Thomas and Wirth showed that directed \(k\)-DSP is W[1]-hard when parameterized by \(k\), even on acyclic digraphs with unit weights~\cite{ctw24}. Under the standard assumption that W[1]-hard problems are not fixed-parameter tractable, this excludes a running time of the form \(f(k)\cdot n^{O(1)}\). Our function \(g\) is doubly exponential in \(k\).

The algorithm builds the paths of an unknown solution one vertex at a time, in the spirit of the algorithm of Fortune, Hopcroft and Wyllie for disjoint paths in a DAG, which moves one pointer along each path in a topological order~\cite{fhw80}. Our algorithm is free in the order in which it adds the vertices: it may continue a path at either end of a stretch that it has already built, or start a new stretch in the middle of a path. As it does not know the solution, it searches through the possible ways of doing this. It stores only a few of the vertices it has used. It forgets a used vertex as soon as no other path can still reach it, where reachability is taken in the shortest-path DAG of that other pair, the union of all shortest paths between its two terminals. For a bound \(p\) on the number of stored vertices, not counting the terminals, we call this algorithm the \emph{sweep with memory \(p\)}. It is always correct, and it runs in \(n^{p+O(1)}\) time when \(k\) and \(p\) are constants. It finds a solution whenever the vertices of some solution can be added in an order along which it never has to store more than \(p\) vertices; we then say that the solution can be \emph{built with \(p\) stored vertices} (\cref{thm:sweep}). The theorem then follows from a statement about solutions alone.

\begin{theorem}[Informal version of \cref{thm:general-memory}]\label{thm:memory-informal}
For every \(k\), every solution of an instance of directed \(k\)-DSP with positive weights can be built with \(g(k)+1\) stored vertices, where \(g(k)=2^{O(2^kk\log k)}\).
\end{theorem}

The proof of \cref{thm:memory-informal} turns reachability into an order of numbers. Let \(\pi_i(v)\) be the distance from \(s_i\) to \(v\), capped at \(\dist(s_i,t_i)\). Along a shortest \(s_i\)-\(t_i\) path, \(\pi_i\) grows by the weight of each arc and every other \(\pi_j\) by at most that weight, so no difference \(\pi_i-\pi_j\) decreases along the path of the pair \(i\). For two pairs, \(\pi_1-\pi_2\) therefore does not decrease along one path and does not increase along the other, and building both paths in the order of this difference stores only a few vertices (\cref{sec:ov-two}). For three or more pairs, no linear function of the \(\pi_i\) orders all paths in this way (\cref{sec:ov-three}).

We therefore split the pairs into two groups. For each group we look for one function \(L\) that agrees, up to a constant, with \(\pi_i\) along the path of every pair \(i\) of the group, and whose increase from any vertex \(u\) to any vertex \(v\) is at most \(\max_j(\pi_j(v)-\pi_j(u))\). The difference of the two functions then orders all paths. The constants of such a function are a solution of a system of difference constraints, so it exists unless exchanging the ends of some paths of the group, in a cycle, decreases their total length, where the length from \(u\) to \(v\) is \(\max_j(\pi_j(v)-\pi_j(u))\). One such exchange cannot decrease the length on many pieces of the solution that lie one after another along every path. This bounds by a function of \(k\) the number of such pieces on which the functions do not exist. An induction on this number cuts the solution, by spheres and grids in the space of the potentials, into parts with fewer such pieces, and each round of cuts costs a number of stored vertices polynomial in \(k\).

These group functions, the bound on repeated exchanges, and the induction, in which every vertex is released by a few fixed vertices of the other paths, shared by nearby vertices, are specific to this paper. From earlier work we use the pointer idea of Fortune, Hopcroft and Wyllie~\cite{fhw80}, the solution of difference constraints by shortest paths~\cite[Section~24.4]{clrs09}, and the theorem of Erd\H{o}s and Szekeres~\cite{es35}. The spheres are chosen with a random radius, as in probabilistic partitions of metric spaces~\cite{bar96}, and the grids with a random shift. \Cref{sec:overview} gives an overview of the proof. We do not know how small the memory can be made (\cref{sec:four}).

\subsection{Related work}

\paragraph{Two pairs.}
For two pairs in undirected graphs, further polynomial algorithms are due to Gottschau, Kaiser and Waldmann, who also allow zero weights~\cite{gkw19}, and to Kobayashi and Sako~\cite{ks19}, and Akmal, Vassilevska Williams and Wein gave a randomized linear-time algorithm for positive weights~\cite{avw24}. For two pairs in digraphs, Choudhary, Kumar and Saggi decide whether a solution exists in randomized \(O(mn\log n)\) time~\cite{cks26}, and Cao, Qi and Wang find a solution in expected \(O(\min\{mn,n^{\omega+\varepsilon}\})\) time for every \(\varepsilon>0\), where \(\omega<2.372\) is the matrix multiplication exponent, and in expected linear time on DAGs and on undirected graphs with polynomially bounded integer weights~\cite{cqw26}.

\paragraph{A fixed number of pairs.}
The undirected algorithms for fixed \(k\)~\cite{loc21,bnr+21} use that along an edge \(\{u,v\}\) of an undirected graph, the distance from any fixed vertex changes by at most \(w(u,v)\) in both directions. In a digraph only one of the two bounds holds, and our argument uses only this one-sided bound. For acyclic digraphs, the algorithm of Fortune, Hopcroft and Wyllie finds \(k\) disjoint paths in \(O(mn^{k-1})\) time~\cite{fhw80}, and it also finds disjoint shortest paths if every path is restricted to the arcs on shortest paths of its pair~\cite{bk17}. Cao, Qi and Wang improved this bound, for \(k\ge3\) and also for shortest paths, to expected \(O_k(m+n^{k-3+\omega+\varepsilon})\) time~\cite{cqw26}. For planar digraphs, an \(n^{O(k)}\)-time algorithm is known~\cite{bk17}. \Cref{sec:overview} explains why the methods for acyclic digraphs and for two pairs do not extend to three pairs in a digraph with cycles.

\paragraph{Related problems.}
In the problem of \(k\) disjoint paths without the shortest-path requirement, undirected graphs admit polynomial algorithms for fixed \(k\)~\cite{rs95,kkr12}. Akmal and Wein study a relaxation of \(k\)-DSP in which every vertex may lie on a bounded number of the shortest paths~\cite{aw22}.

\section{Technical overview}
\label{sec:overview}

By \cref{thm:sweep}, the sweep with memory \(p\) finds a solution as soon as some solution can be built with \(p\) stored vertices. This section explains why every solution can be built with \(p\) depending on \(k\) alone. We fix a solution and turn the task into a question about numbers on its vertices (\cref{sec:ov-task}). For two paths, one number orders both paths (\cref{sec:ov-two}). For three paths, no linear function of the numbers does (\cref{sec:ov-three}). We then use the difference of two functions, one for each of two groups of paths (\cref{sec:ov-groups}). The functions exist unless a cyclic exchange of path ends is cheap (\cref{sec:ov-selectors}). Cheap exchanges cannot repeat along the paths (\cref{sec:ov-repeat}). Finally, we cut the solution around them, at a cost that depends on \(k\) alone (\cref{sec:ov-cuts,sec:ov-witnesses,sec:ov-share,sec:ov-count}).

\subsection{The task: an order of a fixed solution}
\label{sec:ov-task}

The sweep grows the paths of an unknown solution vertex by vertex. It builds each path in \emph{segments}, stretches of consecutive vertices of the path. It may extend a segment at either end, or start a new segment inside a path. It searches through all states that can arise in this way. A state consists of the stored vertices and of their arrangement into segments. With at most \(p\) stored vertices,
\begin{align*}
\#\text{states}=O_{k,p}(n^p),
\end{align*}
and this is the source of the running time \(n^{p+O(1)}\).

Fix a solution \(P_1,\dots,P_k\). A vertex is \emph{processed} once the sweep has added it to its path. The terminals are processed from the start and are never counted. Write \(u\reach{j}v\) if \(v\) can be reached from \(u\) in the shortest-path DAG of the pair \(j\). The unprocessed vertices of \(P_j\) form stretches. Each stretch lies between two processed vertices \(g^-\) and \(g^+\) of \(P_j\); \cref{sec:sweep} calls \((g^-,g^+)\) a gap. As far as the sweep can tell, the missing part of \(P_j\) may pass through any vertex \(v\) with
\begin{align}\label{eq:ov-release}
g^-\reach{j}v\reach{j}g^+.
\end{align}
So the sweep stores a processed vertex \(v\) in two cases. \Cref{thm:sweep} shows that it may forget all other processed vertices.
\begin{itemize}
\item {\bf Own path.} \(v\) has an unprocessed neighbour on its own path, where the path continues.
\item {\bf Another path.} \(v\) satisfies \eqref{eq:ov-release} for some unprocessed stretch of another path \(P_j\). If the sweep forgot such a vertex, it could use it a second time.
\end{itemize}
Our task is to find, for every solution, an order of its vertices along which at most \(g(k)+1\) vertices are stored at any moment (\cref{thm:general-memory}).

Condition \eqref{eq:ov-release} is about reachability. The analysis replaces it by a weaker condition on numbers. Choose potentials \(\pi_1,\dots,\pi_k\) such that every arc \((u,v)\) has \(\pi_i(v)-\pi_i(u)\le w(u,v)\), with equality on the arcs of shortest \(s_i\)-\(t_i\) paths. The distance from \(s_i\), capped at \(\dist(s_i,t_i)\), will do (\cref{lem:potentials}). On an arc of a shortest \(s_i\)-\(t_i\) path,
\begin{align}\label{eq:ov-arc}
\pi_i(v)-\pi_i(u)=w(u,v)\ge\pi_j(v)-\pi_j(u)\qquad(j\ne i).
\end{align}
We call the differences \(\pi_a-\pi_b\) the \emph{pair coordinates} of a vertex. Those with \(\pi_i\) first, \(\pi_i-\pi_j\) for \(j\ne i\), are the \emph{own coordinates} of the pair \(i\). By \eqref{eq:ov-arc}, no own coordinate of \(i\) decreases along an arc of a shortest \(s_i\)-\(t_i\) path. We write \(u\preceq_iv\) if
\begin{align}\label{eq:ov-order}
(\pi_i-\pi_j)(u)\le(\pi_i-\pi_j)(v)\qquad\text{for all } j\ne i.
\end{align}
Reachability implies this order, as in the first line below. Hence \eqref{eq:ov-release} implies its counterpart for the potentials, the second line:
\begin{align}
u\reach{j}v&\implies u\preceq_jv,\notag\\
g^-\reach{j}v\reach{j}g^+&\implies g^-\preceq_jv\preceq_jg^+.
\label{eq:ov-held}
\end{align}

A processed vertex \(v\) other than a terminal is \emph{held} if it has an unprocessed neighbour on its path, or if some unprocessed stretch of another path \(P_j\), between \(g^-\) and \(g^+\), has
\begin{align}\label{eq:ov-holds}
g^-\preceq_jv\preceq_jg^+.
\end{align}
In the second case, \(P_j\) \emph{holds} \(v\). By \eqref{eq:ov-held}, every stored vertex is held. The \emph{width} of an order of the vertices is the largest number of vertices held at one moment along it. So at every moment,
\begin{align*}
\#\{\text{stored vertices}\}\le\#\{\text{held vertices}\}\le\operatorname{width}.
\end{align*}
The width of a solution is the least width of its orders. It suffices to bound it, up to one correction. The sweep joins two segments of a path in two moves: it extends one segment by the new vertex, and then merges the two. Between these moves, the new vertex is an additional segment end (\cref{lem:orders,lem:transfer}). This is the \(+1\) in \(g(k)+1\).

Whether a vertex is held depends only on the potentials and on the order of the vertices along the paths. We call these data the \emph{configuration} of the solution. From \cref{sec:config} on, the proof works with configurations alone, and the paths are called chains \(X_i\). The formal definitions differ from the ones above in three points (\cref{def:order,def:demands}).

First, different vertices can have the same pair coordinates. Then \eqref{eq:ov-order} holds in both directions, and the formal relation \(\preceq_i\) breaks the tie by \(\pi_1\). \Cref{sec:ov-two} shows how \(\pi_1\) orders such vertices for two paths. Elsewhere in this section, we assume that different vertices have different pair coordinates.

Second, the formal notion of a held vertex refers only to \(v\) and to the vertices of \(P_j\), not to the stretches. Consider
\begin{align*}
\{q\in P_j : q\preceq_jv\}
\qquad\text{and}\qquad
\{q\in P_j : v\preceq_jq\}.
\end{align*}
The vertices of \(P_j\) increase in \(\preceq_j\), so the first set is an initial part of \(P_j\) and the second a final part. If one of them is empty, \(P_j\) never holds \(v\). Otherwise, the \emph{lower bound} \(p^*\) of \(v\) on \(P_j\) is the last vertex of the first set, and the \emph{upper bound} \(q^*\) is the first vertex of the second. A vertex in both sets has the same own coordinates of \(j\) as \(v\). It then has the same pair coordinates, since
\begin{align*}
\pi_a-\pi_b=(\pi_j-\pi_b)-(\pi_j-\pi_a).
\end{align*}
Our assumption excludes this, so \(p^*\) comes before \(q^*\). Formally, \(P_j\) holds \(v\) until some two consecutive vertices from \(p^*\) to \(q^*\) are both processed. This is the \emph{demand} of \(v\) on \(P_j\). That the neighbours of \(v\) on its own path be processed is its \emph{neighbour demand} (\cref{def:demands}).

The formal notion covers \eqref{eq:ov-holds}. If \(P_j\) holds \(v\) through a stretch between \(g^-\) and \(g^+\), then \(g^-\) lies in the first set and \(g^+\) in the second, so
\begin{align*}
g^-\preceq_jp^*\preceq_jv\preceq_jq^*\preceq_jg^+.
\end{align*}
Of any two consecutive vertices from \(p^*\) to \(q^*\), one lies strictly between \(g^-\) and \(g^+\) and is unprocessed. So a vertex that is held in the sense of \eqref{eq:ov-holds} is held formally. It therefore suffices to bound the formal width (\cref{lem:stored-held}).

The formal notion makes sense for a \emph{sub-configuration}: a set of vertices of the solution, with their potentials and their order along the paths. Bounds, neighbours and consecutive vertices are then taken within the set, and no terminals need be present. The width of a sub-configuration is defined in the same way. Sub-configurations are used from \cref{sec:ov-selectors} on; the cells and blocks of \cref{sec:ov-cuts} are sub-configurations.

Third, the formal orders also list the terminals, which can then be held; \cref{lem:transfer} accounts for this. The arguments of this section go through for the formal notions, and the formal sections carry them out.

\subsection{Two paths: one coordinate}
\label{sec:ov-two}

For two paths there is one pair coordinate up to its sign,
\begin{align*}
x\coloneqq\pi_1-\pi_2,
\end{align*}
and \eqref{eq:ov-order} reads
\begin{align}\label{eq:ov-x}
u\preceq_1v\iff x(u)\le x(v),
\qquad
u\preceq_2v\iff x(u)\ge x(v).
\end{align}
So \(x\) does not decrease along \(P_1\) and does not increase along \(P_2\). The order processes \(P_1\) forwards from \(s_1\) and \(P_2\) backwards from \(t_2\). While both paths are incomplete, it advances the path whose current end has the smaller value of \(x\). When one path is complete, it finishes the other.

Let \(h\) and \(g\) be the current ends of \(P_1\) and \(P_2\). The unprocessed stretch of \(P_1\) lies between \(h\) and \(t_1\), and that of \(P_2\) between \(s_2\) and \(g\). Let \(v\) be a processed vertex of \(P_1\) other than \(h\). The case of \(P_2\) is symmetric, with the roles of the two paths, and of \(h\) and \(g\), exchanged. Its neighbours on \(P_1\) are processed, so \(v\) is held only if \(P_2\) holds it, that is, only if \(s_2\preceq_2v\preceq_2g\). In that case
\begin{align*}
x(g)\le x(v)\le x(g).
\end{align*}
The first inequality is \eqref{eq:ov-x} for \(v\preceq_2g\). For the second, \(P_2\) was incomplete when the order moved on from \(v\). At that moment, \(v\) was the end with the smaller value of \(x\), and the values at the ends only grow. So a held vertex of \(P_1\) other than \(h\) has \(x(v)=x(g)\). Symmetrically, a held vertex of \(P_2\) other than \(g\) has \(x(v)=x(h)\). If no two vertices have the same value of \(x\), only \(h\) and \(g\) can be held, and the order has width \(2\).

Vertices with the same value of \(x\) need a second key. Along an arc \((u,v)\) of \(P_1\) or of \(P_2\),
\begin{align*}
x(u)=x(v)
\quad\implies\quad
\pi_1(v)-\pi_1(u)=\pi_2(v)-\pi_2(u)=w(u,v)>0.
\end{align*}
As \(x\) is monotone along each path, the vertices of a path with the same value of \(x\) are consecutive. So \(\pi_1\) increases strictly among them, on both paths, and the order processes them by increasing \(\pi_1\). On \(P_2\), this runs against the direction in which the path is processed. The processed vertices of \(P_2\) then no longer form one stretch at its end. The order corrects this by processing at most three vertices of each path early (\cref{sec:ov-groups}). With them, every configuration with two paths has width at most \(8\) (\cref{lem:sortable-simple,lem:sortable}).

\subsection{Three paths: no coordinate}
\label{sec:ov-three}

For two pairs and for acyclic digraphs, all paths of a solution follow one order: the coordinate \(x\), or a topological order. The algorithm of Fortune, Hopcroft and Wyllie keeps one pointer on each of the \(k\) paths and always advances the lowest pointer in a topological order~\cite{fhw80}. The algebraic algorithms of Cao, Qi and Wang also work along such an order~\cite{cqw26}.

For three or more pairs in a digraph with cycles, there is no such order. The DAGs of different pairs together can contain cycles, so there is no topological order. A pair coordinate orders only its own two paths. Along \(P_3\), for instance, the own coordinates \(\pi_3-\pi_1\) and \(\pi_3-\pi_2\) do not decrease, and
\begin{align*}
x=\pi_1-\pi_2=(\pi_3-\pi_2)-(\pi_3-\pi_1)
\end{align*}
is their difference. So \(x\) can go up and down along \(P_3\).

No linear function of the pair coordinates does better. By \eqref{eq:ov-arc}, an arc on a shortest \(s_i\)-\(t_i\) path only guarantees that no own coordinate of \(i\) decreases. Call a change of the pair coordinates \emph{forward for \(i\)} if it increases every own coordinate of \(i\). For a linear function \(\ell\) of the pair coordinates, let \(I^+\) be the set of pairs \(i\) such that \(\ell\) grows under every change that is forward for \(i\). Let \(I^-\) be the set of pairs \(j\) such that \(\ell\) falls under every change that is forward for \(j\). For \(\ell=x\), these sets are \(\{1\}\) and \(\{2\}\). \Cref{lem:linear-order} shows that no linear function does more:
\begin{align*}
\lvert I^+\rvert\le1,\qquad\lvert I^-\rvert\le1.
\end{align*}
If both sets are non-empty, say \(I^+=\{i\}\) and \(I^-=\{j\}\), then \(\ell\) is a positive multiple of \(\pi_i-\pi_j\). So for two pairs, \(x\) is the only choice up to a factor. A linear function that did for \(k\) pairs what \(x\) does for two would have \(I^+\cup I^-=\{1,\dots,k\}\), and hence
\begin{align*}
k=\lvert I^+\cup I^-\rvert\le\lvert I^+\rvert+\lvert I^-\rvert\le2,
\end{align*}
which is impossible for \(k\ge3\).

\subsection{Two groups of paths}
\label{sec:ov-groups}

We keep the form ``one function minus another'' of the coordinate \(x\). The potential of a single path is replaced by a function for a whole group of paths. Such a function is compared with the \emph{length} from a vertex \(u\) to a vertex \(v\),
\begin{align}\label{eq:ov-len}
\len(u,v)\coloneqq\max_{1\le j\le k}\bigl(\pi_j(v)-\pi_j(u)\bigr).
\end{align}
The length satisfies the triangle inequality and is at least the increase of every potential. If \(u\preceq_iv\), then by \eqref{eq:ov-order} no potential grows by more than \(\pi_i\) from \(u\) to \(v\). So the length is exact:
\begin{align}\label{eq:ov-exact}
\len(u,v)=\pi_i(v)-\pi_i(u)\qquad\text{if } u\preceq_iv.
\end{align}

\begin{definition}[Selector]\label{def:ov-selector}
The \emph{range} of a path \(P_i\) consists of the vertices of \(P_i\) and of the vertices \(v\) of other paths with \(y\preceq_iv\preceq_iy'\) for two vertices \(y,y'\) of \(P_i\). Let \(A\) be a set of indices. A \emph{selector} for \(A\) is a function \(L\) on the vertices of the solution, together with constants \(\theta_i\) for \(i\in A\), its \emph{offsets}, such that the following holds.
\begin{itemize}
\item {\bf Property 1.} \(L(v)-L(u)\le\len(u,v)\) for all vertices \(u\) and \(v\): the selector grows no faster than the length.
\item {\bf Property 2.} \(L(v)=\pi_i(v)+\theta_i\) for every \(i\in A\) and every vertex \(v\) in the range of \(P_i\): on the range of every path of its group, the selector agrees with the potential of that path.
\end{itemize}
\end{definition}

If \(P_i\) holds a vertex, the vertex lies in the range of \(P_i\), by \eqref{eq:ov-holds}. This is why Property~2 is asked on the range. For a group with one path, \(\pi_i\) is a selector with offset \(0\), since the length is at least the increase of \(\pi_i\). A selector for a larger group need not be a potential, nor a linear function of the potentials.

Suppose that the indices split into two groups \(A\) and \(B\) with selectors \(L_A\) and \(L_B\). We then call the solution \emph{sortable}, and define the primary key
\begin{align*}
\kappa\coloneqq L_A-L_B.
\end{align*}
Let \(u\preceq_iv\) be two vertices in the range of \(P_i\), where \(i\in A\). Property~2 of \(L_A\), Property~1 of \(L_B\) and \eqref{eq:ov-exact} give
\begin{align*}
\kappa(v)-\kappa(u)
&=\bigl(\pi_i(v)-\pi_i(u)\bigr)-\bigl(L_B(v)-L_B(u)\bigr)\\
&\ge\bigl(\pi_i(v)-\pi_i(u)\bigr)-\len(u,v)=0.
\end{align*}
For \(i\in B\), the two selectors exchange their roles. So, on the range of \(P_i\),
\begin{align}\label{eq:ov-kappa}
u\preceq_iv\implies
\begin{cases}
\kappa(u)\le\kappa(v) & \text{if } i\in A,\\
\kappa(u)\ge\kappa(v) & \text{if } i\in B.
\end{cases}
\end{align}
For \(A=\{1\}\), \(B=\{2\}\), \(L_A=\pi_1\) and \(L_B=\pi_2\), the key \(\kappa\) is the coordinate \(x\), and \eqref{eq:ov-kappa} is one half of \eqref{eq:ov-x}.

The argument for two paths now applies to all paths at once, with \(\kappa\) in place of \(x\). The order processes the paths of \(A\) forwards and the paths of \(B\) backwards. Among the incomplete paths, it advances one whose current end has the smallest value of \(\kappa\). Let \(h_l\) be the current end of \(P_l\). The unprocessed stretch of \(P_l\) lies between \(h_l\) and \(t_l\) if \(l\in A\), and between \(s_l\) and \(h_l\) if \(l\in B\). So if \(P_l\) holds a processed vertex \(v\), then
\begin{align*}
&h_l\preceq_lv\preceq_lt_l\qquad(l\in A),\\
&s_l\preceq_lv\preceq_lh_l\qquad(l\in B).
\end{align*}
In both cases, \(v\) lies in the range of \(P_l\), and \eqref{eq:ov-kappa} gives
\begin{align*}
\kappa(v)\ge\kappa(h_l).
\end{align*}
As for two paths, the order moved on from \(v\) when \(v\) was the incomplete end with the smallest \(\kappa\), and the values at the ends only grow. So, for a processed vertex \(v\) other than the \(k\) current ends,
\begin{align*}
P_l\text{ holds } v
\quad\implies\quad
\kappa(v)=\kappa(h_l).
\end{align*}

Equal primary keys cannot be avoided: \(\kappa\) can be constant along a long stretch of a path. A secondary key must order such vertices strictly along every path. For two paths, \(\pi_1\) served. Here, \(\pi_1\) need not increase along a stretch of \(P_i\) on which \(\kappa\) is constant: for \(i\ne1\), by \eqref{eq:ov-arc}, an arc of \(P_i\) bounds the change of \(\pi_1\) only from above.

The secondary key therefore first compares
\begin{align*}
G\coloneqq kL_A-\sum_{j=1}^k\pi_j.
\end{align*}
By Property~2, on the range of \(P_i\),
\begin{align*}
L_A=
\begin{cases}
\pi_i+\theta_i & (i\in A),\\
L_B+\kappa=\pi_i+\theta_i+\kappa & (i\in B).
\end{cases}
\end{align*}
Write \(\varphi_i\coloneqq\sum_{j\ne i}(\pi_i-\pi_j)\) for the sum of the own coordinates of \(i\), the \emph{progress} along \(P_i\). Then, on the range of \(P_i\),
\begin{align*}
G=\varphi_i+
\begin{cases}
k\theta_i & \text{if } i\in A,\\
k\theta_i+k\kappa & \text{if } i\in B.
\end{cases}
\end{align*}
So among vertices of \(P_i\) with the same \(\kappa\), the functions \(G\) and \(\varphi_i\) differ by a constant. The progress does not decrease along \(P_i\), by \eqref{eq:ov-arc}. The order takes such vertices by increasing \(G\) on every path, also on the paths of \(B\) (\cref{lem:keys}).

It remains to compare two vertices \(u\preceq_iv\) of \(P_i\) with the same values of \(\kappa\) and \(G\). For them,
\begin{align*}
\sum_{j\ne i}\underbrace{\bigl((\pi_i-\pi_j)(v)-(\pi_i-\pi_j)(u)\bigr)}_{\ge0}=G(v)-G(u)=0.
\end{align*}
So all own coordinates of \(i\) agree at \(u\) and \(v\), and hence all pair coordinates. Under our assumption, \(u=v\). In general, \(\pi_1\) decides between them, as in \cref{sec:ov-two}.

The order therefore goes through the pairs of keys \((\kappa,\kappa')\), where \(\kappa'\coloneqq(G,\pi_1)\), in increasing lexicographic order. This modifies the rule ``advance the end with the smallest \(\kappa\)'': plateaus, defined below, are taken forwards on every path. When the order reaches a pair \((\alpha,\beta)\), the keys have \emph{placed} the vertices
\begin{align*}
K(\alpha,\beta)\coloneqq\{v : \kappa(v)<\alpha\}\cup\{v : \kappa(v)=\alpha,\ \kappa'(v)\le\beta\}.
\end{align*}
The order processes these vertices and a few more. On every path \(P_l\), the vertices with \(\kappa=\alpha\) are consecutive, by \eqref{eq:ov-kappa}. They form the \emph{\(\alpha\)-plateau}, which the secondary key takes forwards. If the plateau is not empty, let \(y_l\) be its first vertex and \(y_l^-\) the predecessor of \(y_l\), if it has one. On a path of \(A\), the placed vertices form an initial part. On a path \(P_l\) of \(B\), they need not form one stretch. There the vertices with \(\kappa<\alpha\) form a final part, and the plateau lies just before it. Of the plateau, the keys have placed only an initial part, possibly none.

Let \(v\in K(\alpha,\beta)\) lie on another path, with bounds \(p^*\) and \(q^*\) on \(P_l\). A placed vertex from \(p^*\) to \(q^*\), processed with its neighbours, meets the demand of \(v\) on \(P_l\): with one of its neighbours, it gives two consecutive processed vertices in this range. By \eqref{eq:ov-kappa}, \(\kappa(v)\) lies between \(\kappa(p^*)\) and \(\kappa(q^*)\), and \(\kappa(v)\le\alpha\). A bound with \(\kappa<\alpha\) is placed. Otherwise \(\kappa(v)=\alpha=\min\{\kappa(p^*),\kappa(q^*)\}\). If \(\kappa(p^*)=\alpha\), the secondary key gives \(\kappa'(p^*)\le\kappa'(v)\le\beta\), so \(p^*\) is placed. There remains
\begin{align*}
\kappa(q^*)=\alpha<\kappa(p^*).
\end{align*}
As \(\kappa\) does not decrease along a path of \(A\), this happens only on a path \(P_l\) of \(B\), with \(p^*\) before the plateau and \(q^*\) in it. If some vertex of the plateau is placed, then \(y_l\) is placed and lies between \(p^*\) and \(q^*\). If none is placed, then no vertex from \(p^*\) to \(q^*\) is placed, but
\begin{align*}
p^*\le y_l^-<y_l\le q^*\qquad\text{in the order of } P_l.
\end{align*}
So the order processes early the vertices of
\begin{align*}
Z(\alpha,\beta)\coloneqq K(\alpha,\beta)\cup N\bigl(K(\alpha,\beta)\bigr)\cup\bigcup_l\{y_l^-,y_l\}.
\end{align*}
Here \(N(\cdot)\) adds the neighbours on the paths, for the neighbour demands, and the union runs over the paths with a non-empty \(\alpha\)-plateau. Beyond \(K(\alpha,\beta)\), the set has at most three vertices on each path. They are one neighbour of the part with \(\kappa<\alpha\), the vertex \(y_l^-\), and the successor of the placed part of the plateau, or \(y_l\) when none of it is placed.

Consider a moment between two consecutive pairs of keys \((\alpha,\beta)\) and \((\alpha',\beta')\), and let \(Q\) be the set of processed vertices, and \(\held(Q)\) the set of held ones. The sets \(Z\) only grow (Part~1 of \cref{lem:stage}), so
\begin{align*}
Z(\alpha,\beta)\subseteq Q\subseteq Z(\alpha',\beta').
\end{align*}
No vertex of \(K(\alpha,\beta)\) is held. Its neighbour demand is met by \(N(K(\alpha,\beta))\). On every other path on which it has bounds, its demand is met by a placed vertex between its bounds, processed with its neighbours, or by the pair \(y_l^-,y_l\) (\cref{lem:stage}). So the held vertices satisfy
\begin{align*}
\held(Q)\subseteq Z(\alpha',\beta')\setminus K(\alpha,\beta)
=\bigl(Z(\alpha',\beta')\setminus K(\alpha',\beta')\bigr)\cup\bigl(K(\alpha',\beta')\setminus K(\alpha,\beta)\bigr).
\end{align*}
The first set has at most three vertices on each path. The second consists of the vertices with the keys \((\alpha',\beta')\), at most one on each path. Hence (\cref{lem:sortable})
\begin{align*}
\operatorname{width}\le\underbrace{3k}_{\text{processed early}}+\underbrace{k}_{\text{keys }(\alpha',\beta')}=4k.
\end{align*}
Our task is therefore to find selectors.

\subsection{When a group has a selector}
\label{sec:ov-selectors}

Let \(f_i\) and \(e_i\) be the first and the last vertex of \(P_i\). For the whole solution, these are \(s_i\) and \(t_i\); the notation also serves for sub-configurations. A natural candidate for a selector for \(A\) measures the length from the first vertex of every path of \(A\), and takes the minimum:
\begin{align*}
L(v)\coloneqq\min_{i\in A}\bigl(\pi_i(f_i)+\theta_i+\len(f_i,v)\bigr).
\end{align*}
Property~1 holds for all offsets, by the triangle inequality for the length: if the term of \(i\) is the smallest at \(u\), then
\begin{align*}
L(v)\le\pi_i(f_i)+\theta_i+\len(f_i,u)+\len(u,v)=L(u)+\len(u,v).
\end{align*}
For Property~2, let \(v\) lie in the range of \(P_j\). Then \(f_j\preceq_jv\preceq_je_j\), and by \eqref{eq:ov-exact} the term of \(j\) is
\begin{align*}
\pi_j(f_j)+\theta_j+\len(f_j,v)=\pi_j(v)+\theta_j.
\end{align*}
So Property~2 asks that the term of \(j\) be the smallest on the range of \(P_j\).

It suffices to ask this at the last vertex \(e_j\). This gives the constraints
\begin{align}\label{eq:ov-constraints}
\theta_j+\pi_j(e_j)\le\theta_i+\pi_i(f_i)+\len(f_i,e_j)\qquad(i,j\in A).
\end{align}
Indeed, for a vertex \(v\) in the range of \(P_j\), the triangle inequality and \eqref{eq:ov-exact} give
\begin{align*}
\len(f_i,e_j)\le\len(f_i,v)+\len(v,e_j)=\len(f_i,v)+\pi_j(e_j)-\pi_j(v),
\end{align*}
and \eqref{eq:ov-constraints} then bounds the term of \(j\) at \(v\) by the term of \(i\).

The constraints \eqref{eq:ov-constraints} are difference constraints on the offsets~\cite[Section~24.4]{clrs09}. They have a solution unless they contradict each other along some cycle \(i\to\sigma(i)\to\sigma^2(i)\to\dots\to i\), where \(\sigma\) is a cyclic permutation of a set \(C\subseteq A\). Adding the constraints along the cycle cancels the offsets. By \eqref{eq:ov-exact}, \(\pi_i(e_i)-\pi_i(f_i)=\len(f_i,e_i)\), so the sum reads
\begin{align*}
\sum_{i\in C}\len(f_i,e_i)\le\sum_{i\in C}\len(f_i,e_{\sigma(i)}).
\end{align*}
The \emph{deficit} of \(\sigma\) is the difference of the two sides,
\begin{align*}
\delta_\sigma\coloneqq\sum_{i\in C}\bigl(\len(f_i,e_{\sigma(i)})-\len(f_i,e_i)\bigr).
\end{align*}
We call \(\sigma\) \emph{cheaper} if \(\delta_\sigma<0\). Then matching the first vertex of every path to the last vertex of the next path is shorter, in length, than matching every path to itself. The constraints contradict each other along a cycle exactly when its permutation is cheaper. So a group without a cheaper cyclic permutation has a selector (\cref{lem:matching}). For two paths, the only cyclic permutation exchanges their ends.

\subsection{Cheaper permutations do not repeat}
\label{sec:ov-repeat}

The rest of the proof deals with configurations that are not sortable. All notions above make sense for a sub-configuration \(Y\) (\cref{sec:ov-task}). Then \(f_i\) and \(e_i\) are its first and last vertex on \(P_i\), and \(\delta_\sigma(Y)\) is the deficit. A sub-configuration may have no vertices on some paths. Such a path asks nothing of a selector, since its range is empty. Cyclic permutations are taken over the paths on which the sub-configuration has vertices.

Two facts carry over. A sub-configuration of a sortable configuration is sortable, because selectors restrict to it. And by \cref{sec:ov-selectors}, a sub-configuration \(Y\) that is not sortable has, for every split of the indices into \(A\) and \(B\), a cyclic permutation \(\sigma\) inside \(A\) or inside \(B\) with
\begin{align*}
\delta_\sigma(Y)<0.
\end{align*}
The key fact is that one permutation cannot be cheaper on many sub-configurations that \emph{follow one another} along its paths. This means that on each of these paths, all vertices of one of them come before all vertices of the next.

The exchange \(\sigma\) of two paths shows why. Let \(P_1\) and \(P_2\) be in one group, and let \(Y'\) come after \(Y\) on both paths. Write \(f_i,e_i\) and \(f_i',e_i'\) for their first and last vertices on \(P_i\). Compare two routes from \(f_1\) to \(e_1'\), adding up the lengths between consecutive vertices of the route. The first route stays on \(P_1\); by \eqref{eq:ov-exact}, its length is the increase of \(\pi_1\). The second changes to \(P_2\) inside \(Y\) and back to \(P_1\) inside \(Y'\); by \eqref{eq:ov-len}, each of its lengths is at least the increase of \(\pi_1\). So
\begin{align*}
\len(f_1,e_1)+\len(e_1,f_1')+\len(f_1',e_1')&=\pi_1(e_1')-\pi_1(f_1),\\
\len(f_1,e_2)+\len(e_2,f_2')+\len(f_2',e_1')&\ge\pi_1(e_1')-\pi_1(f_1).
\end{align*}
The two routes from \(f_2\) to \(e_2'\) satisfy the same relations, with the paths exchanged.

Add the two inequalities and subtract the two equations. The lengths from \(e_1\) to \(f_1'\) and from \(e_2\) to \(f_2'\) cancel, and what remains is
\begin{align*}
&\bigl(\len(f_1,e_2)+\len(f_2,e_1)-\len(f_1,e_1)-\len(f_2,e_2)\bigr)\\
&\qquad+\bigl(\len(f_1',e_2')+\len(f_2',e_1')-\len(f_1',e_1')-\len(f_2',e_2')\bigr)\ge0,
\end{align*}
that is,
\begin{align*}
\delta_\sigma(Y)+\delta_\sigma(Y')\ge0.
\end{align*}
So the exchange is not cheaper on both \(Y\) and \(Y'\).

For a cyclic permutation \(\sigma\) of \(m\) paths, the same argument uses \(m\) sub-configurations \(Y^1,\dots,Y^m\) that follow one another. The route that starts on \(P_i\) moves from \(P_{\sigma^{t-1}(i)}\) to \(P_{\sigma^t(i)}\) inside \(Y^t\), and it is back on \(P_i\) after \(Y^m\). The result is (\cref{lem:repeat})
\begin{align}\label{eq:ov-repeat}
\sum_{t=1}^m\delta_\sigma(Y^t)\ge0.
\end{align}
So \(\sigma\) is not cheaper on all \(m\) of them.

Two sub-configurations are \emph{separated} if, on every path, all vertices of one of them come before all vertices of the other. Which one comes first may differ from path to path. A \emph{packing} is a family of pairwise separated sub-configurations that are not sortable. We call the largest size of a packing in a configuration its \emph{packing number} \(\nu\). It is bounded in terms of \(k\) alone (\cref{lem:packing}):
\begin{align}\label{eq:ov-packing}
\nu<N_k\coloneqq\underbrace{2^k}_{\text{sets of paths}}\cdot\bigl(\underbrace{k^{k+1}}_{\text{members left}}\bigr)^{2^{k-1}}.
\end{align}
The proof starts from a packing with \(N_k\) members and keeps fewer and fewer of them, in three steps:
\begin{align*}
N_k
\ \xrightarrow{\ \text{Step 1}\ }\
\frac{N_k}{2^k}=\bigl(k^{k+1}\bigr)^{2^{k-1}}
\ \xrightarrow{\ \text{Step 2}\ }\
k^{k+1}
\ \xrightarrow{\ \text{Step 3}\ }\
k.
\end{align*}

Step~1 keeps members that have vertices on the same set of paths; there are at most \(2^k\) such sets. Steps~2 and~3 compare the members along these paths. A permutation that is cheaper on several members must involve only paths on which all of them have vertices.

Step~2 makes the members follow one another. List them in their order along one of their paths. The theorem of Erd\H{o}s and Szekeres~\cite{es35} finds, in every sequence of \(N\) distinct numbers, a monotone subsequence of length at least \(\sqrt N\). We apply it once for each of the at most \(k-1\) other paths, to the places of the listed members along that path. Each application keeps at least the square root of the number of members, so at least
\begin{align*}
\Bigl(\frac{N_k}{2^k}\Bigr)^{1/2^{k-1}}=k^{k+1}
\end{align*}
members are left after the at most \(k-1\) applications. They follow one another forwards along some paths and backwards along the others. Put the paths of the first kind into \(A\) and those of the second kind into \(B\). Paths on which these members have no vertices ask nothing of a selector, and we put them into \(A\).

Step~3 keeps members with the same permutation. Every member has a cheaper cyclic permutation inside \(A\) or inside \(B\). There are at most \(k^k\) cyclic permutations of subsets of \(\{1,\dots,k\}\) (\cref{lem:packing}), so some permutation \(\sigma\) is cheaper on
\begin{align*}
\frac{k^{k+1}}{k^k}=k\ge\lvert C\rvert
\end{align*}
of the members, where \(C\) is the set that \(\sigma\) permutes. List these forwards if \(\sigma\) permutes paths of \(A\), and backwards otherwise. Then they follow one another along all paths of \(\sigma\). If \(\sigma\) permutes \(m\le k\) paths, then \(m\) of them contradict \eqref{eq:ov-repeat}.

\subsection{Cutting around a ball that is not sortable}
\label{sec:ov-cuts}

Let \(W(\nu)\) be the largest width of a configuration with \(k\) paths and packing number at most \(\nu\). Here sub-configurations count as configurations, so that the parts cut out below are covered. A configuration with packing number \(0\) is sortable, so \(W(0)\le4k\). The memory bound follows by induction from
\begin{align}\label{eq:ov-step}
W(\nu)\le W(\nu-1)+c_k\qquad(\nu\ge1),
\end{align}
where \(c_k\) is a polynomial in \(k\). To prove \eqref{eq:ov-step}, we cut a configuration with packing number \(\nu\) into cells and process the cells one after another. Three things are needed, and the rest of the overview supplies them in turn.
\begin{itemize}
\item {\bf Smaller cells.} Every cell has packing number at most \(\nu-1\), and hence width at most \(W(\nu-1)\) (this subsection).
\item {\bf A rule for parts.} Processing a set of vertices part by part costs, beyond the width of a single part, twice the number of vertices that have to be processed early at a cut (\cref{sec:ov-witnesses}).
\item {\bf Cheap cuts.} The cuts can be chosen so that this number is polynomial in \(k\) (\cref{sec:ov-share,sec:ov-count}).
\end{itemize}

The cuts are geometric. The \emph{position} of a vertex is the vector of its potentials, up to adding the same number to all of them. The \emph{distance} of two positions is the largest change of a pair coordinate:
\begin{align*}
\lVert y-x\rVert\coloneqq\max_{a\ne b}\bigl\lvert(\pi_a-\pi_b)(y)-(\pi_a-\pi_b)(x)\bigr\rvert.
\end{align*}
Along a path \(P_i\), the distance from a vertex only grows (\cref{lem:distance}). Indeed, from \(y\) to a later vertex \(z\) of \(P_i\), no own coordinate of \(i\) decreases, and every pair coordinate is a difference of own coordinates of \(i\). So \(\lVert z-y\rVert\) is the largest increase of an own coordinate of \(i\), which grows with \(z\). If \(z\) lies between \(y\) and \(y'\) on a path, in this order, then
\begin{align}\label{eq:ov-between}
\lVert z-y\rVert\le\lVert y'-y\rVert.
\end{align}

Choose a vertex \(o\) and a radius \(r\) such that the ball
\begin{align*}
Y^*\coloneqq\{v : \lVert v-o\rVert\le r\}
\end{align*}
is a sub-configuration that is not sortable, with \(r\) as small as possible over all choices of \(o\). Such a ball exists. Indeed, the packing number is at least \(1\), so the configuration is not sortable, and a large ball around any vertex is the whole configuration.

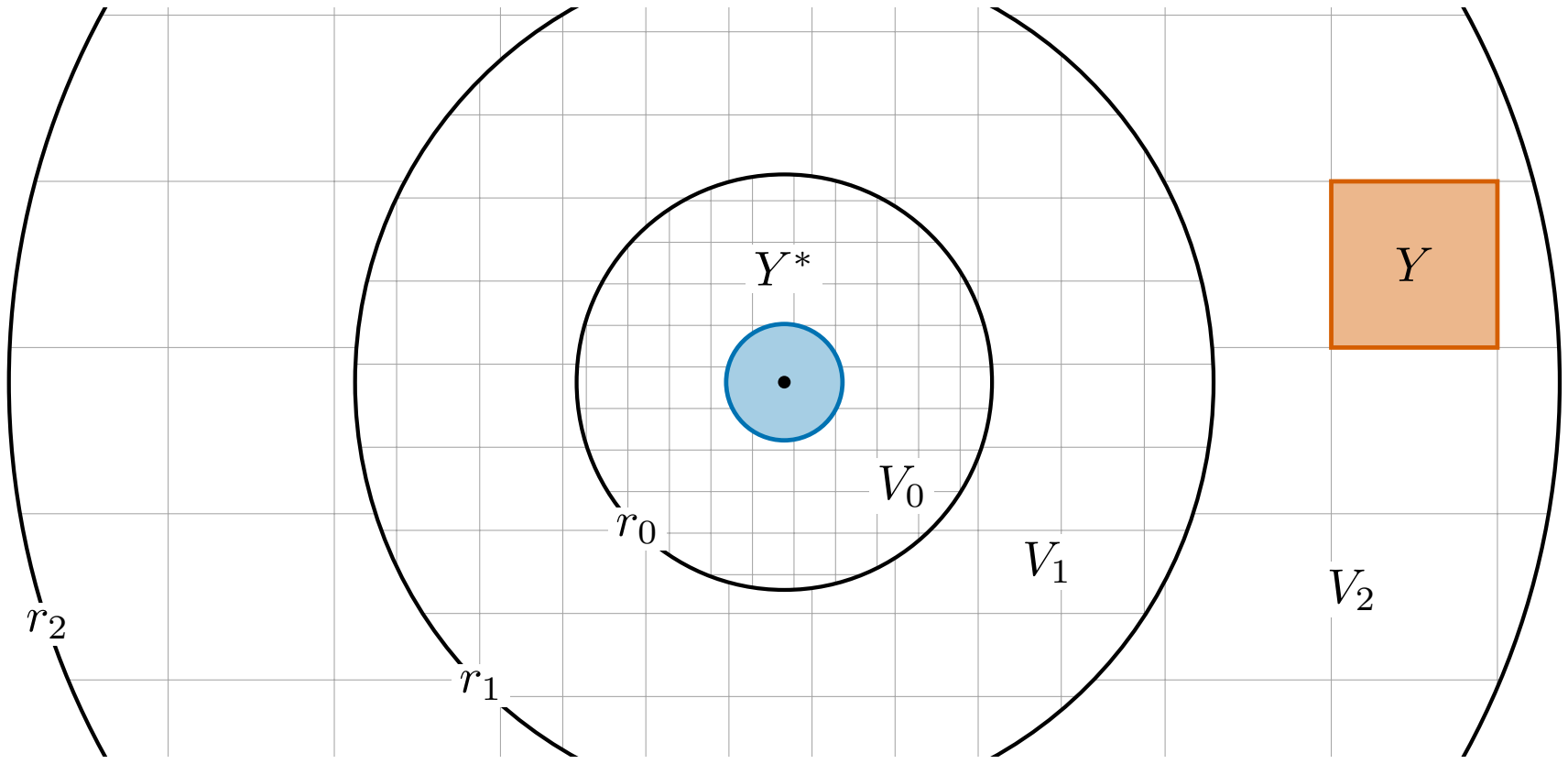
\begin{figure}[t]
\centering
\begin{tikzpicture}[x=1cm,y=1cm,font=\small]
\begin{scope}
\clip (-6.2,-2.7) rectangle (6.2,2.7);
\begin{scope}
\clip (0,0) circle (1.5);
\draw[gray!70,very thin,xshift=0.07cm,yshift=0.11cm,step=0.3] (-2,-2) grid (2,2);
\end{scope}
\begin{scope}[even odd rule]
\clip (0,0) circle (3.1) (0,0) circle (1.5);
\draw[gray!70,very thin,xshift=0.2cm,yshift=0.13cm,step=0.6] (-4.2,-4.2) grid (4.2,4.2);
\end{scope}
\begin{scope}[even odd rule]
\clip (0,0) circle (5.6) (0,0) circle (3.1);
\draw[gray!70,very thin,xshift=0.35cm,yshift=0.25cm,step=1.2] (-7.2,-7.2) grid (7.2,7.2);
\fill[DSPOrange!45] (3.95,0.25) rectangle (5.15,1.45);
\draw[DSPOrange,line width=0.9pt] (3.95,0.25) rectangle (5.15,1.45);
\end{scope}
\draw[line width=0.8pt] (0,0) circle (1.5);
\draw[line width=0.8pt] (0,0) circle (3.1);
\draw[line width=0.8pt] (0,0) circle (5.6);
\fill[DSPBlue!35] (0,0) circle (0.42);
\draw[DSPBlue,line width=0.9pt] (0,0) circle (0.42);
\fill (0,0) circle (1.3pt);
\end{scope}
\node[fill=white,inner sep=1.5pt] at (0,0.82) {\(Y^*\)};
\node[inner sep=1.5pt] at (4.55,0.85) {\(Y\)};
\node[fill=white,inner sep=1.5pt] at (-1.06,-1.06) {\(r_0\)};
\node[fill=white,inner sep=1.5pt] at (-2.19,-2.19) {\(r_1\)};
\node[fill=white,inner sep=1.5pt] at (-5.32,-1.75) {\(r_2\)};
\node[fill=white,inner sep=1.5pt] at (0.85,-0.75) {\(V_0\)};
\node[fill=white,inner sep=1.5pt] at (1.9,-1.3) {\(V_1\)};
\node[fill=white,inner sep=1.5pt] at (4.1,-1.5) {\(V_2\)};
\end{tikzpicture}
\caption{The cuts in the proof of \cref{thm:width-packing}, not to scale. The ball \(Y^*\) around the vertex \(o\) is a smallest ball that is not sortable. Spheres of radii \(r_0<r_1<\dots\) around \(o\) cut the vertices into the block \(V_0\) and the shells \(V_1,V_2,\dots\), and each of them is cut by a grid whose spacing grows with the radius. A cell of \(V_0\) is smaller than \(Y^*\) and therefore sortable. A cell \(Y\) of a shell is small compared with its distance from \(Y^*\), and then \(Y\) and \(Y^*\) are separated.}
\label{fig:shells}
\end{figure}

Spheres around \(o\) cut the vertices into \emph{blocks}, the block \(V_0\) and the shells \(V_1,V_2,\dots\):
\begin{align*}
V_0\coloneqq\{v : \lVert v-o\rVert<r_0\},
\qquad
V_j\coloneqq\{v : r_{j-1}\le\lVert v-o\rVert<r_j\}.
\end{align*}
The radii satisfy \(2^{j+2}r<r_j<2^{j+3}r\), and \cref{sec:ov-count} chooses them exactly. A grid cuts every block into \emph{cells}. Its walls are level sets of the \(k-1\) pair coordinates \(\pi_a-\pi_k\), \(a<k\), at equal spacing \(h\). Since \(\pi_a-\pi_b=(\pi_a-\pi_k)-(\pi_b-\pi_k)\), two vertices of one cell have distance less than \(2h\). The spacing grows with the block,
\begin{align*}
h=
\begin{cases}
r/2 & \text{in } V_0,\\
2^{j-1}r & \text{in } V_j\ (j\ge1),
\end{cases}
\end{align*}
so a cell has diameter less than \(r\) in \(V_0\), and less than \(2^jr\) in \(V_j\) (\cref{fig:shells}).

Every cell \(Y\) has a smaller packing number than the whole configuration. Let \(Y\) be a cell of \(V_0\), and \(y\) one of its vertices. Then, for some \(r'<r\),
\begin{align*}
Y\subseteq\{v : \lVert v-y\rVert\le r'\}.
\end{align*}
This ball is sortable by the choice of \(r\). So \(Y\) is sortable, and \(\nu(Y)=0\). Now let \(Y\) be a cell of a shell \(V_j\). Then, for all \(u\in Y^*\) and \(y,y'\in Y\),
\begin{align*}
\lVert u-o\rVert\le r,
\qquad
\lVert y-o\rVert>2^{j+1}r,
\qquad
\lVert y'-y\rVert<2^jr.
\end{align*}
Suppose that a vertex \(y\) of \(Y\) lay on a path between two vertices \(u,u'\) of \(Y^*\), or a vertex \(u\) of \(Y^*\) between two vertices \(y,y'\) of \(Y\). Then \eqref{eq:ov-between} would give, respectively,
\begin{align*}
\lVert y-o\rVert\le\lVert u'-u\rVert+\lVert u-o\rVert\le3r,
\qquad
\lVert y-o\rVert\le\lVert y'-y\rVert+\lVert u-o\rVert<2^jr+r.
\end{align*}
Both contradict \(\lVert y-o\rVert>2^{j+1}r\). So \(Y\) and \(Y^*\) are separated. A packing in \(Y\), together with \(Y^*\), is then a packing in the whole configuration (\cref{lem:far-separated,thm:width-packing}):
\begin{align*}
\nu(Y)+1\le\nu.
\end{align*}
This is the first of the three requirements.

\subsection{Processing the cells one after another}
\label{sec:ov-witnesses}

When the cells are processed one after another, a processed vertex \(v\) may be held by a path that comes near it only in a much later cell. We therefore fix in advance, for every vertex \(v\) and every other path \(P_l\) whose range contains \(v\), two consecutive vertices \(z,z'\) of \(P_l\) with
\begin{align}\label{eq:ov-witness}
v\npreceq_lz
\qquad\text{and}\qquad
z'\npreceq_lv.
\end{align}
Once \(z\) and \(z'\) are processed, \(P_l\) no longer holds \(v\). Indeed, an unprocessed stretch of \(P_l\) between \(g^-\) and \(g^+\) then lies before \(z\) or after \(z'\). If \(P_l\) held \(v\) through this stretch, then \(g^-\preceq_lv\preceq_lg^+\), and so
\begin{align*}
v\preceq_lg^+\preceq_lz
\qquad\text{or}\qquad
z'\preceq_lg^-\preceq_lv,
\end{align*}
which contradicts \eqref{eq:ov-witness}. In the formal notion, \eqref{eq:ov-witness} says that \(z'\) comes after \(p^*\) and \(z\) before \(q^*\). So \(z\) and \(z'\) are two consecutive vertices from \(p^*\) to \(q^*\). We call \(z\), \(z'\) and the neighbours of \(v\) on its own path the \emph{witnesses} of \(v\) (\cref{def:witness,lem:witness}). \Cref{sec:ov-share} says how \(z\) and \(z'\) are chosen. We write \(\wit(S)\) for the set of witnesses of the vertices of a set \(S\).

Witnesses make it possible to finish a set \(S\) of vertices before the rest. We call the following set the \emph{interface} of \(S\):
\begin{align*}
\partial S\coloneqq\bigl(\wit(S)\setminus S\bigr)\cup\{v\in S : N(v)\not\subseteq S\}.
\end{align*}
Once \(S\) and \(\partial S\) are processed, all witnesses of the vertices of \(S\) are processed. Then no vertex of \(S\) is held, whatever happens outside. The vertices of the second kind mark the places where a path leaves \(S\). At such a place, two consecutive vertices of the path lie one inside \(S\) and one outside. Both belong to \(\partial S\), because the outer one is a neighbour, hence a witness, of the inner one. We need them when we treat one part of a larger set as a configuration of its own.

Now let a set \(X\) of vertices be divided into parts \(U_1,\dots,U_t\), and write
\begin{align*}
S_j\coloneqq U_1\cup\dots\cup U_j,
\qquad
S_0\coloneqq\emptyset.
\end{align*}
Here \(X\) is a configuration of its own: its orders process only vertices of \(X\), and its witnesses are vertices of \(X\). The order goes through the parts. For the part \(U_j\), it first processes the unprocessed vertices of \(\partial S_{j-1}\) and \(\partial S_j\). It then processes the vertices of \(U_j\) in an order of the smallest width.

For a sub-configuration \(Y\) and a set \(Q\) of vertices, \(\held_Y(Q)\) is the set of vertices held when \(Y\) is taken alone and exactly the vertices of \(Q\) are processed. Consider a moment after the two interfaces have been processed; the proof of \cref{lem:serial} treats the moments before. Let \(Q\) be the set of processed vertices, and \(Q'\) the processed initial part of the chosen order of \(U_j\). The other processed vertices of \(U_j\) lie in the two interfaces, and the order of \(U_j\) has the smallest width:
\begin{align*}
Q\cap U_j\subseteq Q'\cup\partial S_{j-1}\cup\partial S_j,
\qquad
\lvert\held_{U_j}(Q')\rvert\le\operatorname{width}(U_j).
\end{align*}
We claim that
\begin{align*}
\held_X(Q)\subseteq\partial S_{j-1}\cup\partial S_j\cup\held_{U_j}(Q').
\end{align*}
Held vertices are processed, so \(\held_X(Q)\subseteq Q\). The claim then follows from three facts, for the earlier parts, the vertices outside \(S_j\), and the current part:
\begin{align*}
\held_X(Q)\cap S_{j-1}&=\emptyset,\\
Q\setminus S_j&\subseteq\partial S_{j-1}\cup\partial S_j,\\
\bigl(\held_X(Q)\cap U_j\bigr)\setminus\bigl(\partial S_{j-1}\cup\partial S_j\bigr)&\subseteq\held_{U_j}(Q').
\end{align*}
The first holds because the vertices of the earlier parts have all their witnesses processed. The second holds because, outside \(S_j\), the order has processed only witnesses of vertices of \(S_{j-1}\), which lie in \(\partial S_{j-1}\), and vertices of \(\partial S_j\).

For the third, let \(v\) be a held vertex of \(U_j\) outside the two interfaces. By the first inclusion above, \(v\in Q'\). Then
\begin{align*}
v\in\held_X(Q)
\quad\implies\quad
v\in\held_{U_j}(Q\cap U_j)
\quad\implies\quad
v\in\held_{U_j}(Q').
\end{align*}
The first step is \cref{lem:local}. It needs two conditions: the witnesses outside \(U_j\) of the vertices of \(U_j\) are processed, and so are the two vertices at every place where a path leaves \(U_j\). Both hold:
\begin{align*}
\wit(U_j)\setminus U_j\subseteq S_{j-1}\cup\partial S_j\subseteq Q,
\end{align*}
and where a path leaves \(U_j\), it leaves \(S_{j-1}\) or \(S_j\), so the two vertices lie in \(\partial S_{j-1}\) or \(\partial S_j\). Under these conditions, a demand of \(v\) that is not met in \(X\) involves only vertices of \(U_j\). Indeed, an unprocessed witness of \(v\) lies in \(U_j\), and so does the whole stretch from \(p^*\) to \(q^*\) around it, where \(p^*\) and \(q^*\) are the bounds of \(v\) in \(X\). The stretch could leave \(U_j\) only at two consecutive vertices that are both processed, and an unmet demand has no two such vertices. The second step holds because \(v\in Q'\subseteq Q\cap U_j\). A processed vertex is held as long as certain other vertices are unprocessed, so it stays held when fewer vertices are processed.

The claim gives (\cref{lem:serial})
\begin{align}\label{eq:ov-serial}
\lvert\partial S_j\rvert\le s\ \text{ for all } j
\quad\implies\quad
\operatorname{width}(X)\le\max_j\operatorname{width}(U_j)+2s.
\end{align}
The bound does not grow with the number of parts. This is the second requirement.

We use \eqref{eq:ov-serial} twice. For the first use, \(X\) is one block and the parts are its cells. The witnesses chosen for the whole configuration can lie outside the block. So we choose witnesses afresh among the vertices of the block, by the rule of \cref{sec:ov-share}. Nothing else changes: the width is defined without witnesses, which only serve to construct an order. For the second use, \(X\) is the whole configuration and the parts are the blocks. The sets \(S_j\) are then the balls around \(o\) that the spheres bound. Suppose that every union of cells of a block has an interface of at most \(q\) vertices, and every such ball one of at most \(s\) vertices. Every cell has width at most \(W(\nu-1)\). So the two uses give, for every block \(V_j\) and then for the configuration,
\begin{align*}
\operatorname{width}(V_j)&\le W(\nu-1)+2q,\\
W(\nu)&\le\max_j\operatorname{width}(V_j)+2s\le W(\nu-1)+\underbrace{2q}_{\text{cells of a block}}+\underbrace{2s}_{\text{blocks}}.
\end{align*}
This is \eqref{eq:ov-step}, once spheres and grids are found for which \(q\) and \(s\) are polynomials in \(k\). That is the third requirement.

\subsection{Witnesses that nearby vertices share}
\label{sec:ov-share}

The size of the interface of a ball depends on where the witnesses are placed. We place them so that vertices close to one another have the same witnesses.

Draw \(P_l\) as a polygonal curve through the positions of its vertices. A point of the curve has pair coordinates, by linear interpolation along its segment. So it has a progress \(\varphi_l\) (\cref{sec:ov-groups}) and distances to other points. No own coordinate of \(l\) decreases along the curve, and the position changes only if some own coordinate of \(l\) changes, since \(\pi_a-\pi_b=(\pi_l-\pi_b)-(\pi_l-\pi_a)\). So the progress increases along the curve, strictly wherever the position changes. Let \(v\) be a vertex of another path in the range of \(P_l\), so that \(y\preceq_lv\preceq_ly'\) for two vertices \(y,y'\) of \(P_l\). The progress is a sum of own coordinates of \(l\), so
\begin{align*}
\varphi_l(y)\le\varphi_l(v)\le\varphi_l(y').
\end{align*}
Hence the curve has exactly one point \(\hat v\) with the progress of \(v\). We call \(\hat v\) the \emph{passing point} of \(v\), and
\begin{align*}
d_v\coloneqq\lVert\hat v-v\rVert
\end{align*}
its \emph{passing distance}, for the fixed path \(P_l\). If \(d_v=0\), then \(v\) lies on the curve. By our assumption that different vertices have different pair coordinates (\cref{sec:ov-task}), it lies strictly between two consecutive vertices of \(P_l\). These two vertices are its witnesses. They satisfy \eqref{eq:ov-witness}, since no own coordinate of \(l\) decreases along the curve and the three positions differ.

Now let \(d_v>0\). Let \(\epsilon_j\) be the change of the own coordinate \(\pi_l-\pi_j\) from \(v\) to \(\hat v\), and \(\epsilon_l\coloneqq0\). A pair coordinate \(\pi_a-\pi_b=(\pi_l-\pi_b)-(\pi_l-\pi_a)\) then changes by \(\epsilon_b-\epsilon_a\). The points \(v\) and \(\hat v\) have the same progress and distance \(d_v\), which reads
\begin{align*}
\sum_j\epsilon_j=0,
\qquad
\max_j\epsilon_j-\min_j\epsilon_j=d_v.
\end{align*}
Bounding \(k-1\) of the \(k\) numbers by the largest and one by the smallest gives
\begin{align*}
0=\sum_j\epsilon_j\le(k-1)\max_j\epsilon_j+\min_j\epsilon_j=k\max_j\epsilon_j-d_v.
\end{align*}
So some \(\epsilon_a\) is at least \(d_v/k\). In the same way, some \(\epsilon_b\) is at most \(-d_v/k\). Here \(a,b\ne l\), because \(\epsilon_l=0\).

Let \(z\) and \(z'\) be two consecutive vertices of \(P_l\) with \(\hat v\) on the segment between them. Let \(u\) be a vertex with \(\lVert u-v\rVert<d_v/k\). No own coordinate of \(l\) decreases from \(z\) to \(\hat v\) to \(z'\). Every pair coordinate differs by less than \(d_v/k\) between \(u\) and \(v\). So
\begin{align*}
(\pi_l-\pi_a)(z')&\ge(\pi_l-\pi_a)(\hat v)\ge(\pi_l-\pi_a)(v)+d_v/k>(\pi_l-\pi_a)(u),\\
(\pi_l-\pi_b)(z)&\le(\pi_l-\pi_b)(\hat v)\le(\pi_l-\pi_b)(v)-d_v/k<(\pi_l-\pi_b)(u).
\end{align*}
This is \eqref{eq:ov-witness} for \(u\) (\cref{lem:passing}):
\begin{align}\label{eq:ov-share}
\lVert u-v\rVert<d_v/k
\quad\implies\quad
u\npreceq_lz
\quad\text{and}\quad
z'\npreceq_lu.
\end{align}

Nearby vertices can therefore share their witnesses, and they have to. If every vertex used the witnesses at its own passing point, many vertices close to one another could have different witnesses across the same cut. All of them would enter the interface, and the term \(2s\) of \eqref{eq:ov-serial}. For every two paths \(P_i\) and \(P_l\), we divide the vertices of \(P_i\) that lie in the range of \(P_l\), with a positive passing distance, into \emph{clusters}. All members of a cluster use the witnesses at the passing point \(\hat w\) of one member \(w\), its \emph{representative}. Every such vertex \(v\) gets an interval of the progress along its own path \(P_i\),
\begin{align}\label{eq:ov-room}
I(v)\coloneqq\Bigl[\varphi_i(v)-\frac{d_v}{4k},\ \varphi_i(v)+\frac{d_v}{4k}\Bigr].
\end{align}
The clusters are formed greedily, by decreasing passing distance. A vertex becomes a representative if its interval is disjoint from the intervals of all earlier representatives. Otherwise, it joins the cluster of an earlier representative \(w\) whose interval meets its own (\cref{sec:witness-choice}). This gives two properties.

First, the intervals of the representatives are disjoint. So each cluster has an interval of the progress along \(P_i\) to itself, of length proportional to its passing distance. \Cref{sec:ov-count} uses this room to count the clusters near a sphere.

Second, a member \(v\) of the cluster of \(w\) has \(d_v\le d_w\), and its interval meets \(I(w)\). Along \(P_i\), the distance is at most the change of the progress (\cref{lem:distance}). So
\begin{align}\label{eq:ov-cluster}
\lVert v-w\rVert\le\lvert\varphi_i(v)-\varphi_i(w)\rvert\le\frac{d_v}{4k}+\frac{d_w}{4k}\le\frac{d_w}{2k}<\frac{d_w}{k}.
\end{align}
By \eqref{eq:ov-share} for \(w\), the witnesses at \(\hat w\) serve every member of its cluster.

\subsection{A sphere with a small interface}
\label{sec:ov-count}

It remains to find the cuts. Fix \(R>0\), and let \(S_c\) be the set of vertices within distance less than \(c\) of \(o\). We choose \(c\) uniformly at random between \(R\) and \(2R\), and bound the expected size of \(\partial S_c\). Its vertices are of two kinds (\cref{lem:across}).

The first kind are the ends of segments of the curves that cross the sphere of radius \(c\): they have a point inside the sphere and a point outside it, where outside means not inside. Such a segment meets the sphere. Three sorts of interface vertices are of this kind.
\begin{itemize}
\item {\bf Neighbours.} A vertex inside the sphere and a neighbour of it outside are the two ends of the segment between them.
\item {\bf Vertices on a curve.} A witness outside of a vertex inside that lies on a curve (\(d_v=0\)) is an end of the segment between the two witnesses, which passes through the vertex.
\item {\bf Passing points inside.} A witness outside of a cluster whose passing point \(\hat w\) lies inside is an end of the segment between the two witnesses, which passes through \(\hat w\).
\end{itemize}

We count the segments that meet the sphere. For two points \(a,b\) of the curve of \(P_l\), by the triangle inequality and \cref{lem:distance},
\begin{align*}
\bigl\lvert\lVert a-o\rVert-\lVert b-o\rVert\bigr\rvert\le\lVert a-b\rVert\le\lvert\varphi_l(a)-\varphi_l(b)\rvert.
\end{align*}
For a segment \(e\) of this curve, let \(\lambda_e\) be the increase of the progress along the part of \(e\) inside the ball of radius \(2R\). Let \(p_e\) be the probability that the sphere meets \(e\). This happens only if \(c\) is the distance from \(o\) to a point of this part. By the display, these distances fill an interval of length at most \(\lambda_e\). As \(c\) is uniform on an interval of length \(R\),
\begin{align*}
p_e\le\frac{\lambda_e}{R}.
\end{align*}
Any two points in the ball have distance at most \(4R\). So their progress differs by at most \(4(k-1)R\) (\cref{lem:distance}). The progress increases along the curve, so the parts of different segments inside the ball have ranges of progress that overlap only at their ends. Hence, for the segments \(e\) of one path,
\begin{align*}
\sum_ep_e\le\frac1R\sum_e\lambda_e\le4(k-1).
\end{align*}
Over the \(k\) paths, the expected number of segments that meet the sphere is at most \(4k(k-1)\).

The second kind are the witnesses, at \(\hat w\), of the clusters that have a member \(v\) inside the sphere and \(\hat w\) outside. By \eqref{eq:ov-cluster},
\begin{align*}
\lVert v-\hat w\rVert\le\lVert v-w\rVert+\lVert w-\hat w\rVert\le\frac{d_w}{2k}+d_w\le\tfrac32d_w.
\end{align*}
So this happens only for \(c\) in an interval of length \(\tfrac32d_w\), that is, with probability at most
\begin{align*}
\min\Bigl\{\frac{3d_w}{2R},\ 1\Bigr\}.
\end{align*}
The clusters for which it can happen have a member in the ball of radius \(2R\). The disjoint intervals \eqref{eq:ov-room} limit how many there are of each size (\cref{lem:room}). For every two paths,
\begin{align*}
\sum_{d_w\le4kR}d_w=O(k^2R),
\qquad
\#\{w : d_w>4kR\}=O(k).
\end{align*}

For the first bound, take one member inside the ball from each of these clusters. Their progress values lie within \(4(k-1)R\) of one another. By \eqref{eq:ov-cluster}, the progress of \(w\) differs from that of a member of its cluster by at most \(d_w/(2k)\le2R\). And \(I(w)\) has length \(d_w/(2k)\le2R\). So the disjoint intervals of these clusters lie in one range, and their lengths add up to at most its length:
\begin{align*}
\sum_{d_w\le4kR}\frac{d_w}{2k}\le\underbrace{4(k-1)R}_{\text{members}}+\underbrace{2\cdot2R}_{\text{from a member to }w}+\underbrace{2R}_{\text{half-lengths}}=(4k+2)R.
\end{align*}

For the second bound, the members inside the ball have distance at most \(4R\) from one another. The passing distance changes by at most \(k\) times the distance (\cref{lem:passing}), so their passing distances differ by at most \(4kR\). The same fact and \eqref{eq:ov-cluster} give \(d_v\ge d_w/2\) for a member \(v\) of the cluster of \(w\), and \(d_v\le d_w\) by the greedy order. Let \(d_{\max}=d_w\) and \(d_{\min}=d_{w'}\) be the largest and the smallest passing distance of these clusters, with chosen members \(v\) and \(v'\). As \(d_{\min}>4kR\),
\begin{align*}
d_{\max}\le2d_v\le2(d_{v'}+4kR)\le2(d_{\min}+4kR)<4d_{\min}.
\end{align*} Their disjoint intervals have length at least \(d_{\min}/(2k)\), and lie in one range of length at most \(4(k-1)R+3d_{\max}/(2k)\). So their number is less than
\begin{align*}
\frac{4(k-1)R+3d_{\max}/(2k)}{d_{\min}/(2k)}=\frac{8k(k-1)R}{d_{\min}}+\frac{3d_{\max}}{d_{\min}}<2(k-1)+12.
\end{align*}
Together, for every two paths, the expected number of these clusters is at most
\begin{align*}
\frac{3}{2R}\sum_{d_w\le4kR}d_w+\#\{w : d_w>4kR\}=O(k^2).
\end{align*}

Now add the two kinds. A segment has two ends, a cluster has two witnesses, and there are \(k(k-1)\) ordered pairs of paths. So
\begin{align*}
\mathbb E\bigl[\lvert\partial S_c\rvert\bigr]\le\underbrace{2\cdot4k(k-1)}_{\text{segments}}+\underbrace{2\cdot k(k-1)\cdot O(k^2)}_{\text{clusters}}=O(k^4).
\end{align*}
Some radius between \(R\) and \(2R\) therefore gives a ball whose interface has at most \(s_k\coloneqq32(k+1)^4\) vertices (\cref{lem:sphere}). In the proof, \(R=2^{j+2}r\), which gives the radii \(r_j\) of \cref{sec:ov-cuts}.

Grids are treated in the same way. A wall of the grid is a level set of one of the \(k-1\) pair coordinates \(\pi_a-\pi_k\), as a sphere is a level set of the distance from \(o\). A pair coordinate also changes by at most the distance, and \cref{lem:levels} treats both cases at once. Whichever cells form a union, a vertex of its interface is an end of a segment that meets a wall, or a witness of a cluster that a wall separates from its passing point. So the set of all these vertices, over all walls, contains the interface of every union of cells.

We bound the size of this set for a random shift of the grid, in a block of diameter at most \(\Delta\). Fix one of the \(k-1\) coordinates. Let \(\mu(c)\) be the number of these vertices at the wall where the coordinate has the value \(c\). The walls in this direction are at the values \(c\in\vartheta+h\mathbb Z\), with \(\vartheta\) uniform in \([0,h)\). So
\begin{align*}
\mathbb E\Bigl[\sum_{c\in\vartheta+h\mathbb Z}\mu(c)\Bigr]=\frac1h\int \mu(c)\mathrm{d}c.
\end{align*}
The integral is bounded as for spheres. A segment is counted for the values that the coordinate takes on it. A cluster is counted only within \(\tfrac32d_w\) of the value at its passing point. On the block, the coordinate varies by at most \(\Delta\). With \(\Delta\) in place of the diameter \(4R\) of the ball, the two counts give
\begin{align*}
\int \mu(c)\mathrm{d}c\le\underbrace{2\cdot k(k-1)\Delta}_{\text{segments}}+\underbrace{2\cdot k(k-1)\cdot O(k^2\Delta)}_{\text{clusters}}=O(k^4\Delta).
\end{align*}
Over the \(k-1\) coordinates, the expected number of these vertices is at most
\begin{align*}
\underbrace{(k-1)}_{\text{directions}}\cdot\frac1h\cdot O(k^4\Delta)=O(k^5)\cdot\frac{\Delta}{h}.
\end{align*}
So a suitable shift gives a set of at most \(q_k\Delta/h\) vertices, with \(q_k\coloneqq12k^5\) (\cref{lem:grid}). In the proof, \(\Delta/h<32\).

So the third requirement holds with \(s=s_k\) and \(q=32q_k\). Hence \eqref{eq:ov-step} holds with
\begin{align*}
c_k\coloneqq1000(k+1)^5\ge\underbrace{2\cdot32q_k}_{\text{cells of a block}}+\underbrace{2s_k}_{\text{blocks}}.
\end{align*}
With \(W(0)\le4k\), the step \eqref{eq:ov-step} and the bound \eqref{eq:ov-packing}, every configuration has width at most
\begin{align*}
g(k)\coloneqq\underbrace{4k}_{\text{sortable cell}}+\underbrace{c_k}_{\text{one round of cuts}}\cdot\underbrace{N_k}_{\text{rounds}},
\end{align*}
which is \cref{thm:width}. One more stored vertex, for the moment when two segments of a path are joined, turns an order of the configuration into a run of the sweep (\cref{lem:transfer}). This gives \cref{thm:general-memory}.

\section{Preliminaries}
\label{sec:prelim}

This section fixes the model and introduces two objects that the rest of the paper works with. The shortest-path DAG of each terminal pair is all the algorithm uses. The potentials place every vertex at a point, and only the analysis uses them.

\subsection{Model and shortest-path DAGs}

The input is a digraph \(G=(V,E)\) with \(n\) vertices and \(m\) arcs, weights \(w\colon E\to\R_{>0}\), and terminal pairs \((s_1,t_1),\dots,(s_k,t_k)\). A \emph{solution} consists of pairwise vertex-disjoint paths \(P_1,\dots,P_k\) such that every \(P_i\) is a shortest \(s_i\)-\(t_i\) path. We count additions and comparisons of weights at unit cost.

Two simple reductions normalize the input. Of several parallel arcs, only a lightest one can lie on a shortest path. So we keep only that one, and \(m\le n^2\). If the terminals of two distinct pairs coincide, there is no solution. So we assume that the terminals of distinct pairs are distinct. We still allow \(s_i=t_i\).

We write \(\dist(u,v)\) for the distance from \(u\) to \(v\). The \emph{shortest-path DAG} of the pair \(i\) is the set of arcs
\begin{align*}
D_i
&\coloneqq\{(u,v)\in E : \dist(s_i,u)+w(u,v)+\dist(v,t_i)=\dist(s_i,t_i)<\infty\}.
\end{align*}
For every arc \((u,v)\), the triangle inequality gives
\begin{align*}
\dist(s_i,v)&\le\dist(s_i,u)+w(u,v),\\
\dist(s_i,t_i)&\le\dist(s_i,v)+\dist(v,t_i).
\end{align*}
For an arc of \(D_i\), all these distances are finite. Adding the two inequalities and cancelling \(\dist(s_i,v)\) gives
\begin{align*}
\dist(s_i,t_i)\le\dist(s_i,u)+w(u,v)+\dist(v,t_i).
\end{align*}
An arc of \(D_i\) attains equality here by definition. So it attains equality in each of the two inequalities. In particular, an arc \((u,v)\) of \(D_i\) satisfies
\begin{align*}
\dist(s_i,v)&=\dist(s_i,u)+w(u,v).
\end{align*}
Hence \(\dist(s_i,\cdot)\) increases strictly along the arcs of \(D_i\), and \(D_i\) is acyclic.

An \(s_i\)-\(t_i\) path is shortest exactly when all its arcs lie in \(D_i\). Indeed, every part of a shortest path is a shortest path between its ends. So every arc of a shortest \(s_i\)-\(t_i\) path satisfies the equation that defines \(D_i\). Conversely, along an \(s_i\)-\(t_i\) path with all arcs in \(D_i\), the equations of the last display add up to \(\dist(s_i,t_i)\).

We write \(u\reach{i}v\) if \(v\) is reachable from \(u\) by arcs of \(D_i\), including \(u=v\). As \(D_i\) is acyclic, this is a partial order on \(V\). The DAGs \(D_1,\dots,D_k\) and the relations \(\reach{1},\dots,\reach{k}\) are computed in \(O(kn(n+m))\) time by Dijkstra's algorithm and a search from every vertex.

\subsection{Potentials and positions}

The distances \(\dist(s_i,\cdot)\) are natural potentials, except that they are infinite at the vertices that \(s_i\) cannot reach. We cut them off at \(\dist(s_i,t_i)\). We say that an arc \((u,v)\) is \emph{tight} for a function \(f\) of the vertices if \(f(v)-f(u)=w(u,v)\).

\begin{lemma}[Potentials]\label{lem:potentials}
Suppose that \(t_i\) is reachable from \(s_i\), and put
\begin{align*}
\pi_i(v)\coloneqq\min\bigl\{\dist(s_i,v),\dist(s_i,t_i)\bigr\}.
\end{align*}
Then every arc \((u,v)\) satisfies \(\pi_i(v)-\pi_i(u)\le w(u,v)\), with equality if \((u,v)\in D_i\); that is, the arcs of \(D_i\) are tight for \(\pi_i\).
\end{lemma}

\begin{proof}
Let \(\Lambda\coloneqq\dist(s_i,t_i)\), and let \((u,v)\) be an arc. We first check the inequality. If \(\dist(s_i,u)\ge\Lambda\), then \(\pi_i(u)=\Lambda\ge\pi_i(v)\), and the inequality holds because \(w(u,v)>0\). Otherwise \(\pi_i(u)=\dist(s_i,u)\), and the triangle inequality gives
\begin{align*}
\pi_i(v)\le\dist(s_i,v)\le\dist(s_i,u)+w(u,v)=\pi_i(u)+w(u,v).
\end{align*}

It remains to show that the arcs of \(D_i\) are tight. For \((u,v)\in D_i\), the definition of \(D_i\) gives
\begin{align*}
\dist(s_i,u)+w(u,v)=\dist(s_i,v)\le\Lambda-\dist(v,t_i)\le\Lambda.
\end{align*}
So \(\pi_i=\dist(s_i,\cdot)\) at both ends, and \(\pi_i(v)-\pi_i(u)=w(u,v)\).
\end{proof}

If some \(t_i\) is not reachable from \(s_i\), there is no solution, and the algorithm of \cref{sec:sweep} rejects at once. So we assume from now on that every \(t_i\) is reachable from \(s_i\). We let \(\pi_1,\dots,\pi_k\) be the potentials of \cref{lem:potentials}. The \emph{position} of a vertex \(v\) is its vector of potentials \((\pi_1(v),\dots,\pi_k(v))\), modulo adding the same number to all coordinates. It is described by the \emph{pair coordinates}
\begin{align*}
\pi_{ij}&\coloneqq\pi_i-\pi_j\qquad(i\ne j).
\end{align*}
The \emph{displacement} from a vertex \(u\) to a vertex \(v\) is the position of \(v\) minus the position of \(u\). For an arc \((u,v)\), we call it the displacement of the arc. Like a position, a displacement is a vector \(\Delta\in\R^k\) modulo adding the same number to all coordinates. The \emph{cone} of the pair \(i\) is the set of displacements along which no pair coordinate \(\pi_{ij}\) with \(j\ne i\) decreases:
\begin{align*}
C_i
&\coloneqq\{\Delta : \Delta_i\ge\Delta_j \text{ for all } j\}.
\end{align*}
The condition does not depend on the representative of \(\Delta\). The cones are used in \cref{lem:cones} below. They are used once more in \cref{lem:linear-order}, to show that no linear function of the positions orders three pairs.

\begin{lemma}[Cones]\label{lem:cones}
Let \((u,v)\) be an arc of \(D_i\). For every \(j\ne i\), the pair coordinate \(\pi_{ij}\) does not decrease along the arc, and it is constant if \((u,v)\in D_j\). In particular the displacement of the arc lies in \(C_i\), and if \(u\reach{i}v\), then the displacement from \(u\) to \(v\) lies in \(C_i\).
\end{lemma}

\begin{proof}
Fix \(j\ne i\). The arc is tight for \(\pi_i\) and satisfies the inequality of \cref{lem:potentials} for \(\pi_j\). So
\begin{align*}
\pi_{ij}(v)-\pi_{ij}(u)&=w(u,v)-\bigl(\pi_j(v)-\pi_j(u)\bigr)\ge0.
\end{align*}
Equality holds if the arc is tight for \(\pi_j\), in particular if it lies in \(D_j\). The last statement holds because \(C_i\) is closed under addition. Indeed, if \(u\reach{i}v\), the displacement from \(u\) to \(v\) is the sum of the displacements of the arcs of a path from \(u\) to \(v\) in \(D_i\).
\end{proof}

So along a path \(P_i\) of a solution, the \(k-1\) coordinates \(\pi_{ij}\) do not decrease. Along another path \(P_l\), the coordinate \(\pi_{il}=-\pi_{li}\) does not increase. The coordinates \(\pi_{ij}\) with \(j\ne l\) are not controlled along \(P_l\).

Consider an arc \((u,v)\) of some \(D_i\) with displacement \(0\). Along it, all potentials increase by as much as \(\pi_i\), that is, by the positive weight of the arc:
\begin{align*}
\pi_j(v)-\pi_j(u)=\pi_i(v)-\pi_i(u)=w(u,v)>0\qquad(1\le j\le k).
\end{align*}
In particular, \(\pi_1\) increases strictly. The two ends of such an arc have the same position. This remark is what allows \(\pi_1\) to tell them apart in the orders of \cref{sec:config}.

\section{The sweep with bounded memory}
\label{sec:sweep}

This section describes the algorithm, for any number \(k\) of pairs. We prove that it never accepts an instance without a solution. We bound its running time by \(n^{p+O(1)}\) when it may store at most \(p\) vertices. Finally, we reduce its completeness to a property of solutions. The question is whether one of them can be built in an order that never needs more than \(p\) stored vertices.

The algorithm of Fortune, Hopcroft and Wyllie~\cite{fhw80} advances one pointer per path in a topological order. It need not remember a vertex below all pointers, which can never be reached again. Here the DAGs \(D_j\) of different pairs together may contain cycles. So the condition for forgetting refers to each of them separately.

\subsection{The algorithm}

For vertices \(a,b\) and a pair \(j\) let
\begin{align*}
\spn_j(a,b)
&\coloneqq\{u\in V\setminus\{a,b\} : a\reach{j}u\reach{j}b\}
\end{align*}
be the set of vertices strictly between \(a\) and \(b\) in the DAG \(D_j\).

\emph{States.} A state records, for every pair \(j\), a sequence of \emph{segments}
\begin{align*}
[a_1,b_1],\ [a_2,b_2],\ \dots,\ [a_r,b_r],
\qquad
a_1=s_j,\quad b_r=t_j,\quad b_l\reach{j}a_{l+1}\ \text{for } l<r.
\end{align*}
A segment \([a,b]\) stands for a processed subpath of the \(j\)-th path from \(a\) to \(b\). The state stores its two ends and those of its inner vertices that the forgetting rule below keeps. Between two consecutive segments lies the \emph{gap} \((b_l,a_{l+1})\). We write a gap as \((g^-,g^+)\) when its index does not matter. We write \(\Gamma_j(\xi)\) for the set of gaps of the pair \(j\) in a state \(\xi\).

The \emph{stored} vertices are the ends of all segments and the stored inner vertices. Let \(\Term\coloneqq\{s_1,t_1,\dots,s_k,t_k\}\) be the set of terminals. The \emph{memory} \(\mem(\xi)\) of a state \(\xi\) is the number of its stored vertices outside \(\Term\).

A segment \([u,u]\) with one vertex is also written \(\{u\}\). In the initial state, pair \(j\) has the two segments \(\{s_j\}\) and \(\{t_j\}\) with the gap \((s_j,t_j)\), or the single segment \(\{s_j\}\) if \(s_j=t_j\). If some \(t_j\) is not reachable from \(s_j\), the algorithm rejects at once.

\emph{Moves.} A move acts on one gap \((g^-,g^+)\) of one pair \(j\) and adds one vertex \(u\) to the \(j\)-th path. There are three kinds:
\begin{itemize}
\item {\bf Extend forward.} \((g^-,u)\in D_j\) and \(u\in\spn_j(g^-,g^+)\cup\{g^+\}\). The vertex \(u\) becomes the last vertex of the segment that ends at \(g^-\). The vertex \(g^-\) becomes an inner vertex unless it is also the first vertex of this segment. If \(u=g^+\), the two segments merge. In this case \(u\) also becomes an inner vertex unless it is the last vertex of the merged segment.
\item {\bf Extend backward.} \((u,g^+)\in D_j\) and \(u\in\spn_j(g^-,g^+)\cup\{g^-\}\), symmetrically.
\item {\bf Open.} \(u\in\spn_j(g^-,g^+)\) becomes a new segment \(\{u\}\) between the two.
\end{itemize}
Except when it merges two segments, a move must add a vertex that is not stored. A vertex that becomes an inner vertex of a segment stays stored until the forgetting rule below deletes it.

\emph{Forgetting.} After every move, let \(\xi\) be the new state. The algorithm deletes from \(\xi\) every stored inner vertex \(v\) of a segment of a pair \(i\) such that
\begin{align*}
v\notin\spn_j(g^-,g^+)\qquad\bigl(j\ne i,\ (g^-,g^+)\in\Gamma_j(\xi)\bigr).
\end{align*}
Such a vertex is called \emph{forgettable}. Segment ends are never deleted. The rule does not look at the gaps of the pair \(i\) itself. Indeed, the \(i\)-th path is built from arcs of the acyclic graph \(D_i\). So it cannot return to one of its own vertices (see the proof of \cref{thm:sweep}).

\emph{The sweep \(\Sweep_p\).} A move is \emph{allowed} if the memory after forgetting is at most \(p\). A state is \emph{accepting} if every pair has a single segment. The sweep \(\Sweep_p\) searches the states reachable from the initial state by allowed moves, for instance breadth first. With every state, it stores one predecessor and the move that led to it. If it reaches no accepting state, it reports that there is no solution. Otherwise it replays the moves that lead to the accepting state.

During the replay, the sweep keeps a list for every pair \(j\). The list of \(j\) holds all vertices added to the \(j\)-th path so far, forgotten ones included, segment after segment. The list starts as \(s_j,t_j\), or as \(s_j\) if \(s_j=t_j\). A move that does not merge two segments inserts its vertex between the two ends of its gap. These two ends are adjacent in the list. A merge inserts nothing. At the accepting state, the sweep outputs the \(k\) lists as the paths.

\subsection{Soundness, running time and memory}

A \emph{run} is a sequence of allowed moves starting at the initial state. A vertex is \emph{processed} once a move has added it to a path. The terminals are processed from the start. For a solution \(P\) and vertices \(a,b\) of \(P_j\), let \(P_j(a,b)\) be the set of vertices of \(P_j\) strictly between \(a\) and \(b\). A move on a gap \((g^-,g^+)\) of a pair \(j\) is \emph{along} \(P\) if it adds a vertex of \(P_j\) as follows. A forward extension must add the successor of \(g^-\), and a backward extension the predecessor of \(g^+\). An opening must add a vertex of \(P_j(g^-,g^+)\). A run is \emph{along} \(P\) if all its moves are along \(P\).

Consider a run along \(P\). In every state \(\xi\) of the run, the segments of a pair \(j\) are subpaths of \(P_j\) in the order of \(P_j\). Moreover, if \(Q\) is the set of processed vertices, then
\begin{align*}
V(P_j)\setminus Q\subseteq\bigcup_{(g^-,g^+)\in\Gamma_j(\xi)}P_j(g^-,g^+).
\end{align*}
Both statements hold in the initial state, and every move of the run preserves them. An extension adds to a segment the next vertex of \(P_j\). When the move merges two segments, this vertex is the other end of the gap. An opened vertex is a vertex of \(P_j\) between the two ends of its gap.

A solution \(P\) can be \emph{swept with memory \(p\)} if some run along \(P\) of \(\Sweep_p\) reaches an accepting state.

\begin{theorem}[The sweep]\label{thm:sweep}
Let \(p\ge0\) and \(k\ge1\).
\begin{itemize}
\item {\bf Part 1.} If \(\Sweep_p\) outputs paths, they form a solution.
\item {\bf Part 2.} If some solution can be swept with memory \(p\), then \(\Sweep_p\) outputs a solution.
\item {\bf Part 3.} \(\Sweep_p\) runs in \(F(k,p)n^{p+1}+O(kn(n+m))\) time, where \(F(k,p)\le2^k(ck(p+1))^{c(p+1)}\) for an absolute constant \(c\).
\end{itemize}
\end{theorem}

\begin{proof}
For Part~1, consider the replayed run and fix a pair \(j\). We first show that the \(j\)-th path is a shortest \(s_j\)-\(t_j\) path. Let \(x_0,x_1,\dots,x_L\) be the list of the pair \(j\) at the end of the replay. Throughout the replay, two vertices that are next to each other in the list and belong to the same segment are joined by an arc of \(D_j\). Indeed, an extension adds its vertex next to a segment end along such an arc. A merge joins the last vertex of one segment to the first vertex of the next along such an arc. At the accepting state, the pair has a single segment. It begins with \(s_j\) and ends with \(t_j\). So
\begin{align*}
x_0=s_j,\qquad x_L=t_j,\qquad (x_{h-1},x_h)\in D_j\quad(1\le h\le L).
\end{align*}
That is, these vertices form a walk from \(s_j\) to \(t_j\) in \(D_j\). As \(D_j\) is acyclic, the walk is a path. It is shortest because all its arcs lie in \(D_j\).

It remains to show that the paths are disjoint. Let \(S_t\) be the set of vertices stored after the \(t\)-th move of the replayed run and its forgetting step. Let \(S_0\) be the set stored in the initial state. Suppose a vertex \(u\) lies on the paths of two pairs \(i\ne j\). The terminals of distinct pairs are distinct, so \(u\) is not in the initial state for both pairs. Name the pairs so that \(u\) is on the path of \(j\) first. If \(u\) is on this path from the initial state on, we put \(a\coloneqq0\). Otherwise \(u\) is on this path from a move \(a\) on. Let \(b>a\) be the move that first adds \(u\) to the path of \(i\). Then \(u\) was forgotten in between. Indeed, we show below that
\begin{align*}
u\in S_a,\qquad u\notin S_{b-1}.
\end{align*}
So some move \(c\) with \(a<c<b\) deleted \(u\), that is, \(u\in S_{c-1}\setminus S_c\).
Indeed, \(u\) was stored once it was on the path of \(j\), so \(u\in S_a\). A merge adds no new vertex to a path, so move \(b\) does not merge. A move that does not merge must add a vertex that is not stored. So \(u\) was not stored when move \(b\) added it to the path of \(i\), that is, \(u\notin S_{b-1}\). In every state, the two terminals of a pair are the first end of its first segment and the last end of its last segment. Segment ends are never deleted. So \(u\), which was deleted, is not a terminal. The vertices that are on a path in the initial state are terminals, so \(a\ge1\).

We need one fact about the lists of the replay. Fix a state and a pair, and let a vertex be added to the path of the pair after this state. Then, on the final path, it lies strictly between the two ends of a gap of this state. Indeed, a move inserts its vertex, if any, between the two ends of its gap. It keeps the order of the listed vertices. It replaces its gap by at most two gaps, whose ends lie in the list between the two old ends or at them. The list at the end is the path.

Consider the move \(c\) that forgot \(u\). As \(c<b\), the vertex \(u\) was not yet on the path of \(i\) at that move. By the fact, on the final path of \(i\) it lay strictly between the two ends of some gap \((g^-,g^+)\) of \(i\) in the state after move \(c\). The arcs of this path lie in \(D_i\), so \(g^-\reach{i}u\reach{i}g^+\) and \(u\notin\{g^-,g^+\}\). By the definition of the span,
\begin{align*}
u\in\spn_i(g^-,g^+).
\end{align*}
Since \(i\ne j\), the vertex \(u\) was not forgettable at move \(c\), a contradiction.

For Part~2, the search visits every state reachable by allowed moves. In particular, it visits the accepting state of a run along the solution. So it outputs paths, and by Part~1 they form a solution.

For Part~3, we bound the number of states, the number of moves from a state, and the time per move. A state is determined by the set of its at most \(p\) stored non-terminal vertices together with three data:
\begin{itemize}
\item the pair of each of these vertices and its role in its segment: first end, last end, both ends, or inner vertex;
\item for each pair, the order of its stored non-terminal vertices along the segments, with the stored inner vertices of a segment in a fixed order between its ends;
\item for each pair without a stored non-terminal vertex, whether its two terminals lie in one segment.
\end{itemize}

The data give the segments. Indeed, fix a pair \(j\) with stored non-terminal vertices. List \(s_j\), these vertices in their order, and \(t_j\). The terminals carry no role. Let \(x,y\) be two consecutive entries of this list. They lie in different segments exactly when \(x\) is a last end or both ends, or \(y\) is a first end or both ends. Counting the choices,
\begin{align}
\#\text{states}
&\le\underbrace{(p+1)\cdot n^p}_{\text{stored vertices}}\cdot\underbrace{(4k)^p}_{\text{pairs, roles}}\cdot\underbrace{p!}_{\text{orders}}\cdot\underbrace{2^k}_{\text{terminals}}.
\label{eq:states}
\end{align}

Next, we count the moves from one state. Every segment other than the first and the last of its pair has a non-terminal end. Hence a state has at most \(p\) such segments. A move chooses a gap, one of three kinds and a vertex. So
\begin{align*}
\#\text{gaps}\le k+p,
\qquad
\#\text{moves from one state}\le3(k+p)n.
\end{align*}

It remains to bound the time per move. The relations \(\reach{j}\) are computed once, in \(O(kn(n+m))\) time (\cref{sec:prelim}). After that, checking the conditions of a move and forgetting take time polynomial in \(k\) and \(p\). The visited states are kept in a table indexed by the stored vertices and the data. The table has one entry for each combination counted in \eqref{eq:states}. It is initialised in time proportional to this bound. The index of a state is computed in time polynomial in \(k\) and \(p\). The predecessor of a state is stored with its entry. The replay follows at most one move per state, in \(O(n)\) time each. The replay and the initialisation of the table take at most \(\#\text{states}\cdot O(n)\) time. So the running time is at most
\begin{align*}
&\#\text{states}\cdot\bigl(3(k+p)n\cdot(k+p+1)^{O(1)}+O(n)\bigr)+O(kn(n+m))\\
&\qquad\le2^k(ck(p+1))^{c(p+1)}n^{p+1}+O(kn(n+m))
\end{align*}
for an absolute constant \(c\). This is the bound in Part~3.
\end{proof}

Along a run, a processed vertex \(v\) of a pair \(i\) that is not a segment end is \emph{exposed} at a state \(\xi\) if
\begin{align*}
v\in\bigcup_{j\ne i}\ \bigcup_{(g^-,g^+)\in\Gamma_j(\xi)}\spn_j(g^-,g^+).
\end{align*}
The next lemma says that a run along a solution stores exactly the segment ends and the exposed vertices. It is used in the proof of \cref{lem:orders}.

\begin{lemma}[Memory]\label{lem:memory}
Consider a run along a solution. If a processed vertex of a pair \(i\) lies in the span of no gap of another pair at some state, then it lies in no such span at any later state. The memory of every state equals the number of non-terminal segment ends plus the number of exposed vertices.
\end{lemma}

\begin{proof}
The paths of a solution are disjoint. So along the run, every processed vertex belongs to exactly one pair. This is the only use of the solution. Let \(\xi_0,\xi_1,\dots\) be the states of the run. Here \(\xi_t\) is the state after the \(t\)-th move and the forgetting step that follows it. For a pair \(j\), the union of the spans of the gaps of \(j\) is
\begin{align*}
U_j(\xi_t)
&\coloneqq\bigcup_{(g^-,g^+)\in\Gamma_j(\xi_t)}\spn_j(g^-,g^+).
\end{align*}
We first show that it never grows. A move replaces a gap \((g^-,g^+)\) of \(j\) by at most two gaps \((h^-,h^+)\) with \(g^-\reach{j}h^-\) and \(h^+\reach{j}g^+\). A merge removes a gap. So by transitivity and the acyclicity of \(D_j\),
\begin{align*}
\spn_j(h^-,h^+)\subseteq\spn_j(g^-,g^+).
\end{align*}
Hence the union of the spans never grows:
\begin{align*}
U_j(\xi_t)\subseteq U_j(\xi_{t-1})\qquad(t\ge1).
\end{align*}
A processed vertex of a pair \(i\) lies in the span of no gap of another pair at \(\xi_s\) exactly when it lies outside \(\bigcup_{j\ne i}U_j(\xi_s)\). By the last display,
\begin{align*}
\bigcup_{j\ne i}U_j(\xi_t)\subseteq\bigcup_{j\ne i}U_j(\xi_s)\qquad(t\ge s).
\end{align*}
So such a vertex also lies outside \(\bigcup_{j\ne i}U_j(\xi_t)\) for every \(t\ge s\). This proves the first statement.

For the second statement, fix a state \(\xi_t\) and a processed vertex \(v\) of a pair \(i\) that is an inner vertex of its segment. We show that \(v\) is stored if and only if it is exposed.
\begin{itemize}
\item {\bf Case 1.} Suppose that \(v\) is exposed at \(\xi_t\). By the first statement, it lay in the span of a gap of another pair at every earlier state:
\begin{align*}
v\in\bigcup_{j\ne i}U_j(\xi_s)\qquad(s\le t).
\end{align*}
In particular, this holds at every state since \(v\) became an inner vertex. So \(v\) was never forgettable, and it is still stored.
\item {\bf Case 2.} Suppose that \(v\) is not exposed at \(\xi_t\), that is, \(v\notin\bigcup_{j\ne i}U_j(\xi_t)\). Then \(v\) is forgettable, and the forgetting step after the \(t\)-th move deleted it.
\end{itemize}
So the stored vertices are the segment ends and the exposed vertices.
\end{proof}

From \cref{sec:config} on, the analysis specifies only the order in which the non-terminal vertices of a solution are processed. The next lemma turns such an order into a run. Its hypothesis counts the vertices that a set of processed vertices forces the sweep to store. We now define these vertices without a run.

Let \(Q\) be a set of vertices of a solution \(P\) with \(\Term\subseteq Q\), regarded as processed. For a pair \(j\), the set \(\Gamma_j(Q)\) consists of the pairs \((g^-,g^+)\) of vertices of \(Q\cap V(P_j)\) such that \(g^-\) comes before \(g^+\) on \(P_j\) and
\begin{align*}
P_j(g^-,g^+)\ne\emptyset,
\qquad
P_j(g^-,g^+)\subseteq V(P_j)\setminus Q.
\end{align*}
These pairs play the part of the gaps. A processed non-terminal vertex is \emph{stored under \(Q\)} if it has a neighbour outside \(Q\) on its path. It is also stored under \(Q\) if it lies in the span of such a pair of another path. We write \(N_P(v)\) for the set of neighbours of \(v\) on its path, and \(\stored(Q)\) for the set of vertices stored under \(Q\):
\begin{align}
\stored(Q)
\coloneqq\{v\in Q\setminus\Term : N_P(v)\not\subseteq Q\}
\cup\bigcup_{i\ne j}\ \bigcup_{(g^-,g^+)\in\Gamma_j(Q)}\bigl(Q\cap V(P_i)\cap\spn_j(g^-,g^+)\bigr)\setminus\Term.
\label{eq:stored}
\end{align}

\begin{lemma}[Orders and runs]\label{lem:orders}
Let \(P\) be a solution and \(v_1,\dots,v_N\) an order of its non-terminal vertices. For \(0\le r\le N\), let \(Q_r\coloneqq\{v_1,\dots,v_r\}\cup\Term\), and suppose that
\begin{align*}
\lvert\stored(Q_r)\rvert\le p\qquad(0\le r\le N).
\end{align*}
Then \(P\) can be swept with memory \(p+1\).
\end{lemma}

\begin{proof}
A path \(P_j\) with \(s_j\ne t_j\) that consists of the single arc \((s_j,t_j)\) has no non-terminal vertex. The run begins with these pairs. For each of them, it merges the segments \(\{s_j\}\) and \(\{t_j\}\) by one forward extension, which stores no non-terminal vertex. From then on, the segments of every pair are the maximal stretches of processed vertices of its path.

The run then processes \(v_1,\dots,v_N\) in this order. Let \(v_r\) lie on \(P_j\), with predecessor \(x^-\) and successor \(x^+\) on \(P_j\). If \(x^-\in Q_{r-1}\), the run extends the segment that ends at \(x^-\) forwards by \(v_r\). If also \(x^+\in Q_{r-1}\), it then merges the two segments by one more forward extension, from \(v_r\) to \(x^+\). If only \(x^+\in Q_{r-1}\), it extends the segment that begins at \(x^+\) backwards by \(v_r\). If neither, it opens \(v_r\).

After these moves, the segments of every pair are again the maximal stretches of processed vertices of its path. Indeed, an extension adds \(v_r\) to the stretch of its processed neighbour. A merge joins two stretches that have become one. An opened vertex is a stretch by itself. So the state \(\xi_r\) after the moves for \(v_r\) has the gaps
\begin{align*}
\Gamma_j(\xi_r)=\Gamma_j(Q_r)\qquad(1\le j\le k).
\end{align*}

All these moves are along \(P\), and therefore satisfy the conditions of their kinds. The added vertex lies in \(P_j(g^-,g^+)\subseteq\spn_j(g^-,g^+)\) or, for a merge, is the other end of the gap. An extension adds it along an arc of \(P_j\), which lies in \(D_j\). The vertex \(v_r\) lies outside \(Q_{r-1}\). So it is not stored, as a move that does not merge requires. The merge adds the stored vertex \(x^+\), which a merge may. After \(v_N\), every vertex of \(P\) is processed and every pair has a single segment. So the state is accepting. It remains to show that the memory never exceeds \(p+1\).

We first bound \(\mem(\xi_r)\). We may apply \cref{lem:memory} here, because its proof does not use that the moves are allowed. By this lemma, \(\mem(\xi_r)\) counts the non-terminal segment ends and the exposed vertices. A processed non-terminal vertex is a segment end exactly when it has an unprocessed neighbour on its path. Moreover, \(\Gamma_j(\xi_r)=\Gamma_j(Q_r)\) for every pair \(j\). So the non-terminal segment ends and the exposed vertices are the vertices of the two sets in \eqref{eq:stored}. Hence
\begin{align*}
\mem(\xi_r)=\lvert\stored(Q_r)\rvert\le p.
\end{align*}

It remains to bound the memory of the state \(\xi'\) between an extension and the merge that follows it. Let \(\xi\) be the state before the extension. For \(r\ge2\), this is \(\xi_{r-1}\). For \(r=1\), it is the state after the initial merges, which stores no non-terminal vertex. The state \(\xi'\) differs from \(\xi\) only in the pair \(j\). There, \(v_r\) is a new segment end, and \(x^-\) may have become an inner vertex. The gap \((g^-,g^+)=(x^-,x^+)\) is replaced by \((h^-,h^+)=(v_r,x^+)\), with
\begin{align*}
\spn_j(h^-,h^+)\subseteq\spn_j(g^-,g^+).
\end{align*}
So a vertex exposed in \(\xi'\) was exposed or a segment end in \(\xi\). Moreover, \(x^-\) was stored in \(\xi\) as a segment end. Therefore every vertex other than \(v_r\) that is stored in \(\xi'\) was stored in \(\xi\). Hence
\begin{align*}
\mem(\xi')\le\mem(\xi)+1\le p+1.
\end{align*}
\end{proof}

\section{Configurations and width}
\label{sec:config}

This section removes the graph from the argument. We replace reachability in the DAGs \(D_l\) by an order of positions that it implies. We then turn the memory of a run into the \emph{width} of an order of \(k\) chains of points (\cref{lem:transfer}). The section ends with a bound on this width (\cref{thm:width}) and derives \cref{thm:general-formal} from it. From here on, \(k\ge2\) and \(P=(P_1,\dots,P_k)\) is a fixed solution, except in \cref{thm:general-memory,thm:general-formal}.

The order has to compare also two vertices with the same position (\cref{sec:prelim}). Such vertices differ in all potentials by the same amount. Let \(u\) and \(v\) be two of them, and let a path in a DAG \(D_l\) lead from \(u\) to \(v\). Then \(\pi_l\) increases along it by its positive weight, and every potential changes by the same amount:
\begin{align*}
\pi_m(v)-\pi_m(u)=\pi_l(v)-\pi_l(u)>0\qquad(1\le m\le k).
\end{align*}
So one of them can reach the other only if its potentials are smaller. The potential \(\pi_1\) is therefore used to break ties of position.

\begin{definition}[The order of a pair]\label{def:order}
For a pair \(l\) and vertices \(u,v\), write \(u\preceq_lv\) if \(\pi_{lj}(u)\le\pi_{lj}(v)\) for all \(j\ne l\), and, in case all these inequalities are equalities, \(\pi_1(u)\le\pi_1(v)\). Two vertices are \emph{tied} if all their potentials coincide, that is, if \(\pi(u)=\pi(v)\) for the vector \(\pi\coloneqq(\pi_1,\dots,\pi_k)\).
\end{definition}

As in \cref{sec:overview}, the pair coordinates \(\pi_{lj}\), \(j\ne l\), are the own coordinates of the pair \(l\). Their sum is the progress of a vertex along this pair. Together with \(\pi_1\), the progress gives a linear order that extends \(\preceq_l\), by \eqref{eq:order-psi} below:
\begin{align*}
\varphi_l\coloneqq\sum_{j\ne l}\pi_{lj},
\qquad
\psi_l\coloneqq(\varphi_l,\pi_1),
\end{align*}
where pairs are compared lexicographically. This order is used for the secondary key in \cref{sec:keys} and for the choice of witnesses in \cref{sec:witnesses}.

\begin{lemma}[Order]\label{lem:order}
For every pair \(l\), the relation \(\preceq_l\) is transitive, and for all vertices \(u,v\)
\begin{align}
u\preceq_lv\preceq_lu&\iff\pi(u)=\pi(v),
\label{eq:order-tied}\\
u\reach{l}v&\implies u\preceq_lv,
\label{eq:order-reach}\\
u\preceq_lv&\implies\psi_l(u)\le\psi_l(v),
\label{eq:order-psi}\\
u\preceq_lv,\ \psi_l(u)=\psi_l(v)&\implies\pi(u)=\pi(v).
\label{eq:order-psi-tied}
\end{align}
Moreover, along \(P_l\) the vertices increase strictly in \(\preceq_l\): if \(u\) comes before \(v\) on \(P_l\), then \(u\preceq_lv\) and \(\pi(u)\ne\pi(v)\).
\end{lemma}

\begin{proof}
We first show transitivity. Let \(u\preceq_lv\preceq_lw\). Then
\begin{align*}
\pi_{lj}(u)\le\pi_{lj}(v)\le\pi_{lj}(w)\qquad(j\ne l).
\end{align*}
If \(\pi_{lj}(u)=\pi_{lj}(w)\) for all \(j\ne l\), all these inequalities are equalities. So the definition gives \(\pi_1(u)\le\pi_1(v)\le\pi_1(w)\). Hence \(u\preceq_lw\), and \(\preceq_l\) is transitive.

For \eqref{eq:order-tied}, note that vertices with the same pair coordinates and the same \(\pi_1\) have the same potentials. So
\begin{align*}
u\preceq_lv\preceq_lu
&\iff\pi_{lj}(u)=\pi_{lj}(v)\ (j\ne l),\quad\pi_1(u)=\pi_1(v)\\
&\iff\pi(u)=\pi(v).
\end{align*}

For \eqref{eq:order-reach}, let \(u\reach{l}v\). The pair coordinates \(\pi_{lj}\) do not decrease along a path in \(D_l\) (\cref{lem:cones}), so
\begin{align*}
\pi_{lj}(u)\le\pi_{lj}(v)\qquad\text{for all } j\ne l.
\end{align*}
It remains to compare \(\pi_1\) when all these inequalities are equalities. Then \(u\) and \(v\) have the same position, so every potential changes by the same amount as \(\pi_l\). Since \(D_l\) is tight for \(\pi_l\) and the weights are positive, \(\pi_l\) increases strictly along a non-trivial path in \(D_l\). So does \(\pi_1\):
\begin{align*}
\pi_1(v)-\pi_1(u)=\pi_l(v)-\pi_l(u)>0\qquad(u\ne v).
\end{align*}

The last statement is \eqref{eq:order-reach} for the part of \(P_l\) from \(u\) to \(v\). The last argument also shows that \(u\) and \(v\) are not tied.

For \eqref{eq:order-psi} and \eqref{eq:order-psi-tied}, let \(u\preceq_lv\). No own coordinate \(\pi_{lj}\) is larger at \(u\) than at \(v\), so
\begin{align*}
\varphi_l(u)=\sum_{j\ne l}\pi_{lj}(u)\le\sum_{j\ne l}\pi_{lj}(v)=\varphi_l(v).
\end{align*}
If the two have the same progress, all these inequalities are equalities. The definition of \(\preceq_l\) then compares \(\pi_1\):
\begin{align*}
\pi_{lj}(u)=\pi_{lj}(v)\quad(j\ne l),
\qquad
\pi_1(u)\le\pi_1(v).
\end{align*}
So \(\psi_l(u)\le\psi_l(v)\). If \(\psi_l(u)=\psi_l(v)\), then \(u\) and \(v\) have the same position and the same value of \(\pi_1\). So all their potentials coincide.
\end{proof}

\begin{definition}[Configuration]\label{def:configuration}
A \emph{configuration} is a finite set \(X\) of vertices, each with potentials \(\pi_1(v),\dots,\pi_k(v)\in\R\), together with a partition of \(X\) into \emph{chains} \(X_1,\dots,X_k\). Each chain is listed in an order that increases strictly in its own relation: if \(u\) comes before \(v\) on \(X_l\), then \(u\preceq_lv\) and the two are not tied. Chains may be empty. A subset \(Y\subseteq X\) with the chains \(Y_l\coloneqq Y\cap X_l\), listed in the same order, is again a configuration, a \emph{sub-configuration} of \(X\).
\end{definition}

By the last statement of \cref{lem:order}, the vertices of \(P\), terminals included, form a configuration whose chains are the paths. We call it the configuration of \(P\). The neighbours of a vertex of a configuration are its predecessor and its successor on its chain, when they exist. For a set \(Y\), \(N(Y)\) denotes the set of neighbours of the vertices of \(Y\). Transitivity, \eqref{eq:order-tied}, \eqref{eq:order-psi} and \eqref{eq:order-psi-tied} were proved from \cref{def:order} alone, without the graph. So they hold for the vertices of every configuration. In particular, \(\psi_l\) increases strictly along the chain \(X_l\) of a configuration.

We now say what a processed vertex of a configuration waits for. Let \(v\) be a vertex and \(X_l\) a chain with \(v\notin X_l\), and put
\begin{align}
X_l^{\preceq v}\coloneqq\{q\in X_l : q\preceq_lv\},
\qquad
X_l^{\succeq v}\coloneqq\{q\in X_l : v\preceq_lq\}.
\label{eq:parts}
\end{align}
The chain increases in \(\preceq_l\), and \(\preceq_l\) is transitive (\cref{lem:order}). So the first set is an initial part of \(X_l\), and the second a final part. If both are non-empty, \(X_l\) \emph{encloses} \(v\). We then call the last vertex \(p^*\) of \(X_l^{\preceq v}\) the \emph{lower bound} of \(v\) on \(X_l\), and the first vertex \(q^*\) of \(X_l^{\succeq v}\) the \emph{upper bound}. A vertex in both sets is tied with \(v\) by \eqref{eq:order-tied}. No two vertices of a chain are tied. So
\begin{align*}
\bigl\lvert X_l^{\preceq v}\cap X_l^{\succeq v}\bigr\rvert\le1.
\end{align*}
Hence \(p^*\) is \(q^*\) or comes before it on \(X_l\), and \(p^*=q^*\) exactly when \(v\) is tied with a vertex of \(X_l\). We write \([p^*,q^*]\) for the set of vertices of \(X_l\) from \(p^*\) to \(q^*\). The vertices strictly between the bounds lie in neither set of \eqref{eq:parts}.

The demands of the next definition are chosen so that every vertex that the sweep must store is held (\cref{lem:stored-held}). A vertex waits for its neighbours on its own chain. On every chain that encloses it, it waits for the vertices between its bounds.

\begin{definition}[Demands and width]\label{def:demands}
Let \(Q\subseteq X\) be a set of \emph{processed} vertices. A vertex \(v\) of a chain \(X_i\) has the following \emph{demands}.
\begin{itemize}
\item {\bf Neighbour demand.} It is \emph{met} under \(Q\) if \(N(v)\subseteq Q\).
\item {\bf Demand on a chain.} For every chain \(X_l\), \(l\ne i\), that encloses \(v\), with bounds \(p^*\) and \(q^*\): it is met under \(Q\) if \(Q\) contains two consecutive vertices of \([p^*,q^*]\), or if \(p^*=q^*\in Q\).
\end{itemize}
A vertex \(v\in Q\) is \emph{held} under \(Q\) if one of its demands is not met, and \(\held(Q)\) is the set of these vertices. The \emph{width} of an order \(v_1,\dots,v_N\) of all vertices of \(X\) is \(\max_r\lvert\held(\{v_1,\dots,v_r\})\rvert\), and the width of \(X\) is the least width of its orders.
\end{definition}

A demand that is met under \(Q\) is met under every larger set, so
\begin{align}
\held(Q\cup E)\setminus E\subseteq\held(Q)\qquad(Q,E\subseteq X).
\label{eq:held-monotone}
\end{align}

In the configuration of \(P\), the terminals are vertices like all others. An order of the configuration lists them, and they can be held. The sweep treats the terminals as processed from the start; \cref{lem:transfer} accounts for this difference.

The demand on a chain is often met through the following condition, which the proof of \cref{lem:stage} uses.

\begin{lemma}[Meeting the demand on a chain]\label{lem:meet}
Let \(X_l\) enclose \(v\), with bounds \(p^*\) and \(q^*\), and let \(Q\) be a set of vertices. If \(\{y\}\cup N(y)\subseteq Q\) for some \(y\in[p^*,q^*]\), then the demand of \(v\) on \(X_l\) is met under \(Q\).
\end{lemma}

\begin{proof} If \(p^*=q^*\), then \(y=p^*\in Q\). If \(p^*\ne q^*\), then \(y\) has a neighbour in \([p^*,q^*]\), namely
\begin{align*}
y'
&\coloneqq
\begin{cases}
\text{the successor of } y & \text{if } y\ne q^*,\\
\text{the predecessor of } y & \text{if } y=q^*,
\end{cases}
\end{align*}
and \(y\) and \(y'\) are two consecutive vertices of \([p^*,q^*]\) that lie in \(Q\).
\end{proof}

Apart from remarks, the argument uses the graph again only in the next two lemmas and, at the very end, in \cref{thm:general-memory,thm:general-formal}. The first says that the demands detect every vertex that a run must store, in the sense defined before \cref{lem:orders}.

\begin{lemma}[Stored vertices are held]\label{lem:stored-held}
Let \(Q\) be a set of vertices of \(P\) with \(\Term\subseteq Q\). Then \(\stored(Q)\subseteq\held(Q)\), where \(\held(Q)\) is taken in the configuration of \(P\).
\end{lemma}

\begin{proof}
Let \(v\in\stored(Q)\). If \(N_P(v)\not\subseteq Q\), the neighbour demand of \(v\) is not met. Otherwise, by \eqref{eq:stored}, \(v\in\spn_l(g^-,g^+)\) for a path \(P_l\) other than its own and a pair \((g^-,g^+)\in\Gamma_l(Q)\). So the set \(U\coloneqq P_l(g^-,g^+)\) is non-empty and disjoint from \(Q\). With \(P_l=(x_0,x_1,\dots,x_L)\), \(g^-=x_a\) and \(g^+=x_b\), this says
\begin{align*}
U=\{x_{a+1},\dots,x_{b-1}\},\qquad b\ge a+2,\qquad U\cap Q=\emptyset.
\end{align*}

We first locate the bounds of \(v\) on \(P_l\). By \eqref{eq:order-reach},
\begin{align*}
g^-\preceq_lv\preceq_lg^+,
\end{align*}
so the chain \(P_l\) encloses \(v\). The vertices \(g^-\) and \(g^+\) lie in the first and the second set of \eqref{eq:parts}, respectively. So, on \(P_l\), the vertex \(g^-\) is the lower bound \(p^*\) or comes before it, and \(g^+\) is the upper bound \(q^*\) or comes after it. With \(p^*=x_\alpha\) and \(q^*=x_\beta\), this says
\begin{align*}
a\le\alpha\le\beta\le b
\quad\implies\quad
[p^*,q^*]=\{x_\alpha,\dots,x_\beta\}\subseteq\{g^-\}\cup U\cup\{g^+\}.
\end{align*}

It remains to show that the demand of \(v\) on \(P_l\) is not met. Suppose first that \(p^*=q^*\), that is \(\alpha=\beta\). Then \(v\) is tied with \(x_\alpha\). Since \(v\notin\{g^-,g^+\}\), the paths in \(D_l\) from \(g^-\) to \(v\) and from \(v\) to \(g^+\) are non-trivial. As \(D_l\) is tight for \(\pi_l\) and the weights are positive, \(\pi_l\) increases strictly along them:
\begin{align*}
\pi_l(g^-)<\pi_l(v)=\pi_l(x_\alpha)<\pi_l(g^+).
\end{align*}
So \(x_\alpha\) is neither \(g^-\) nor \(g^+\), that is, \(a<\alpha<b\). Hence \(p^*=x_\alpha\) lies in \(U\), outside \(Q\).

Suppose now that \(p^*\ne q^*\). Two consecutive vertices of \([p^*,q^*]\) are \(x_c\) and \(x_{c+1}\) for some \(c\) with \(a\le c<b\). Since \(b\ge a+2\), they are not \(x_a\) and \(x_b\), so one of them lies in \(U\):
\begin{align*}
\{x_c,x_{c+1}\}\cap U\ne\emptyset\qquad(a\le c<b).
\end{align*}
So \(Q\) does not contain two consecutive vertices of \([p^*,q^*]\).
\end{proof}

The second lemma turns an order of the configuration into a run.

\begin{lemma}[From orders of the configuration to runs]\label{lem:transfer}
If the configuration of \(P\) has width at most \(\omega\), then \(P\) can be swept with memory \(\omega+1\).
\end{lemma}

\begin{proof}
Take an order of the configuration of \(P\) of width at most \(\omega\). Deleting the terminals from it gives an order \(v_1,\dots,v_N\) of the non-terminal vertices of \(P\). By \cref{lem:orders}, it suffices to show, for every \(r\) and the set \(Q\) defined below,
\begin{align*}
\lvert\stored(Q)\rvert\le\omega.
\end{align*}
Then \(P\) can be swept with memory \(\omega+1\).

Fix \(r\) and let \(Q\coloneqq\{v_1,\dots,v_r\}\cup\Term\). If \(r=0\), then \(\stored(Q)=\emptyset\), as \(\stored(Q)\subseteq Q\setminus\Term\). If \(r\ge1\), let \(Q'\subseteq Q\) be the initial part of the original order that ends with \(v_r\). A vertex stored under \(Q\) is one of \(v_1,\dots,v_r\), so it lies in \(Q'\). By \cref{lem:stored-held}, and by \eqref{eq:held-monotone} applied to the sets \(Q'\) and \(E=Q\setminus Q'\),
\begin{align*}
\stored(Q)\subseteq\held(Q)\cap Q'\subseteq\held(Q').
\end{align*}
The original order has width at most \(\omega\), so \(\lvert\held(Q')\rvert\le\omega\). Hence
\begin{align*}
\lvert\stored(Q)\rvert\le\lvert\held(Q')\rvert\le\omega.
\end{align*}
\end{proof}

So it suffices to bound the width of configurations. \Cref{sec:keys,sec:matching,sec:witnesses,sec:cutting,sec:general-proof} prove the following bound.

\begin{theorem}[Width of configurations]\label{thm:width}
Every configuration with \(k\ge2\) chains has width at most
\begin{align*}
g(k)\coloneqq4k+1000(k+1)^5 2^k k^{(k+1)2^{k-1}}.
\end{align*}
\end{theorem}

The proof is an induction on the packing number, the largest number of pairwise separated sub-configurations that are not sortable (\cref{def:packing}). \Cref{sec:keys} bounds the width of a sortable configuration by \(4k\). \Cref{sec:matching} bounds the packing number by a function of \(k\). \Cref{sec:witnesses,sec:cutting} cut a configuration into parts at a cost polynomial in \(k\). \Cref{sec:general-proof} combines these into the induction.

The main results follow from \cref{thm:width}.

\begin{theorem}[Memory for every number of pairs]\label{thm:general-memory}
Every solution of an instance of directed \(k\)-DSP with positive weights can be swept with memory \(g(k)+1\), where \(g\) is the function of \cref{thm:width}.
\end{theorem}

\begin{proof}
For \(k=1\) a solution is a single shortest path. Consider the run that extends the first segment forwards from \(s_1\) along this path, one vertex at a time. With a single pair there is no gap of another pair. So every inner vertex is forgettable. It is deleted by the forgetting step after the move that makes it an inner vertex (\cref{sec:sweep}). Hence the run stores only the current end of the segment besides the terminals, and its memory is at most \(1\). For \(k\ge2\), the configuration of the solution has width at most \(g(k)\) by \cref{thm:width}. Then \cref{lem:transfer} gives the run.
\end{proof}

\begin{theorem}[Formal version of \cref{thm:general}]\label{thm:general-formal}
For every \(k\), the sweep \(\Sweep_{g(k)+1}\) solves directed \(k\)-DSP with positive real weights deterministically in \(O_k(n^{g(k)+2})\) time: it outputs a solution if one exists, and reports that there is none otherwise.
\end{theorem}

\begin{proof}
Let \(p\coloneqq g(k)+1\). By Part~1 of \cref{thm:sweep}, every output of \(\Sweep_p\) is a solution. If a solution exists, it can be swept with memory \(p\) by \cref{thm:general-memory}. Then \(\Sweep_p\) outputs a solution by Part~2 of \cref{thm:sweep}.

By Part~3 of \cref{thm:sweep}, the running time is
\begin{align*}
F(k,p)n^{p+1}+O(kn(n+m))=O(n^{g(k)+2})
\end{align*}
for fixed \(k\), as \(p+1=g(k)+2\) and \(m\le n^2\).
\end{proof}

\section{Sorting by two keys}
\label{sec:keys}

This section shows that a configuration has small width if its indices split into two sets \(A\) and \(B\) with selectors \(L_A\) and \(L_B\). Selectors are functions that behave like the potential \(\pi_i\) on every chain \(X_i\) of their set. An order of small width then sorts the vertices by \(L_A-L_B\), as two chains are sorted by \(\pi_{12}=\pi_1-\pi_2\). Selectors need not exist, and \cref{sec:matching} examines when they do.

For two chains, \(\pi_{12}\) does not decrease along \(X_1\) and does not increase along \(X_2\), by \cref{def:order,def:configuration}. Along a third chain, it is not controlled. The next lemma shows that no linear function of the positions does better. The cone of the pair \(i\) (\cref{sec:prelim}) and its interior are, modulo the all-ones vector,
\begin{align*}
C_i=\{\Delta : \Delta_i\ge\Delta_j\text{ for all } j\},
\qquad
\operatorname{int}C_i=\{\Delta : \Delta_i>\Delta_j\text{ for all } j\ne i\}.
\end{align*}
The displacement of every arc of \(D_i\) lies in \(C_i\) (\cref{lem:cones}). This is all that \cref{lem:cones} says about it. A linear function \(\ell\) of the displacements is well defined modulo the all-ones vector if its coefficients add up to zero. It increases along an arc with displacement \(\Delta\) exactly when \(\ell(\Delta)>0\). Consider a linear function that does not decrease along any arc whose displacement lies in \(C_i\), as \(\pi_{12}\) does not along the arcs of \(D_1\). Such a function is non-negative on \(C_i\), and hence positive on \(\operatorname{int}C_i\) unless it is zero.

\begin{lemma}[One pair per linear order]\label{lem:linear-order}
Let \(\ell(\Delta)=\sum_ta_t\Delta_t\) with \(\sum_ta_t=0\). Let \(I^+\) be the set of indices \(i\) such that \(\ell>0\) on the interior of \(C_i\), and \(I^-\) the set of indices \(j\) such that \(\ell<0\) on the interior of \(C_j\). Then \(\lvert I^+\rvert\le1\) and \(\lvert I^-\rvert\le1\), and if \(I^+=\{i\}\) and \(I^-=\{j\}\), then \(\ell\) is a positive multiple of \(\Delta_i-\Delta_j\).
\end{lemma}

\begin{proof}
Let \(e_t\) be the \(t\)-th unit vector. Subtracting \(\Delta_i\) from all coordinates of a displacement in \(C_i\) gives the representative
\begin{align*}
-\sum_{t\ne i}c_te_t,
\qquad
c_t=\Delta_i-\Delta_t\ge0.
\end{align*}
So, modulo the all-ones vector, \(C_i\) is the cone generated by the vectors \(-e_t\) with \(t\ne i\). Its interior consists of their combinations with positive coefficients \(c_t\). On such a combination,
\begin{align*}
\ell\Bigl(-\sum_{t\ne i}c_te_t\Bigr)=-\sum_{t\ne i}c_ta_t.
\end{align*}
This is positive for all positive \(c_t\) exactly when no \(a_t\) with \(t\ne i\) is positive and not all of them are zero. So, by the definitions of \(I^+\) and \(I^-\), for all indices \(i\) and \(j\),
\begin{align*}
i\in I^+
&\iff a_t\le0\ (t\ne i),\quad(a_t)_{t\ne i}\ne0,\\
j\in I^-
&\iff a_t\ge0\ (t\ne j),\quad(a_t)_{t\ne j}\ne0.
\end{align*}

We next show \(\lvert I^+\rvert\le1\) and \(\lvert I^-\rvert\le1\). Two indices in \(I^+\) would give \(a_t\le0\) for all \(t\). Then \(\sum_ta_t=0\) forces \(a=0\), contrary to the condition that not all \(a_t\) are zero. The same holds for \(I^-\).

Finally, let \(I^+=\{i\}\) and \(I^-=\{j\}\). Then \(i\ne j\): for \(i=j\), the two conditions would give \(a_t=0\) for all \(t\ne i\), which they exclude. For \(t\notin\{i,j\}\), the two conditions give \(a_t\le0\) and \(a_t\ge0\), so \(a_t=0\). They also give \(a_j\le0\le a_i\). As all other coefficients vanish,
\begin{align*}
a_i+a_j=\sum_ta_t=0.
\end{align*}
With \(a\ne0\), this gives \(a_i=-a_j>0\). So \(\ell(\Delta)=a_i(\Delta_i-\Delta_j)\).
\end{proof}

Suppose that a linear function is to sort all \(k\) pairs as \(\pi_{12}\) sorts two. That is, for every pair \(i\), it increases along all displacements in the interior of \(C_i\), or it decreases along all of them. Then every index lies in \(I^+\) or in \(I^-\), and the lemma gives
\begin{align*}
k\le\lvert I^+\rvert+\lvert I^-\rvert\le2.
\end{align*}
So the function that sorts three or more chains cannot be linear. The function \(L_A-L_B\) used below is not linear. Indeed, the selectors of \cref{sec:matching} are minima of functions that are maxima of differences of potentials.

All comparisons are made with the following quantity. For vertices \(u,v\), the \emph{length} from \(u\) to \(v\) is
\begin{align*}
\len(u,v)\coloneqq\max_{1\le j\le k}\bigl(\pi_j(v)-\pi_j(u)\bigr).
\end{align*}
The length can be negative. For all vertices \(u,v,z\) and every \(j\), it satisfies
\begin{align}
\len(u,z)&\le\len(u,v)+\len(v,z),\notag\\
\len(u,v)&\ge\pi_j(v)-\pi_j(u),\notag\\
\len(u,v)&=\pi_i(v)-\pi_i(u)\qquad\text{if } u\preceq_iv.
\label{eq:len}
\end{align}
The first relation holds because a maximum of sums is at most the sum of the maxima. The second holds because a maximum is at least each of its terms. For the third, \(u\preceq_iv\) gives \(\pi_{ij}(u)\le\pi_{ij}(v)\) for every \(j\), that is,
\begin{align*}
\pi_j(v)-\pi_j(u)\le\pi_i(v)-\pi_i(u)\qquad(1\le j\le k).
\end{align*}
So the maximum in \(\len(u,v)\) is attained at \(j=i\).

\begin{definition}[Selector]\label{def:selector}
For an index \(i\), the \emph{range} \(X_i^+\) of the chain \(X_i\) consists of the vertices of \(X_i\) and the vertices that \(X_i\) encloses. Let \(A\subseteq\{1,\dots,k\}\). A \emph{selector} for \(A\) is a function \(L\colon X\to\R\) with \emph{offsets} \(\theta_i\in\R\), \(i\in A\), such that
\begin{itemize}
\item {\bf Property 1.} \(L(v)-L(u)\le\len(u,v)\) for all \(u,v\in X\): the selector grows no faster than the length.
\item {\bf Property 2.} \(L(v)=\pi_i(v)+\theta_i\) for every \(i\in A\) and every \(v\in X_i^+\): on the range of every chain of its set, the selector agrees with the potential of that chain up to a constant.
\end{itemize}
A configuration is \emph{sortable} if \(\{1,\dots,k\}\) can be partitioned into two sets \(A\) and \(B\), one of which may be empty, that both have selectors.
\end{definition}

Property~2 covers the whole range, not only the chain \(X_i\). The reason is that the order has to meet the demands on \(X_i\) of the vertices that \(X_i\) encloses (\cref{def:demands}). The next lemma records the cases in which selectors exist for free. \Cref{sec:matching,sec:general-proof} use Part~2 throughout. They use Part~3 once, to see that a smallest ball that is not sortable has a positive radius.

\begin{lemma}[Simple sortable configurations]\label{lem:sortable-simple}
The following hold.
\begin{itemize}
\item {\bf Part 1.} The empty set of indices has a selector, and every configuration with two chains is sortable.
\item {\bf Part 2.} Every sub-configuration of a sortable configuration is sortable.
\item {\bf Part 3.} A configuration whose vertices all have the same position is sortable.
\end{itemize}
\end{lemma}

\begin{proof}
For Part~1, Property~2 is void for the empty set, and \(\pi_1\) has Property~1 by the second relation of \eqref{eq:len}. For two chains, \(\pi_1\) is a selector for \(\{1\}\) and \(\pi_2\) one for \(\{2\}\), with offsets zero. Property~2 holds by definition, and Property~1 by the second relation of \eqref{eq:len}.

For Part~2, let \(Y\) be a sub-configuration of a sortable configuration \(X\). A vertex that the chain \(Y\cap X_i\) encloses lies between two vertices of \(X_i\) in \(\preceq_i\). So \(X_i\) encloses it too, and the ranges satisfy
\begin{align*}
(Y\cap X_i)^+\subseteq X_i^+.
\end{align*}
Hence the selectors of \(X\), restricted to \(Y\), are selectors of \(Y\).

For Part~3, all potentials of two vertices \(u,v\) with the same position differ by the same amount. So
\begin{align*}
\len(u,v)=\pi_1(v)-\pi_1(u),
\qquad
\pi_1(v)-\pi_i(v)=\pi_1(u)-\pi_i(u)\quad(1\le i\le k).
\end{align*}
Hence \(\pi_1\) has Property~1, and Property~2 for every set of indices with offsets \(\theta_i=\pi_1-\pi_i\).
\end{proof}

For the rest of this section, fix a sortable configuration and selectors \(L_A\) and \(L_B\) for the two sets of a partition, with offsets \(\theta_i\). The order will sort the vertices by the \emph{primary key} \(\kappa\) and, among vertices with the same primary key, by the \emph{secondary key} \(\kappa'\), where
\begin{align*}
\kappa\coloneqq L_A-L_B,
\qquad
\kappa'\coloneqq(G,\pi_1),
\qquad
G\coloneqq kL_A-\sum_{j=1}^k\pi_j,
\end{align*}
and pairs are compared lexicographically. The function \(G\) expresses the progress. At a vertex of \(X_i^+\), Property~2 gives
\begin{align*}
L_A&=\pi_i+\theta_i&&\text{if } i\in A,\\
L_A&=L_B+\kappa=\pi_i+\theta_i+\kappa&&\text{if } i\in B.
\end{align*}
We insert this into the definition of \(G\) and use
\begin{align*}
k\pi_i-\sum_j\pi_j=\sum_{j\ne i}(\pi_i-\pi_j)=\varphi_i.
\end{align*}
This gives, on \(X_i^+\),
\begin{align}
G&=\varphi_i+k\theta_i&&\text{if } i\in A,\notag\\
G&=\varphi_i+k\theta_i+k\kappa&&\text{if } i\in B.
\label{eq:G}
\end{align}
The next lemma draws the consequences for the two keys along a chain.

\begin{lemma}[The keys along a chain]\label{lem:keys}
Let \(u\preceq_iv\) with \(u,v\in X_i^+\).
\begin{itemize}
\item {\bf Part 1.} If \(i\in A\), then \(\kappa(u)\le\kappa(v)\). If \(i\in B\), then \(\kappa(u)\ge\kappa(v)\).
\item {\bf Part 2.} If \(\kappa(u)=\kappa(v)\), then \(\kappa'(u)\le\kappa'(v)\), with equality only if \(u\) and \(v\) are tied.
\end{itemize}
\end{lemma}

\begin{proof}
For Part~1, let \(i\in A\). By Property~2, \(L_A=\pi_i+\theta_i\) at \(u\) and at \(v\). Property~1 for \(L_B\) and the third relation of \eqref{eq:len}, \(\len(u,v)=\pi_i(v)-\pi_i(u)\), then give
\begin{align*}
\kappa(v)-\kappa(u)
&=\bigl(\pi_i(v)-\pi_i(u)\bigr)-\bigl(L_B(v)-L_B(u)\bigr)\\
&\ge\bigl(\pi_i(v)-\pi_i(u)\bigr)-\len(u,v)=0.
\end{align*}
For \(i\in B\) the roles of the two selectors are exchanged. Now \(L_B=\pi_i+\theta_i\) at both vertices, and Property~1 for \(L_A\) gives
\begin{align*}
\kappa(v)-\kappa(u)
=\bigl(L_A(v)-L_A(u)\bigr)-\bigl(\pi_i(v)-\pi_i(u)\bigr)
\le\len(u,v)-\bigl(\pi_i(v)-\pi_i(u)\bigr)=0.
\end{align*}

For Part~2, let \(\kappa(u)=\kappa(v)\). By \eqref{eq:G}, \(G\) and \(\varphi_i\) then differ by the same constant at \(u\) and at \(v\):
\begin{align*}
G(v)-G(u)=\varphi_i(v)-\varphi_i(u).
\end{align*}
So the pairs \(\kappa'=(G,\pi_1)\) of the two vertices compare as their pairs \(\psi_i=(\varphi_i,\pi_1)\) do. Part~2 therefore follows from \eqref{eq:order-psi} and \eqref{eq:order-psi-tied}.
\end{proof}

\paragraph{The shape of a chain.}
Apply \cref{lem:keys} to two vertices of one chain \(X_l\). By Part~1, the primary key does not decrease along \(X_l\) if \(l\in A\), and does not increase along it if \(l\in B\). So, for every value \(\alpha\), the vertices of \(X_l\) with \(\kappa<\alpha\) form an initial part of the chain if \(l\in A\), and a final part if \(l\in B\). For the same reason, the \emph{\(\alpha\)-plateau}
\begin{align*}
\{x\in X_l : \kappa(x)=\alpha\}
\end{align*}
consists of consecutive vertices of \(X_l\). By Part~2, inside a plateau the secondary key increases strictly in the order of the chain, for the chains of both groups.

\paragraph{The order.}
The order sorts the vertices by the primary key and then by the secondary key. So, as the primary key grows, the order takes the chains of \(A\) forwards and the chains of \(B\) backwards. It takes every plateau forwards. A \emph{stage} is a pair \((\alpha,\beta)\) such that some vertex has \(\kappa=\alpha\) and \(\kappa'=\beta\). Stages, like the values of \(\kappa'=(G,\pi_1)\), are ordered lexicographically. The vertices that the two keys place at or before the stage \((\alpha,\beta)\) are
\begin{align*}
K(\alpha,\beta)\coloneqq\{v : \kappa(v)<\alpha\}\cup\{v : \kappa(v)=\alpha,\ \kappa'(v)\le\beta\}.
\end{align*}
The order processes a few more vertices at every stage. For a chain \(X_l\) with a non-empty \(\alpha\)-plateau, let \(y_l\) be the first vertex of the plateau. Let \(y_l^-\) be its predecessor on \(X_l\), omitted if \(y_l\) is the first vertex of the chain. Put
\begin{align}
Z(\alpha,\beta)\coloneqq K(\alpha,\beta)\cup N\bigl(K(\alpha,\beta)\bigr)\cup\bigcup_{l\colon\text{the }\alpha\text{-plateau of }X_l\ne\emptyset}\{y_l^-,y_l\}.
\label{eq:Z}
\end{align}
The order goes through the stages. At each stage, it processes the vertices of \(Z(\alpha,\beta)\) that are not yet processed, in any order. The second term of \eqref{eq:Z} meets the neighbour demands of \(K(\alpha,\beta)\). The third is needed on the chains of \(B\), which the primary key takes backwards. There a vertex of \(K(\alpha,\beta)\) can have its lower bound before the \(\alpha\)-plateau and its upper bound in it. In this case, \(y_l^-,y_l\) are two consecutive vertices between its bounds (Case~2 in the proof of \cref{lem:stage}). The next lemma collects what the bound on the width uses.

\begin{lemma}[A stage]\label{lem:stage}
Let \((\alpha,\beta)\) be a stage.
\begin{itemize}
\item {\bf Part 1.} \(Z(\alpha,\beta)\subseteq Z(\alpha',\beta')\) for every later stage \((\alpha',\beta')\).
\item {\bf Part 2.} \(K(\alpha,\beta)\cap\held\bigl(Z(\alpha,\beta)\bigr)=\emptyset\).
\item {\bf Part 3.} \(\lvert X_l\cap(Z(\alpha,\beta)\setminus K(\alpha,\beta))\rvert\le3\) for every chain \(X_l\).
\end{itemize}
\end{lemma}

\begin{proof}
For Part~1, let \((\alpha',\beta')\) be a later stage. The lexicographic order of the stages gives
\begin{align*}
K(\alpha,\beta)\subseteq K(\alpha',\beta'),
\end{align*}
so the first two terms of \eqref{eq:Z} grow. It remains to place the vertices \(y_l\) and \(y_l^-\) of the \(\alpha\)-plateaus. If \(\alpha'=\alpha\), the two stages have the same plateaus. If \(\alpha'>\alpha\), then \(\kappa(y_l)=\alpha<\alpha'\). So \(y_l\in K(\alpha',\beta')\) and \(y_l^-\in N(K(\alpha',\beta'))\).

For Part~2, let \(v\in K(\alpha,\beta)\). By \eqref{eq:Z}, \(N(v)\subseteq Z(\alpha,\beta)\), so the neighbour demand of \(v\) is met. Let \(X_l\) be a chain that encloses \(v\), with bounds \(p^*\) and \(q^*\). Every \(u\in K(\alpha,\beta)\) has \(\{u\}\cup N(u)\subseteq Z(\alpha,\beta)\). So, by \cref{lem:meet} and \cref{def:demands}, it suffices to find in \([p^*,q^*]\) a vertex of \(K(\alpha,\beta)\), or two consecutive vertices of \(Z(\alpha,\beta)\).

Since \(v\in X_l^+\), Part~1 of \cref{lem:keys} applies to \(v\) and its bounds. It places \(\kappa(v)\) between \(\kappa(p^*)\) and \(\kappa(q^*)\). As \(v\in K(\alpha,\beta)\), also \(\kappa(v)\le\alpha\), so
\begin{align*}
p^*\preceq_lv\preceq_lq^*
\quad\implies\quad
\min\{\kappa(p^*),\kappa(q^*)\}\le\kappa(v)\le\alpha.
\end{align*}
If a bound has \(\kappa<\alpha\), it lies in \(K(\alpha,\beta)\). Otherwise both bounds have \(\kappa\ge\alpha\), so
\begin{align*}
\min\{\kappa(p^*),\kappa(q^*)\}=\kappa(v)=\alpha,
\qquad
\kappa'(v)\le\beta.
\end{align*}
The last inequality holds because \(v\) lies in \(K(\alpha,\beta)\) and has \(\kappa=\alpha\). We distinguish which bound has \(\kappa=\alpha\).
\begin{itemize}
\item {\bf Case 1.} \(\kappa(p^*)=\alpha\). Then Part~2 of \cref{lem:keys} for \(p^*\preceq_lv\) gives
\begin{align*}
\kappa'(p^*)\le\kappa'(v)\le\beta,
\end{align*}
so \(p^*\in K(\alpha,\beta)\).
\item {\bf Case 2.} \(\kappa(q^*)=\alpha<\kappa(p^*)\). This case occurs only for \(l\in B\), since the primary key does not decrease along a chain of \(A\). The vertex \(q^*\) lies in the \(\alpha\)-plateau of \(X_l\). The lower bound \(p^*\) is not in the plateau, and it comes before \(q^*\) (see after \eqref{eq:parts}). As the plateau consists of consecutive vertices, \(p^*\) comes before the whole plateau. So, in the order of the chain,
\begin{align*}
p^*\le y_l^-<y_l\le q^*,
\end{align*}
and \(y_l^-,y_l\) are two consecutive vertices of \([p^*,q^*]\) that lie in \(Z(\alpha,\beta)\) by \eqref{eq:Z}.
\end{itemize}

For Part~3, fix a chain \(X_l\). If the \(\alpha\)-plateau of \(X_l\) is non-empty, write \(y\coloneqq y_l\) and \(y^-\coloneqq y_l^-\). Split the vertices of \(K(\alpha,\beta)\) on \(X_l\) into
\begin{align*}
K^<\coloneqq\{u\in X_l : \kappa(u)<\alpha\},
\qquad
K^=\coloneqq\{u\in X_l : \kappa(u)=\alpha,\ \kappa'(u)\le\beta\}.
\end{align*}
Neighbours of vertices of \(X_l\) lie on \(X_l\), so \eqref{eq:Z} gives
\begin{align*}
X_l\cap\bigl(Z(\alpha,\beta)\setminus K(\alpha,\beta)\bigr)\subseteq\bigl(N(K^<)\cup N(K^=)\cup\{y^-,y\}\bigr)\setminus(K^<\cup K^=).
\end{align*}

The set \(K^<\) is an initial or a final part of the chain. So \(N(K^<)\setminus K^<\) has at most one vertex. We call it \(x\). The set \(K^=\) is an initial part of the \(\alpha\)-plateau, because the secondary key increases along the plateau. If \(K^=\) is not empty, it contains \(y\), and \(N(K^=)\setminus K^=\subseteq\{y^-,z\}\), where \(z\) is the successor of its last vertex. If \(K^=\) is empty, \(N(K^=)\) is empty. So, where a vertex that does not exist is omitted,
\begin{align*}
X_l\cap\bigl(Z(\alpha,\beta)\setminus K(\alpha,\beta)\bigr)\subseteq
\begin{cases}
\{x,y^-,z\} & \text{if } K^=\ne\emptyset,\\
\{x,y^-,y\} & \text{if } K^==\emptyset,
\end{cases}
\end{align*}
which are at most three vertices in both cases.
\end{proof}

\begin{lemma}[Sortable configurations]\label{lem:sortable}
A sortable configuration with \(k\) chains has width at most \(4k\).
\end{lemma}

\begin{proof}
Consider a moment of the order at which it works on a stage \(s'\) or has just completed it. Let \(Q\) be the set of processed vertices at this moment, and let \(s\) be the stage before \(s'\). By Part~1 of \cref{lem:stage}, the vertices processed up to the end of a stage are those of its set \(Z\), so
\begin{align*}
Z(s)\subseteq Q\subseteq Z(s').
\end{align*}
By Part~2 of \cref{lem:stage}, no vertex of \(K(s)\) is held under \(Z(s)\). By \eqref{eq:held-monotone} applied to the sets \(Z(s)\) and \(E=Q\setminus Z(s)\), no vertex of \(K(s)\subseteq Z(s)\) is held under \(Q\) either. As held vertices are processed, this gives
\begin{align*}
\held(Q)\subseteq Q\setminus K(s)\subseteq Z(s')\setminus K(s)=\bigl(Z(s')\setminus K(s')\bigr)\cup\bigl(K(s')\setminus K(s)\bigr).
\end{align*}

We count the two sets. The first has at most \(3k\) vertices by Part~3 of \cref{lem:stage}. The second consists of the vertices whose pair of keys is \(s'\). Indeed, every vertex has the keys of some stage, and no stage lies strictly between \(s\) and \(s'\). The second set contains at most one vertex of every chain, since the secondary key increases strictly inside a plateau. So the number of held vertices is at most
\begin{align*}
\underbrace{3k}_{Z(s')\setminus K(s')}
+\underbrace{k}_{K(s')\setminus K(s)}
=4k.
\end{align*}
If \(s'\) is the first stage, the same count applies with \(Z(s)\) and \(K(s)\) replaced by the empty set.
\end{proof}

\section{Chains without selectors do not repeat}
\label{sec:matching}

For three or more chains, nothing guarantees that a configuration is sortable. This section finds what can prevent it. The obstruction is a matching of the first vertices of some chains to the last vertices of other chains that is cheaper than matching every chain to itself. A set of indices has a selector exactly when no such matching exists among its chains (\cref{lem:matching} and the remark after its proof). The section then shows that this obstruction cannot occur many times in a row along the chains (\cref{lem:repeat,lem:packing}).

For a non-empty chain \(X_i\), let \(f_i\) be its first and \(e_i\) its last vertex. Let \(\sigma\) be a cyclic permutation of a set \(C\) of at least two indices whose chains are non-empty. The \emph{deficit} of \(\sigma\) on \(X\) is
\begin{align*}
\delta_\sigma(X)\coloneqq\sum_{i\in C}\bigl(\len(f_i,e_{\sigma(i)})-\len(f_i,e_i)\bigr).
\end{align*}
We call \(\sigma\) \emph{cheaper} on \(X\) if \(\delta_\sigma(X)<0\). By \eqref{eq:len}, \(\len(f_i,e_i)=\pi_i(e_i)-\pi_i(f_i)\) is the increase of the own potential along the chain \(X_i\).

\begin{lemma}[Selectors from matchings]\label{lem:matching}
Let \(A\subseteq\{1,\dots,k\}\). If no cyclic permutation of a subset of \(A\) is cheaper on \(X\), then \(A\) has a selector.
\end{lemma}

\begin{proof}
Let \(A'\) be the set of indices in \(A\) with non-empty chains. If \(A'\) is empty, the function \(\pi_1\) is a selector by the second relation of \eqref{eq:len}:
\begin{align*}
\len(u,v)\ge\pi_1(v)-\pi_1(u).
\end{align*}
So let \(A'\) be non-empty.

We first attach a weight to every pair of indices of \(A'\). For \(i,j\in A'\) put
\begin{align*}
w_{ij}\coloneqq\pi_i(f_i)+\len(f_i,e_j)-\pi_j(e_j).
\end{align*}
Then \(w_{ii}=0\), because \(\len(f_i,e_i)=\pi_i(e_i)-\pi_i(f_i)\). Now let \(\sigma\) be a cyclic permutation of some \(C\subseteq A'\). Since \(\sigma\) permutes \(C\), and since \(\len(f_i,e_i)=\pi_i(e_i)-\pi_i(f_i)\), the terms \(\pi_i(f_i)\) and \(-\pi_j(e_j)\) of the weights \(w_{i\sigma(i)}\), \(i\in C\), add up to
\begin{align*}
\sum_{i\in C}\bigl(\pi_i(f_i)-\pi_i(e_i)\bigr)=-\sum_{i\in C}\len(f_i,e_i).
\end{align*}
So the hypothesis gives
\begin{align*}
\sum_{i\in C}w_{i\sigma(i)}
=\sum_{i\in C}\len(f_i,e_{\sigma(i)})-\sum_{i\in C}\len(f_i,e_i)
=\delta_\sigma(X)\ge0.
\end{align*}

Next we obtain the offsets from these weights. Consider the complete digraph on \(A'\) with arc weights \(w_{ij}\). Its cycles are the loops and these cyclic permutations, so it has no cycle of negative weight. For \(i\in A'\), let \(\theta_i\) be the least weight of a walk in this digraph that ends at \(i\). The walk may start at any vertex, and the walk without arcs has weight \(0\). The minimum exists. Indeed, removing a cycle from a walk does not increase its weight. So the least weight is attained by one of the finitely many paths. A lightest walk to \(i\) followed by the arc from \(i\) to \(j\) is a walk to \(j\), so
\begin{align}
\theta_j\le\theta_i+w_{ij}\qquad\text{for all } i,j\in A'.
\label{eq:offsets}
\end{align}

With these offsets, define
\begin{align*}
L(v)\coloneqq\min_{i\in A'}\bigl(\pi_i(f_i)+\theta_i+\len(f_i,v)\bigr).
\end{align*}
Property~1 follows from the triangle inequality for the length, the first relation of \eqref{eq:len}. If \(i\) attains the minimum at \(u\), then
\begin{align*}
L(v)\le\pi_i(f_i)+\theta_i+\len(f_i,u)+\len(u,v)=L(u)+\len(u,v).
\end{align*}

For Property~2, let \(j\in A'\) and \(v\in X_j^+\). A vertex of \(X_j\) lies between the ends of its chain. A vertex enclosed by \(X_j\) lies between its two bounds. So in both cases
\begin{align*}
f_j\preceq_jv\preceq_je_j.
\end{align*}
The third relation of \eqref{eq:len} gives \(\len(u,v)=\pi_j(v)-\pi_j(u)\) whenever \(u\preceq_jv\). So the term of \(j\) in the minimum and the length from \(v\) to \(e_j\) are
\begin{align*}
\pi_j(f_j)+\theta_j+\len(f_j,v)=\pi_j(v)+\theta_j,
\qquad
\len(v,e_j)=\pi_j(e_j)-\pi_j(v).
\end{align*}

It remains to show that every term of the minimum is at least \(\pi_j(v)+\theta_j\). For every \(i\in A'\), \eqref{eq:offsets}, the triangle inequality for the length and the value of \(\len(v,e_j)\) just found give
\begin{align*}
\pi_j(e_j)+\theta_j
&\le\pi_i(f_i)+\theta_i+\len(f_i,e_j)\\
&\le\pi_i(f_i)+\theta_i+\len(f_i,v)+\len(v,e_j)\\
&=\pi_i(f_i)+\theta_i+\len(f_i,v)+\pi_j(e_j)-\pi_j(v).
\end{align*}
So the term of \(i\) is at least \(\pi_j(v)+\theta_j\). Hence \(L(v)=\pi_j(v)+\theta_j\). For indices of \(A\) with empty chains, Property~2 is void.
\end{proof}

The converse of \cref{lem:matching} holds as well, although we do not use it. A set \(A\) with a selector \(L\) has no cheaper cyclic permutation. Indeed, let \(\sigma\) be a cyclic permutation of \(C\subseteq A\). Property~1, then Property~2 at the ends of the chains, give
\begin{align*}
\sum_{i\in C}\len(f_i,e_{\sigma(i)})
\ge\sum_{i\in C}\bigl(L(e_{\sigma(i)})-L(f_i)\bigr)
=\sum_{i\in C}\bigl(\pi_i(e_i)-\pi_i(f_i)\bigr)
=\sum_{i\in C}\len(f_i,e_i).
\end{align*}
The offsets cancel because \(\sigma\) permutes \(C\). So \(\delta_\sigma(X)\ge0\).

Hence, by \cref{lem:matching}, a configuration that is not sortable has the following property. For every partition of the indices into \(A\) and \(B\), it has a cheaper cyclic permutation inside \(A\) or inside \(B\). The next lemma shows that a cyclic permutation of \(m\) indices cannot be cheaper on \(m\) sub-configurations that follow one another along the chains. \Cref{lem:packing} below rests on it. For sets \(U,U'\subseteq X_i\), write \(U<_iU'\) if every vertex of \(U\) comes before every vertex of \(U'\) on \(X_i\).

\begin{lemma}[A cheaper permutation does not repeat]\label{lem:repeat}
Let \(\sigma\) be a cyclic permutation of a set \(C\) of \(m\ge2\) indices. Let \(Y^1,\dots,Y^m\) be sub-configurations of \(X\) in which all chains with index in \(C\) are non-empty. Suppose that
\begin{align*}
Y^1_i<_iY^2_i<_i\dots<_iY^m_i\qquad\text{for every } i\in C.
\end{align*}
Then
\begin{align*}
\sum_{t=1}^m\delta_\sigma(Y^t)\ge0.
\end{align*}
In particular \(\sigma\) is not cheaper on all of \(Y^1,\dots,Y^m\).
\end{lemma}

\begin{proof}
We first add up the lengths along a single chain. For \(i\in C\), let \(f_i^t\) and \(e_i^t\) be the first and the last vertex of \(Y^t_i\). The vertices \(f_i^1,e_i^1,f_i^2,e_i^2,\dots,f_i^m,e_i^m\) come in this order on \(X_i\), so they increase in \(\preceq_i\). By the third relation of \eqref{eq:len}, the length between two consecutive ones is the difference of their values of \(\pi_i\). So these lengths add up to the increase of \(\pi_i\):
\begin{align}
\sum_{t=1}^m\len(f_i^t,e_i^t)+\sum_{t=1}^{m-1}\len(e_i^t,f_i^{t+1})=\pi_i(e_i^m)-\pi_i(f_i^1).
\label{eq:own-route}
\end{align}

Next, fix \(i\in C\). Consider a second sequence of vertices that starts at \(f_i^1\) and changes its chain inside every \(Y^t\). With \(i_t\coloneqq\sigma^{t-1}(i)\), it is
\begin{align*}
f_{i_1}^1,\ e_{i_2}^1,\ f_{i_2}^2,\ e_{i_3}^2,\ \dots,\ f_{i_m}^m,\ e_{i_{m+1}}^m.
\end{align*}
Since \(\sigma\) is cyclic of length \(m\), we have \(i_{m+1}=\sigma^m(i)=i\). So the sequence ends at \(e_i^m\).

By the second relation of \eqref{eq:len}, each length between consecutive vertices of the sequence is at least the difference of their values of \(\pi_i\). So these lengths add up to at least the increase of \(\pi_i\):
\begin{align}
\sum_{t=1}^m\len(f_{i_t}^t,e_{i_{t+1}}^t)+\sum_{t=1}^{m-1}\len(e_{i_{t+1}}^t,f_{i_{t+1}}^{t+1})\ge\pi_i(e_i^m)-\pi_i(f_i^1).
\label{eq:other-route}
\end{align}

It remains to sum over \(i\in C\). For fixed \(t\), the index \(i_t\) runs over \(C\) once as \(i\) runs over \(C\). Moreover, \(i_{t+1}=\sigma(i_t)\). Substituting \(j=i_t\) in the first sums of \eqref{eq:other-route} and \(j=i_{t+1}\) in the second sums gives
\begin{align*}
\sum_{i\in C}\len(f_{i_t}^t,e_{i_{t+1}}^t)&=\sum_{j\in C}\len(f_j^t,e_{\sigma(j)}^t)&&(t\le m),\\
\sum_{i\in C}\len(e_{i_{t+1}}^t,f_{i_{t+1}}^{t+1})&=\sum_{j\in C}\len(e_j^t,f_j^{t+1})&&(t<m).
\end{align*}
Summing \eqref{eq:other-route} over \(i\in C\) and exchanging the order of summation therefore gives
\begin{align*}
\sum_{t=1}^m\sum_{j\in C}\len(f_j^t,e_{\sigma(j)}^t)+\sum_{t=1}^{m-1}\sum_{j\in C}\len(e_j^t,f_j^{t+1})\ge\sum_{i\in C}\bigl(\pi_i(e_i^m)-\pi_i(f_i^1)\bigr).
\end{align*}

The sum of \eqref{eq:own-route} over \(i\in C\) has the same right-hand side. The same lengths \(\len(e_j^t,f_j^{t+1})\) occur in it. Subtracting it leaves
\begin{align*}
\sum_{t=1}^m\sum_{j\in C}\bigl(\len(f_j^t,e_{\sigma(j)}^t)-\len(f_j^t,e_j^t)\bigr)\ge0.
\end{align*}
By the definition of the deficit, the left-hand side is \(\sum_t\delta_\sigma(Y^t)\).
\end{proof}

The lemma uses no strictness: lengths may be zero or negative, and vertices may share positions.

\Cref{sec:general-proof} proves the width bound by an induction. The induction cuts a configuration that is not sortable into cells. Each cell must contain fewer parts that are not sortable than the whole. We now count such parts. Two sub-configurations \(Y,Y'\) of \(X\) are \emph{separated} if
\begin{align*}
Y_i<_iY'_i\quad\text{or}\quad Y'_i<_iY_i\qquad\text{for every } i \text{ with } Y_i\ne\emptyset\ne Y'_i.
\end{align*}
The order may differ from chain to chain. This weak form is the one that the cuts of \cref{sec:general-proof} provide (\cref{lem:far-separated}). However, \cref{lem:repeat} needs sub-configurations that follow one another in the same order on all chains of the permutation. \Cref{lem:monotone-list,lem:packing} below extract such sub-configurations from many pairwise separated ones.

\begin{definition}[Packing number]\label{def:packing}
A \emph{packing} in \(X\) is a family of pairwise separated sub-configurations of \(X\), none of which is sortable. The \emph{packing number} \(\nu(X)\) is the largest size of a packing in \(X\).
\end{definition}

So \(\nu(X)=0\) exactly when \(X\) is sortable. Indeed, by Part~2 of \cref{lem:sortable-simple}, a sortable configuration has no packing with a member. A configuration that is not sortable is a packing by itself. A packing in a sub-configuration \(Y\) of \(X\) is also a packing in \(X\), so \(\nu(Y)\le\nu(X)\).

To bound the packing number, we first extract many members from a large packing. They are chosen so that they follow one another in one direction along every chain they use. A list \(Y^1,\dots,Y^M\) of sub-configurations of \(X\) is \emph{increasing} along \(X_i\) if its chains \(Y^1_i,\dots,Y^M_i\) are non-empty and
\begin{align*}
Y^1_i<_iY^2_i<_i\dots<_iY^M_i.
\end{align*}
It is \emph{decreasing} along \(X_i\) if the reversed list is increasing along \(X_i\).

\begin{lemma}[Monotone lists]\label{lem:monotone-list}
Let \(M\ge1\) be an integer, and suppose that \(X\) has a packing with \(2^kM^{2^{k-1}}\) members. Then there are a non-empty set \(T\) of indices and a list of \(M\) members of the packing with two properties. First, \(T\) is the set of indices of the non-empty chains of every member of the list. Second, the list is increasing or decreasing along every \(X_i\) with \(i\in T\).
\end{lemma}

\begin{proof}
We first find many members with the same non-empty chains. A member of the packing is not sortable, so it is not empty. Indeed, if all chains are empty, \(\pi_1\) is a selector for every set of indices. There are at most \(2^k\) sets of indices, so at least
\begin{align*}
\frac{2^kM^{2^{k-1}}}{2^k}=M^{2^{k-1}}
\end{align*}
members have the same set \(T\ne\emptyset\) of indices of non-empty chains. On every \(X_i\) with \(i\in T\), the chains of these members occupy pairwise disjoint stretches. So they are linearly ordered along \(X_i\).

Next we make the list monotone along every chain of \(T\). Fix \(i_0\in T\), and list these members in their order along \(X_{i_0}\). Then the list is increasing along \(X_{i_0}\). By the theorem of Erd\H{o}s and Szekeres~\cite{es35}, every sequence \(a_1,\dots,a_N\) of \(N\) distinct numbers has a monotone subsequence \(a_{t_1},\dots,a_{t_r}\) of length \(r\ge\sqrt N\):
\begin{align*}
a_{t_1}<\dots<a_{t_r}\qquad\text{or}\qquad a_{t_1}>\dots>a_{t_r}.
\end{align*}

Apply this once for each of the at most \(k-1\) other indices \(i\) of \(T\). Number the members by their place in the order along \(X_i\); these numbers are distinct. Read these numbers in the order of the current list, and apply the theorem to this sequence. The list is replaced by the sublist that the theorem provides, which is increasing or decreasing along \(X_i\). Each application keeps at least the square root of the number of members. A sublist of a monotone list stays monotone along the chains treated before. There are at most \(k-1\) applications, and
\begin{align*}
\bigl(M^{2^{k-1}}\bigr)^{1/2^{k-1}}=M.
\end{align*}
So a list of \(M\) members remains that is increasing or decreasing along every \(X_i\), \(i\in T\).
\end{proof}

\begin{lemma}[Packings are bounded]\label{lem:packing}
Every configuration \(X\) with \(k\) chains has
\begin{align*}
\nu(X)<N_k\coloneqq2^k k^{(k+1)2^{k-1}}.
\end{align*}
\end{lemma}

\begin{proof}
Suppose that \(X\) has a packing with \(N_k\) members, and put \(M\coloneqq k^{k+1}\). Then
\begin{align*}
N_k=2^k k^{(k+1)2^{k-1}}=2^kM^{2^{k-1}}.
\end{align*}
By \cref{lem:monotone-list}, some list of \(M\) members of the packing is increasing or decreasing along every \(X_i\) with \(i\in T\). Here \(T\ne\emptyset\) is the set of indices of the non-empty chains of every member of the list.

We split the indices by the direction of this list. The chains with an index outside \(T\) are empty in every member of the list. So it does not matter where these indices go, and we put them into \(A\). Let \(T^{\uparrow}\subseteq T\) be the set of indices \(i\in T\) such that the list is increasing along \(X_i\), and put
\begin{align*}
A\coloneqq T^{\uparrow}\cup\bigl(\{1,\dots,k\}\setminus T\bigr),
\qquad
B\coloneqq\{1,\dots,k\}\setminus A.
\end{align*}

Every member of the list has a cheaper cyclic permutation inside \(A\) or inside \(B\). Indeed, no member \(Y\) of the list is sortable, so \(A\) or \(B\) has no selector in \(Y\). By \cref{lem:matching}, some cyclic permutation \(\sigma\) of a set \(C\) of at least two indices inside \(A\) or inside \(B\) is cheaper on \(Y\). The deficit is defined only for indices whose chains in \(Y\) are non-empty. So \(C\subseteq T\), and
\begin{align*}
C\subseteq A\cap T\quad\text{or}\quad C\subseteq B\cap T.
\end{align*}
Choose one such permutation for every member.

Next we find a permutation that is chosen for many members. The number of cyclic permutations of subsets of \(\{1,\dots,k\}\) with at least two elements is
\begin{align*}
\sum_{m=2}^k\binom{k}{m}(m-1)!\le\sum_{m=2}^k\frac{k^m}{m}\le k^k.
\end{align*}
The first step holds because \(\binom{k}{m}(m-1)!=k(k-1)\cdots(k-m+1)/m\). The second holds because every term is at most \(k^m/2\), and
\begin{align*}
\sum_{m\le k}k^m\le k^k\cdot k/(k-1)\le2k^k.
\end{align*}
There are \(M\) members and at most \(k^k\) such permutations, and
\begin{align*}
M=k^{k+1}>(k-1)k^k.
\end{align*}
So some permutation \(\sigma\) of a set \(C\) is chosen for at least \(k\ge\lvert C\rvert\) members.

Finally, we arrange these members for \cref{lem:repeat}. Take \(\lvert C\rvert\) of them, \(Y^1,\dots,Y^{\lvert C\rvert}\), in the order of the list if \(C\subseteq A\) and in the reverse order if \(C\subseteq B\). Then
\begin{align*}
Y^1_i<_i\dots<_iY^{\lvert C\rvert}_i\qquad\text{for every } i\in C.
\end{align*}
Moreover, \(\sigma\) is cheaper on each of them, so
\begin{align*}
\sum_{t=1}^{\lvert C\rvert}\delta_\sigma(Y^t)<0.
\end{align*}
This contradicts \cref{lem:repeat}.
\end{proof}

\section{Witnesses and interfaces}
\label{sec:witnesses}

\Cref{sec:keys,sec:matching} dealt with sortable configurations and with families of separated configurations that are not sortable. This section and the next cut an arbitrary configuration into sets of vertices that are processed one after another. This section prepares the cuts.

Suppose that a set \(S\) is processed before the rest. A vertex of \(S\) may then wait for vertices of a chain that encloses it (\cref{def:demands}), and these vertices may lie far outside \(S\). We therefore make a choice in advance. For every vertex \(v\) and every chain \(X_l\) that encloses it, we fix at most two vertices of \(X_l\) that meet the demand of \(v\) on \(X_l\). They are the witnesses of \(v\) on \(X_l\) (\cref{def:witness,lem:witness}). The order processes early the witnesses that the vertices of \(S\) have outside \(S\), together with the vertices at which chains leave \(S\). Their number is added to the width. It is the cost of processing \(S\) first, and we say that these witnesses lie \emph{across the cut}. \Cref{sec:blocks} makes this precise. It shows that a partition can be processed part by part at twice this cost (\cref{lem:serial}). \Cref{sec:cutting} needs sets of small cost. For that, vertices that lie close to one another must share their witnesses. \Cref{sec:passing} introduces the geometry that the choice uses, and \cref{sec:witness-choice} makes the choice.

\subsection{Distance, progress and passing points}
\label{sec:passing}

The witnesses of \(v\) on \(X_l\) are taken near the place where the chain \(X_l\), drawn as a curve, has made the same progress as \(v\). This subsection defines that place and measures how far it is from \(v\).

A \emph{point} is a vector \(x\) of \(\R^k\) modulo adding the same number to all coordinates. Its pair coordinates are \(\pi_{ab}(x)\coloneqq x_a-x_b\). They do not depend on the representative. The position of a vertex \(v\) is a point, which we also write \(v\). Its pair coordinates as a point are those it has as a vertex. The distance between two points \(x,y\) is
\begin{align*}
\lVert y-x\rVert\coloneqq\max_{a\ne b}\bigl\lvert\pi_{ab}(y)-\pi_{ab}(x)\bigr\rvert.
\end{align*}
This is a norm of their difference. Indeed, it is a maximum of absolute values of linear functions of \(y-x\). It vanishes only if all coordinates of \(y-x\) are equal, that is, if \(x\) and \(y\) are the same point.

The progress \(\varphi_l=\sum_{j\ne l}\pi_{lj}\) of \cref{sec:config} depends only on the position, so it is defined for points. Along a chain \(X_l\) no own coordinate \(\pi_{lj}\) decreases, so the progress \(\varphi_l\) does not decrease.

The next lemma shows that in the direction of a chain, distance and progress agree up to the factor \(k-1\). So a stretch of a chain can be measured by its progress. Relation \eqref{eq:dist-monotone} is used only in \cref{lem:far-separated}.

\begin{lemma}[Distance and progress]\label{lem:distance}
Let \(x,y,z\) be points and \(l\) an index. Then
\begin{align}
\lvert\varphi_l(y)-\varphi_l(x)\rvert\le(k-1)\lVert y-x\rVert.
\label{eq:progress-lipschitz}
\end{align}
If \(\pi_{lj}(x)\le\pi_{lj}(y)\) for all \(j\ne l\), then
\begin{align}
\lVert y-x\rVert\le\varphi_l(y)-\varphi_l(x),
\label{eq:dist-progress}
\end{align}
and if moreover \(\pi_{lj}(x)\le\pi_{lj}(z)\le\pi_{lj}(y)\) for all \(j\ne l\), then
\begin{align}
\lVert z-x\rVert\le\lVert y-x\rVert.
\label{eq:dist-monotone}
\end{align}
\end{lemma}

\begin{proof}
Put \(\eta_j\coloneqq\pi_{lj}(y)-\pi_{lj}(x)\) for \(j\ne l\) and \(\eta_l\coloneqq0\). Every pair coordinate is a difference of two own coordinates of \(l\). So it changes by \(\eta_b-\eta_a\) from \(x\) to \(y\):
\begin{align*}
\pi_{ab}=\pi_{lb}-\pi_{la},
\qquad
\pi_{ab}(y)-\pi_{ab}(x)=\eta_b-\eta_a.
\end{align*}
Hence
\begin{align}
\lVert y-x\rVert=\max_j\eta_j-\min_j\eta_j,
\qquad
\varphi_l(y)-\varphi_l(x)=\sum_j\eta_j.
\label{eq:norm-eps}
\end{align}

For \eqref{eq:progress-lipschitz}, \eqref{eq:norm-eps} bounds the difference of any two of the numbers \(\eta_j\) by \(\lVert y-x\rVert\). The sum in \eqref{eq:norm-eps} has, besides \(\eta_l=0\), the \(k-1\) terms with \(j\ne l\). So
\begin{align*}
\lvert\eta_j\rvert=\lvert\eta_j-\eta_l\rvert\le\lVert y-x\rVert,
\qquad
\lvert\varphi_l(y)-\varphi_l(x)\rvert\le\sum_{j\ne l}\lvert\eta_j\rvert\le(k-1)\lVert y-x\rVert.
\end{align*}

For \eqref{eq:dist-progress}, all \(\eta_j\) are non-negative and \(\eta_l=0\). So the minimum is \(0\), and the maximum is at most the sum:
\begin{align*}
\lVert y-x\rVert=\max_j\eta_j-\min_j\eta_j=\max_j\eta_j\le\sum_j\eta_j=\varphi_l(y)-\varphi_l(x).
\end{align*}

For \eqref{eq:dist-monotone}, let \(\eta_j'\) be the corresponding numbers for the pair \(x,z\). The hypothesis gives \(0\le\eta_j'\le\eta_j\) for all \(j\). So both minima are \(0\), and \eqref{eq:norm-eps} for the two pairs gives
\begin{align*}
\lVert z-x\rVert=\max_j\eta_j'\le\max_j\eta_j=\lVert y-x\rVert.
\end{align*}
\end{proof}

So \eqref{eq:dist-progress} and \eqref{eq:dist-monotone} apply to vertices with \(x\preceq_ly\) and \(x\preceq_lz\preceq_ly\), respectively. In particular, they apply to vertices of the chain \(X_l\) in their order.

Join the positions of consecutive vertices of a chain \(X_l\) by straight segments, and call the result the \emph{curve} \(\gamma_l\) of the chain. From here on, a segment is always one of these straight pieces. The segments of the sweep in \cref{sec:sweep} do not occur again. Along \(\gamma_l\) no \(\pi_{lj}\) decreases. By \eqref{eq:dist-progress}, a stretch of the curve on which the progress is constant is a single point. Let \(s\) lie between the progress of the first and of the last vertex of \(X_l\). The progress is continuous and does not decrease along \(\gamma_l\), and by the last sentence it is constant only on single points. So the curve has exactly one point \(\gamma_l(s)\) of progress \(s\). By \eqref{eq:dist-progress} again, two such values \(s,s'\) satisfy
\begin{align}
\lVert\gamma_l(s')-\gamma_l(s)\rVert\le\lvert s'-s\rvert.
\label{eq:path-lipschitz}
\end{align}

Let \(v\) be a vertex enclosed by \(X_l\), with bounds \(p^*\preceq_lv\preceq_lq^*\). By \eqref{eq:order-psi}, the bounds satisfy
\begin{align*}
\varphi_l(p^*)\le\varphi_l(v)\le\varphi_l(q^*).
\end{align*}
So the curve has a point of the same progress as \(v\). We call
\begin{align*}
\hat v_l\coloneqq\gamma_l(\varphi_l(v)),
\qquad
d_l(v)\coloneqq\lVert\hat v_l-v\rVert
\end{align*}
the point at which \(X_l\) \emph{passes} \(v\) and the \emph{passing distance}. If \(d_l(v)=0\), the curve goes through the position of \(v\), and we say that \(X_l\) \emph{touches} \(v\). If \(d_l(v)>0\), we say that \(X_l\) \emph{misses} \(v\).

The witnesses rest on Part~2 of the next lemma. By this part, the two ends of the segment at the passing point of \(v\) lie between the bounds of every vertex close to \(v\). This is what the demand of such a vertex on \(X_l\) asks for.

\begin{lemma}[Passing points]\label{lem:passing}
Let \(u,v\) be vertices enclosed by a chain \(X_l\).
\begin{itemize}
\item {\bf Part 1.} The passing distance changes by at most \(k\) times the distance:
\begin{align}
\lvert d_l(u)-d_l(v)\rvert\le k\lVert u-v\rVert.
\label{eq:passing-lipschitz}
\end{align}
\item {\bf Part 2.} Let \(d_l(v)>0\), let \(z,z'\) be consecutive vertices of \(X_l\) whose segment contains \(\hat v_l\), and let \(\lVert u-v\rVert<d_l(v)/k\). Then \(u\) is not tied with a vertex of \(X_l\). Moreover, \(z\) and \(z'\) both lie in \([p^*,q^*]\), where \(p^*\) and \(q^*\) are the bounds of \(u\) on \(X_l\).
\end{itemize}
\end{lemma}

\begin{proof}
For Part~1, the triangle inequality, \eqref{eq:path-lipschitz} and \eqref{eq:progress-lipschitz} give
\begin{align*}
\lvert d_l(u)-d_l(v)\rvert
\le\lVert u-v\rVert+\lVert\hat u_l-\hat v_l\rVert
\le\lVert u-v\rVert+\lvert\varphi_l(u)-\varphi_l(v)\rvert
\le k\lVert u-v\rVert.
\end{align*}

For Part~2, let \(d\coloneqq d_l(v)\). Let \(\eta_j\coloneqq\pi_{lj}(\hat v_l)-\pi_{lj}(v)\) for \(j\ne l\), and \(\eta_l\coloneqq0\). By \eqref{eq:norm-eps}, and since \(\hat v_l\) has the progress of \(v\),
\begin{align*}
\max_j\eta_j-\min_j\eta_j=d,
\qquad
\sum_j\eta_j=0.
\end{align*}

We first bound the largest and the smallest of these numbers. For the first line of the next display, we bound \(k-1\) of the \(k\) numbers \(\eta_j\) by the largest and one of them by the smallest. For the second line, we bound \(k-1\) of them by the smallest and one by the largest. This gives
\begin{align*}
0=\sum_j\eta_j&\le(k-1)\max_j\eta_j+\min_j\eta_j=k\max_j\eta_j-d,\\
0=\sum_j\eta_j&\ge(k-1)\min_j\eta_j+\max_j\eta_j=k\min_j\eta_j+d.
\end{align*}
The last step in each line uses \(\max_j\eta_j-\min_j\eta_j=d\). Hence
\begin{align*}
\max_j\eta_j\ge d/k,
\qquad
\min_j\eta_j\le-d/k.
\end{align*}

So some own coordinate of \(l\) is larger at \(\hat v_l\) than at \(u\), and another is smaller. Indeed, let \(a\) and \(b\) be indices at which \(\eta_j\) is largest and smallest. Since \(d>0\) and \(\eta_l=0\), both differ from \(l\). So
\begin{align*}
\pi_{la}(\hat v_l)\ge\pi_{la}(v)+d/k>\pi_{la}(u),
\qquad
\pi_{lb}(\hat v_l)\le\pi_{lb}(v)-d/k<\pi_{lb}(u).
\end{align*}
The non-strict inequalities are the bounds on \(\max_j\eta_j\) and \(\min_j\eta_j\). The strict inequalities hold because \(\lVert u-v\rVert<d/k\) and \(\pi_{la}\), \(\pi_{lb}\) are pair coordinates.

Next we pass from \(\hat v_l\) to the vertices \(z\) and \(z'\). Along the segment from \(z\) to \(z'\), no own coordinate of \(l\) decreases. The point \(\hat v_l\) lies on this segment, so
\begin{align*}
\pi_{la}(z')\ge\pi_{la}(\hat v_l)>\pi_{la}(u),
\qquad
\pi_{lb}(z)\le\pi_{lb}(\hat v_l)<\pi_{lb}(u).
\end{align*}

It remains to compare \(z\) and \(z'\) with the bounds of \(u\). By the definition of \(\preceq_l\), the first inequality excludes \(z'\preceq_lu\), and the second excludes \(u\preceq_lz\). In the notation of \eqref{eq:parts},
\begin{align*}
z'\notin X_l^{\preceq u},
\qquad
z\notin X_l^{\succeq u}.
\end{align*}
The first set is an initial part of \(X_l\) ending at \(p^*\), so \(z'\) comes after \(p^*\). The second set is a final part starting at \(q^*\), so \(z\) comes before \(q^*\). As \(z\) and \(z'\) are consecutive, this gives, on \(X_l\),
\begin{align*}
p^*\preceq_lz\preceq_lz'\preceq_lq^*.
\end{align*}
Hence both lie in \([p^*,q^*]\). Moreover, \(p^*\ne q^*\) because \(z\ne z'\). So \(u\) is not tied with a vertex of \(X_l\).
\end{proof}

\subsection{The witnesses of a vertex}
\label{sec:witness-choice}

We now choose the witnesses. The choice has to serve two later arguments. First, the witnesses of \(v\) on \(X_l\) must meet the demand of \(v\) on \(X_l\) (\cref{lem:witness}). Second, for the sets \(S\) that \cref{sec:cutting} cuts out, few witnesses may lie across the cut.

\begin{figure}[t]
\centering
\begin{tikzpicture}[x=1cm,y=1cm,font=\small,
  vx/.style={circle,fill=black,draw=black,inner sep=1.4pt},
  wt/.style={rectangle,draw=black,line width=0.7pt,inner sep=3pt},
  pp/.style={line width=0.8pt}]
\newcommand{\passcross}[2]{\draw[pp] (#1-0.09,#2-0.09) -- (#1+0.09,#2+0.09) (#1-0.09,#2+0.09) -- (#1+0.09,#2-0.09);}
\begin{scope}
\draw[dsp path,DSPBlue] (-3.0,2.25) -- (-1.05,2.55) -- (-0.35,2.66) -- (0.35,2.77) -- (1.05,2.88) -- (3.0,3.2);
\foreach \p in {(-3.0,2.25),(-1.05,2.55),(-0.35,2.66),(0.35,2.77),(1.05,2.88),(3.0,3.2)} \node[vx,DSPBlue] at \p {};
\foreach \p in {(-1.05,2.55),(-0.35,2.66),(0.35,2.77),(1.05,2.88)} \node[wt] at \p {};
\node[DSPBlue,above] at (2.6,3.2) {\(X_l\)};
\draw[dsp path,DSPOrange,dashed] (-3.0,0.2) -- (-0.7,0.45) -- (0,0.55) -- (0.7,0.65) -- (3.0,0.9);
\foreach \p in {(-3.0,0.2),(3.0,0.9)} \node[vx,DSPOrange] at \p {};
\foreach \p in {(-0.7,0.45),(0,0.55),(0.7,0.65)} \node[vx] at \p {};
\node[below=2pt] at (0,0.55) {\(v\)};
\node[DSPOrange,below] at (2.6,0.85) {\(X_i\)};
\foreach \x/\y/\yy in {-0.7/0.45/2.605,0/0.55/2.715,0.7/0.65/2.825} {
  \draw[densely dotted,line width=0.7pt] (\x,\y) -- (\x,\yy);
  \passcross{\x}{\yy}
}
\draw[densely dashed,gray,line width=0.8pt] (-3.1,1.25) .. controls (-1,1.75) and (1,1.75) .. (3.1,1.45);
\node[gray] at (-2.55,1.62) {cut};
\end{scope}
\begin{scope}[xshift=7.3cm]
\draw[dsp path,DSPBlue] (-3.0,2.25) -- (-1.05,2.55) -- (-0.35,2.66) -- (0.35,2.77) -- (1.05,2.88) -- (3.0,3.2);
\foreach \p in {(-3.0,2.25),(-1.05,2.55),(-0.35,2.66),(0.35,2.77),(1.05,2.88),(3.0,3.2)} \node[vx,DSPBlue] at \p {};
\foreach \p in {(-0.35,2.66),(0.35,2.77)} \node[wt] at \p {};
\node[above=3pt] at (-0.42,2.66) {\(z\)};
\node[above=3pt] at (0.42,2.77) {\(z'\)};
\node[DSPBlue,above] at (2.6,3.2) {\(X_l\)};
\draw[dsp path,DSPOrange,dashed] (-3.0,0.2) -- (-0.7,0.45) -- (0,0.55) -- (0.7,0.65) -- (3.0,0.9);
\foreach \p in {(-3.0,0.2),(3.0,0.9)} \node[vx,DSPOrange] at \p {};
\foreach \p in {(-0.7,0.45),(0,0.55),(0.7,0.65)} \node[vx] at \p {};
\node[below=2pt] at (0,0.55) {\(v\)};
\node[DSPOrange,below] at (2.6,0.85) {\(X_i\)};
\draw[densely dotted,line width=0.7pt] (0,0.55) -- (0,2.715) node[pos=0.72,right] {\(d\)};
\passcross{0}{2.715}
\node[right=2pt] at (0.05,2.45) {\(\hat v_l\)};
\draw[densely dashed,gray] (0,0.55) circle (1.0);
\node[gray] at (-1.85,-0.3) {radius \(d/k\)};
\draw[densely dashed,gray,line width=0.8pt] (-3.1,1.25) .. controls (-1,1.75) and (1,1.75) .. (3.1,1.45);
\end{scope}
\end{tikzpicture}
\caption{Three vertices of \(X_i\) that the chain \(X_l\) misses. The set \(S\) that is processed first consists of the vertices below the gray curve. Left: if every vertex took the two ends of the segment at its own passing point (crosses), four witnesses (squares) would lie across the cut. With more vertices, more witnesses would lie across it. Right: \(X_l\) passes \(v\) at the point \(\hat v_l\), at distance \(d\). The ends \(z\) and \(z'\) of the segment at \(\hat v_l\) serve every vertex within distance \(d/k\) of \(v\) (\cref{lem:passing}). So two witnesses suffice. The drawing is schematic: positions have \(k-1\) coordinates, and the balls of the distance are polytopes.}
\label{fig:passing}
\end{figure}
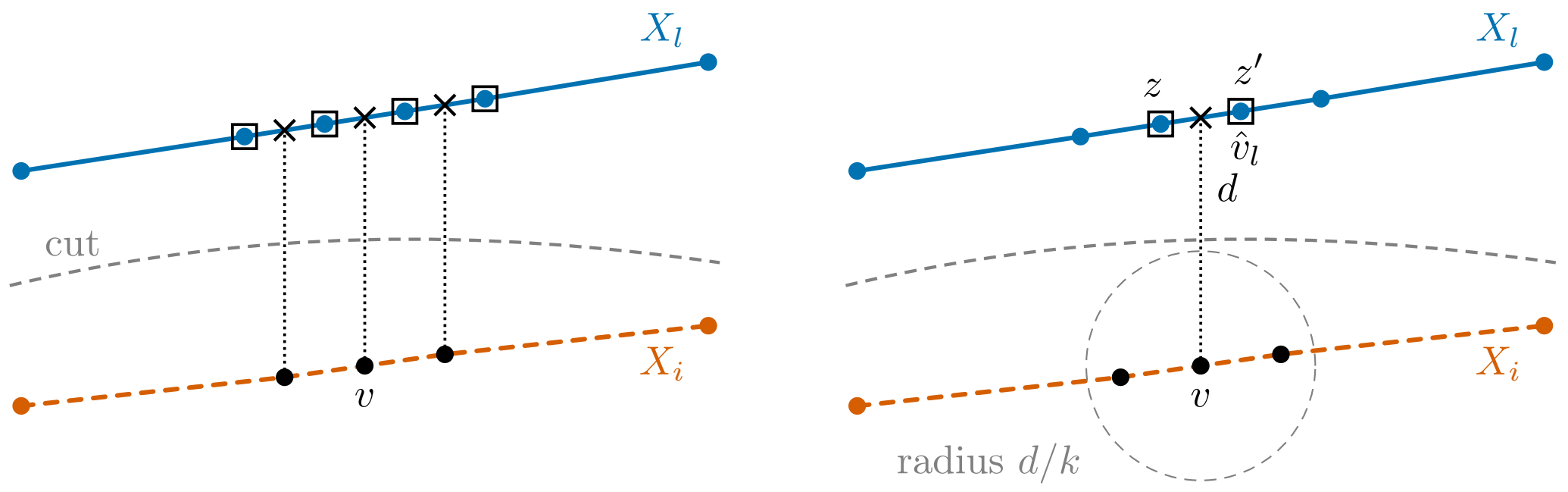

\paragraph{Why witnesses are shared.}
Let \(X_l\) miss \(v\). By Part~2 of \cref{lem:passing} with \(u=v\), the two ends of the segment at the passing point \(\hat v_l\) meet the demand of \(v\) on \(X_l\). Taking these for every vertex is too expensive. Suppose that many vertices of \(X_i\) lie close to \(v\) and that \(X_l\) has many short segments near \(\hat v_l\). Their passing points can then lie on different segments. A set \(S\) that contains these vertices but not these segments then has all the ends of these segments across the cut (\cref{fig:passing}, left). By the same Part~2, the ends of the segment at \(\hat v_l\) serve every vertex within distance \(d_l(v)/k\) of \(v\). So all these vertices can share them (\cref{fig:passing}, right).

\paragraph{Clusters.}
Fix an ordered pair of indices \(i\ne l\). We divide the vertices of \(X_i\) that \(X_l\) misses into \emph{clusters}. Each cluster has a \emph{representative} \(w\), one of its members. The passing point of \(w\) provides the witnesses of all members. By the passing distance of the cluster we mean \(d_l(w)\). The clusters have to meet two requirements.
\begin{itemize}
\item {\bf Sharing.} Every member \(v\) of the cluster of \(w\) has \(\lVert v-w\rVert<d_l(w)/k\), so that Part~2 of \cref{lem:passing} applies to \(w\) and \(v\).
\item {\bf Room.} Every representative \(w\) has an interval of values of the progress \(\varphi_i\), of length proportional to \(d_l(w)\). The intervals of different representatives are disjoint. \Cref{lem:room} turns this into bounds on the clusters that have members close to one another. \Cref{sec:cutting} uses these bounds to count the witnesses across a cut.
\end{itemize}

Both are obtained from intervals of progress along \(X_i\). The reason is that \eqref{eq:dist-progress} bounds distances along \(X_i\) by differences of the progress \(\varphi_i\):
\begin{align*}
\lVert u-v\rVert\le\lvert\varphi_i(u)-\varphi_i(v)\rvert\qquad(u,v\in X_i).
\end{align*}
Let \(\varepsilon\coloneqq1/(4k)\). Give every vertex \(v\) of \(X_i\) that \(X_l\) misses the following interval of values of the progress along its own chain:
\begin{align*}
I(v)\coloneqq\bigl[\varphi_i(v)-\varepsilon d_l(v),\ \varphi_i(v)+\varepsilon d_l(v)\bigr].
\end{align*}
Go through these vertices in order of decreasing passing distance. A vertex becomes a representative if its interval is disjoint from the intervals of all earlier representatives. Otherwise, it joins the cluster of an earlier representative whose interval meets its own.

The intervals of the representatives are pairwise disjoint, which gives room. The second estimate of \cref{lem:groups} below is sharing. The value of \(\varepsilon\) makes \(2k\varepsilon=1/2\), which \cref{lem:groups} uses twice.

For a representative \(w\), fix consecutive vertices \(z_w,z_w'\) of \(X_l\) whose segment contains \(\hat w_l\). Like the clusters, they depend on the fixed pair \(i,l\). They exist because \(X_l\) has at least two vertices. Indeed, a chain with a single vertex \(x\) encloses \(w\) only if \(x\preceq_lw\preceq_lx\). By \eqref{eq:order-tied}, this holds only if \(x\) is tied with \(w\). Then the curve is the single point \(x=w\), and \(X_l\) touches \(w\). But \(X_l\) misses \(w\).

\paragraph{Vertices on the curve.}
If \(X_l\) touches \(v\), the position of \(v\) lies on \(\gamma_l\). The witnesses are then the vertices of \(X_l\) just before and just after \(v\) on the curve. Positions alone do not determine them. Indeed, consecutive vertices of a chain can have the same position (in the configuration of \(P\), when the arc between them has displacement \(0\)). We therefore compare the pair \(\psi_l=(\varphi_l,\pi_1)\), which increases strictly along \(X_l\) (\cref{lem:order}). These witnesses need no sharing. By \cref{lem:witness}, they are the vertex tied with \(v\), or the two ends of a segment that contains the position of \(v\). \Cref{sec:cutting} counts such segments directly (\cref{lem:across}).

\begin{definition}[Witness sets]\label{def:witness}
Let \(v\) be a vertex of \(X_i\), and put \(\wit_i(v)\coloneqq N(v)\). For a chain \(X_l\), \(l\ne i\), that encloses \(v\), put
\begin{itemize}
\item {\bf Vertices that \(X_l\) misses.} If \(X_l\) misses \(v\) and \(w\) is the representative of the cluster of \(v\), then \(\wit_l(v)\coloneqq\{z_w,z_w'\}\).
\item {\bf Vertices that \(X_l\) touches.} If \(X_l\) touches \(v\), then \(\wit_l(v)\) consists of the last vertex \(z\) of \(X_l\) with \(\psi_l(z)\le\psi_l(v)\) and of the first vertex \(z'\) of \(X_l\) with \(\psi_l(z')\ge\psi_l(v)\).
\end{itemize}
The set \(\wit(v)\) of \emph{witnesses} of \(v\) is the union of these sets, and \(\wit(S)\coloneqq\bigcup_{v\in S}\wit(v)\) for \(S\subseteq X\).
\end{definition}

In the second case, the vertices \(z\) and \(z'\) exist. Indeed, by \eqref{eq:order-psi}, the bounds of \(v\) satisfy
\begin{align*}
\psi_l(p^*)\le\psi_l(v)\le\psi_l(q^*).
\end{align*}
A vertex has at most two witnesses on every chain.

The next lemma collects what the construction gives for a member of a cluster. The second estimate is the sharing requirement. The other three are used in \cref{sec:cutting}.

\begin{lemma}[Clusters]\label{lem:groups}
Let \(v\) be a vertex of \(X_i\) that \(X_l\) misses, in the cluster of the representative \(w\). Then
\begin{align*}
\lvert\varphi_i(v)-\varphi_i(w)\rvert&\le2\varepsilon d_l(w),
&
\lVert v-w\rVert&\le2\varepsilon d_l(w)<d_l(w)/k,\\
\lVert\hat w_l-v\rVert&\le\tfrac32d_l(w),
&
d_l(w)/2&\le d_l(v)\le d_l(w).
\end{align*}
\end{lemma}

\begin{proof}
The intervals of \(v\) and \(w\) meet. Moreover, \(d_l(v)\le d_l(w)\), because \(w\) was treated earlier or \(v=w\). So the centres of the two intervals differ by at most the sum of their radii. This is the first estimate:
\begin{align*}
\lvert\varphi_i(v)-\varphi_i(w)\rvert\le\varepsilon d_l(v)+\varepsilon d_l(w)\le2\varepsilon d_l(w).
\end{align*}

For the second estimate, both vertices lie on \(X_i\), so one precedes the other in \(\preceq_i\). Hence \eqref{eq:dist-progress} bounds their distance by the difference of their progress. With the first estimate and \(2\varepsilon=1/(2k)\), this gives
\begin{align*}
\lVert v-w\rVert\le\lvert\varphi_i(v)-\varphi_i(w)\rvert\le2\varepsilon d_l(w)=d_l(w)/(2k)<d_l(w)/k.
\end{align*}
The last step uses \(d_l(w)>0\), as \(X_l\) misses \(w\).

For the third estimate, the triangle inequality and the second estimate give
\begin{align*}
\lVert\hat w_l-v\rVert\le\lVert\hat w_l-w\rVert+\lVert w-v\rVert\le d_l(w)+2\varepsilon d_l(w)\le\tfrac32d_l(w).
\end{align*}

The upper bound \(d_l(v)\le d_l(w)\) of the fourth estimate was noted above. It remains to show the lower bound. By \eqref{eq:passing-lipschitz} and the second estimate,
\begin{align*}
d_l(v)\ge d_l(w)-k\lVert v-w\rVert\ge d_l(w)-2k\varepsilon d_l(w)=d_l(w)/2.
\end{align*}
\end{proof}

The next lemma shows that these witnesses meet the demands. Its statement about the vertices that \(X_l\) touches is used in \cref{lem:across}. There it shows that their witnesses lie across a cut only together with a segment that crosses it.

\begin{lemma}[Witnesses]\label{lem:witness}
Let \(v\) be a vertex and \(X_l\) a chain that encloses it, with bounds \(p^*\) and \(q^*\).
\begin{itemize}
\item {\bf Part 1.} If \(p^*=q^*\), then \(X_l\) touches \(v\), and \(\wit_l(v)=\{p^*\}\).
\item {\bf Part 2.} If \(p^*\ne q^*\), then \(\wit_l(v)\) consists of two consecutive vertices of \([p^*,q^*]\). If \(X_l\) touches \(v\), the position of \(v\) lies on the segment between them.
\end{itemize}
Consequently, \(v\notin\held(Q)\) for every set \(Q\) with \(\wit(v)\subseteq Q\).
\end{lemma}

\begin{proof}
For Part~1, the vertex \(p^*\) is both bounds of \(v\), so \(p^*\preceq_lv\preceq_lp^*\). By \eqref{eq:order-tied}, \(p^*\) is tied with \(v\). It has the position and the progress of \(v\), so
\begin{align*}
\hat v_l=\gamma_l(\varphi_l(v))=\gamma_l(\varphi_l(p^*))=p^*=v.
\end{align*}
Hence \(X_l\) touches \(v\). As \(\psi_l(p^*)=\psi_l(v)\) and \(\psi_l\) increases strictly along \(X_l\), both vertices \(z\) and \(z'\) of \cref{def:witness} are \(p^*\). So \(\wit_l(v)=\{p^*\}\).

For Part~2, suppose first that \(X_l\) touches \(v\). No vertex of \(X_l\) is tied with \(v\), since \(p^*\ne q^*\). We show that no vertex \(x\) of \(X_l\) has \(\psi_l(x)=\psi_l(v)\). Suppose that \(x\) is such a vertex. Its position is the point of \(\gamma_l\) at its own progress, which is the progress of \(v\). Moreover, \(\hat v_l=v\) because \(X_l\) touches \(v\). So
\begin{align*}
x=\gamma_l(\varphi_l(x))=\gamma_l(\varphi_l(v))=\hat v_l=v.
\end{align*}
With \(\pi_1(x)=\pi_1(v)\), all potentials of \(x\) and \(v\) would coincide. This is impossible, since no vertex of \(X_l\) is tied with \(v\).

Hence the vertex \(z\) of \cref{def:witness} is the last vertex of \(X_l\) with \(\psi_l<\psi_l(v)\), and \(z'\) is the first with \(\psi_l>\psi_l(v)\). So \(z\) and \(z'\) are consecutive. The bounds of \(v\) are not tied with \(v\), so \eqref{eq:order-psi} and \eqref{eq:order-psi-tied} give
\begin{align*}
\psi_l(p^*)<\psi_l(v)<\psi_l(q^*).
\end{align*}
So \(p^*\) is \(z\) or comes before it, and \(q^*\) is \(z'\) or comes after it. That is, \(z\) and \(z'\) lie in \([p^*,q^*]\). Since \(\psi_l\) compares progress first,
\begin{align*}
\varphi_l(z)\le\varphi_l(v)\le\varphi_l(z').
\end{align*}
So the point \(\gamma_l(\varphi_l(v))\), which is the position of \(v\), lies on the segment between \(z\) and \(z'\).

Suppose now that \(X_l\) misses \(v\), and let \(w\) be the representative of the cluster of \(v\). Then \(\lVert v-w\rVert<d_l(w)/k\), by \cref{lem:groups}. So Part~2 of \cref{lem:passing} applies, with \(w\) in the role of \(v\) and \(v\) in the role of \(u\). It shows that
\begin{align*}
z_w,z_w'\in[p^*,q^*].
\end{align*}
By \cref{def:witness}, these two consecutive vertices form \(\wit_l(v)\).

For the last statement, the neighbour demand of \(v\) is met because \(N(v)=\wit_i(v)\subseteq\wit(v)\). The demand of \(v\) on a chain \(X_l\) that encloses it is met by Part~1 if \(p^*=q^*\). It is met by Part~2 otherwise.
\end{proof}

\subsection{Processing a partition part by part}
\label{sec:blocks}

This subsection makes precise what it costs to process a set \(S\) of vertices before the rest of the configuration. By \cref{lem:witness}, no vertex of \(S\) is held once \(\wit(S)\) is processed. So the witnesses outside \(S\) are processed early. So are both vertices of every pair of consecutive vertices of a chain of which exactly one lies in \(S\). When the order later works inside one part of a partition, a vertex of the part must have the same demands as in the whole configuration. This needs both vertices of every such pair (Condition~2 of \cref{lem:local} below). The vertex outside \(S\) of such a pair is a neighbour, hence a witness, of the vertex inside.

\begin{definition}[Interface]\label{def:interface}
The \emph{interface} of a set \(S\subseteq X\) is
\begin{align*}
\partial S\coloneqq\bigl(\wit(S)\setminus S\bigr)\cup\{v\in S : N(v)\not\subseteq S\}.
\end{align*}
\end{definition}

Suppose that \(X\) is partitioned into sets that the order processes one after another. The next lemma lets the order ignore the rest of the configuration while it works inside one part \(V\) of the partition. Assume that the witnesses outside \(V\) and the vertices at which chains enter or leave \(V\) are processed. Then a vertex of \(V\) is held only if it would be held in \(V\) alone. We write \(\held_X\) and \(\held_V\) for the held sets in \(X\) and in the sub-configuration \(V\).

\begin{lemma}[Local demands]\label{lem:local}
Let \(V\subseteq X\) and \(Q\subseteq X\). Suppose that
\begin{itemize}
\item {\bf Condition 1.} \(\wit(V)\setminus V\subseteq Q\).
\item {\bf Condition 2.} \(\{x,y\}\subseteq Q\) for any two consecutive vertices \(x,y\) of a chain of which exactly one lies in \(V\).
\end{itemize}
Then \(\held_X(Q)\cap V\subseteq\held_V(Q\cap V)\).
\end{lemma}

\begin{proof}
Let \(v\in V\) be held under \(Q\) in \(X\). Suppose first that its neighbour demand is not met. Let \(y\) be a neighbour of \(v\) outside \(Q\). It is a witness of \(v\), so Condition~1 gives
\begin{align*}
y\in\wit(v)\setminus Q\subseteq V.
\end{align*}
So \(y\) is a neighbour of \(v\) in \(V\) as well.

Otherwise, let \(X_l\) be a chain that encloses \(v\) in \(X\), with bounds \(p^*\) and \(q^*\), on which the demand of \(v\) is not met. If \(p^*=q^*\), then \(p^*\) is outside \(Q\). By Part~1 of \cref{lem:witness}, it is the witness of \(v\) on \(X_l\), so Condition~1 gives
\begin{align*}
p^*\in\wit(v)\setminus Q\subseteq V.
\end{align*}
In \(V\) the vertex \(p^*\) is still tied with \(v\), so it is both bounds of \(v\). Hence the demand is not met under \(Q\cap V\).

Now let \(p^*\ne q^*\). As the demand is not met, \(Q\) does not contain two consecutive vertices of \([p^*,q^*]\). By Part~2 of \cref{lem:witness}, the two witnesses of \(v\) on \(X_l\) are such vertices. So one of them, \(x\), is outside \(Q\), and by Condition~1
\begin{align*}
x\in\wit(v)\setminus Q\subseteq V.
\end{align*}

We next show that all vertices from \(p^*\) to \(q^*\) lie in \(V\). Let \(y\) and \(y'\) be consecutive vertices of \([p^*,q^*]\) with \(y\in V\). If \(y'\) were outside \(V\), Condition~2 would put both in \(Q\). But \(Q\) contains no two consecutive vertices of \([p^*,q^*]\). So \(y'\in V\). Walking along \(X_l\) from \(x\) to \(p^*\) and from \(x\) to \(q^*\) therefore gives
\begin{align*}
[p^*,q^*]\subseteq V.
\end{align*}

It follows that \(v\) has the same bounds in \(V\). The chain of \(V\) is \(X_l\cap V\) in the order of \(X_l\). In the notation of \eqref{eq:parts},
\begin{align*}
(X_l\cap V)^{\preceq v}=X_l^{\preceq v}\cap V\ni p^*,
\qquad
(X_l\cap V)^{\succeq v}=X_l^{\succeq v}\cap V\ni q^*.
\end{align*}
The vertex \(p^*\) is the last vertex of \(X_l^{\preceq v}\), and \(q^*\) is the first of \(X_l^{\succeq v}\). So they stay the last and the first vertex of these intersections. Hence the chain of \(V\) encloses \(v\), with the bounds \(p^*\) and \(q^*\). Since \([p^*,q^*]\subseteq V\), the vertices from \(p^*\) to \(q^*\) are consecutive on the chain of \(V\) as they are on \(X_l\). No two consecutive ones lie in \(Q\cap V\), so the demand is not met in \(V\).
\end{proof}

With this lemma, the parts of a partition can be processed one after another. While the order works on a part, it holds only two kinds of vertices. The first kind are vertices that an order of this part alone would hold. The second kind are vertices in one of two interfaces: that of the union of the earlier parts, and that of the union of the parts up to the current one. So the cost does not grow with the number of parts.

\begin{lemma}[Serial processing]\label{lem:serial}
Let \(X\) be partitioned into parts \(V_1,\dots,V_t\), and let \(S_j\coloneqq V_1\cup\dots\cup V_j\). Suppose that \(\lvert\partial S_j\rvert\le s\) for every \(j\), and that every \(V_j\), as a sub-configuration, has width at most \(\omega\). Then \(X\) has width at most \(\omega+2s\).
\end{lemma}

\begin{proof}
Put \(S_0\coloneqq\emptyset\). The set of \emph{pins} of the part \(V_j\) is
\begin{align*}
\Pi_j\coloneqq\partial S_{j-1}\cup\partial S_j,
\qquad
\lvert\Pi_j\rvert\le2s.
\end{align*}

The order treats the parts in turn. For the part \(V_j\), it first processes the pins that are not yet processed. It then processes the vertices of \(V_j\) in an order of width at most \(\omega\) of this sub-configuration, skipping those already processed.

We show by induction on \(j\) that when the order has finished the part \(V_j\), the set of processed vertices is
\begin{align*}
S_j\cup\wit(S_j).
\end{align*}
For \(j=0\), before the first part, both sets are empty. Let \(j\ge1\), and consider the set of vertices processed when the part \(V_j\) is finished. By the induction hypothesis for \(j-1\) and the description of the order, it is the set on the left below. By \cref{def:interface}, \(\partial S\subseteq S\cup\wit(S)\) for every set \(S\). Moreover, \(\wit(S_{j-1})\subseteq\wit(S_j)\), because \(S_{j-1}\subseteq S_j\). Hence
\begin{align*}
S_{j-1}\cup\wit(S_{j-1})\cup\Pi_j\cup V_j\subseteq S_j\cup\wit(S_j).
\end{align*}
Conversely, \(S_j=S_{j-1}\cup V_j\) is processed. So is \(\wit(S_j)\), because by the definitions of the interface and of the pins
\begin{align*}
\wit(S_j)\setminus S_j\subseteq\partial S_j\subseteq\Pi_j.
\end{align*}

By \cref{lem:witness}, it follows that no vertex of \(S_j\) is held once the order has finished the part \(V_j\). Indeed, let \(Q\) be the set of processed vertices at any later moment. Then
\begin{align*}
\wit(v)\subseteq\wit(S_j)\subseteq Q\qquad(v\in S_j).
\end{align*}

Next we bound the number of held vertices while the order works on the part \(V_j\). Consider such a moment, and let \(Q\) be the set of processed vertices. No vertex of \(S_{j-1}\) is held. By the claim for \(j-1\), \(Q\) satisfies the first inclusion below. The second follows from it because \(\wit(S_{j-1})\setminus S_{j-1}\subseteq\partial S_{j-1}\):
\begin{align*}
Q&\subseteq S_{j-1}\cup\wit(S_{j-1})\cup\Pi_j\cup V_j,\\
Q\setminus S_j&\subseteq\bigl(\wit(S_{j-1})\setminus S_{j-1}\bigr)\cup\Pi_j\subseteq\partial S_{j-1}\cup\Pi_j=\Pi_j.
\end{align*}

While the pins are being processed, every processed vertex of \(V_j\) is a pin or a witness of \(S_{j-1}\), so
\begin{align*}
Q\cap V_j\subseteq\bigl(\wit(S_{j-1})\setminus S_{j-1}\bigr)\cup\Pi_j=\Pi_j.
\end{align*}
With \(Q\setminus S_j\subseteq\Pi_j\), it follows that
\begin{align*}
Q\setminus S_{j-1}=(Q\cap V_j)\cup(Q\setminus S_j)\subseteq\Pi_j.
\end{align*}
Held vertices are processed, and no vertex of \(S_{j-1}\) is held. So all held vertices are pins:
\begin{align*}
\held_X(Q)\subseteq Q\setminus S_{j-1}\subseteq\Pi_j.
\end{align*}
Hence at most \(\lvert\Pi_j\rvert\) vertices are held.

After the pins, \cref{lem:local} applies to \(V=V_j\), and we check its two conditions. From now on \(S_{j-1}\) and \(\Pi_j\) are processed. For Condition~1, a witness of a vertex of \(V_j\) outside \(V_j\) lies in \(S_{j-1}\) or in \(\wit(S_j)\setminus S_j\subseteq\Pi_j\):
\begin{align*}
\wit(V_j)\setminus V_j\subseteq S_{j-1}\cup\bigl(\wit(S_j)\setminus S_j\bigr)\subseteq S_{j-1}\cup\Pi_j\subseteq Q.
\end{align*}

For Condition~2, let exactly one of two consecutive vertices of a chain lie in \(V_j\), say \(x\in V_j\) and \(y\notin V_j\). The other one, \(y\), lies in \(S_{j-1}\) or outside \(S_j\):
\begin{align*}
y\in X\setminus V_j=S_{j-1}\cup(X\setminus S_j).
\end{align*}
So for \(S=S_{j-1}\) or \(S=S_j\), one of the two lies in \(S\) and the other outside. The one in \(S\) has a neighbour outside \(S\). The other is a neighbour, hence a witness, of a vertex of \(S\). Hence, by \cref{def:interface},
\begin{align*}
\{x,y\}\subseteq\partial S\subseteq\Pi_j\subseteq Q.
\end{align*}

It remains to count the held vertices after the pins. Let \(Q'\) consist of the vertices of the order of \(V_j\) up to its current vertex. All its vertices are processed. The other processed vertices of \(V_j\) are pins or witnesses of \(S_{j-1}\), hence in \(\Pi_j\):
\begin{align*}
Q'\subseteq Q\cap V_j\subseteq Q'\cup\Pi_j.
\end{align*}
By \cref{lem:local}, and by the monotonicity \eqref{eq:held-monotone} in \(V_j\), applied to the sets \(Q'\) and \(E=(Q\cap V_j)\setminus Q'\subseteq\Pi_j\),
\begin{align*}
\held_X(Q)\cap V_j\setminus\Pi_j\subseteq\held_{V_j}(Q\cap V_j)\setminus E\subseteq\held_{V_j}(Q').
\end{align*}
As the order of \(V_j\) has width at most \(\omega\), the right-hand side has at most \(\omega\) vertices. No vertex of \(S_{j-1}\) is held, and \(Q\setminus S_j\subseteq\Pi_j\). So
\begin{align*}
\held_X(Q)\subseteq\bigl(\held_X(Q)\cap V_j\setminus\Pi_j\bigr)\cup\Pi_j,
\end{align*}
and the number of held vertices is at most
\begin{align*}
\underbrace{\omega}_{\text{held inside } V_j}
+\underbrace{\lvert\Pi_j\rvert}_{\text{pins}}
\le\omega+2s.
\end{align*}
\end{proof}

\section{Cutting by spheres and grids}
\label{sec:cutting}

By \cref{lem:serial}, a partition can be processed part by part if, for every \(j\), the union of the first \(j\) parts has a small interface. This section finds such partitions, and \cref{sec:general-proof} uses them to prove \cref{thm:width}. The parts are cut out by spheres around a point and by a grid. We show that some sphere with a radius in a given range has an interface whose size is polynomial in \(k\) (\cref{lem:sphere}). Consider a grid whose spacing is a fixed fraction of the largest distance between two vertices. Some shift of this grid has such interfaces for all unions of its cells (\cref{lem:grid}). Spheres and grids are both level sets of functions \(\rho\) of points with \(\lvert\rho(y)-\rho(x)\rvert\le\lVert y-x\rVert\). One lemma about the average level set of such a function yields both (\cref{lem:levels}).

\subsection{What a cut costs}
\label{sec:cut-cost}

Spheres and grids both cut out the vertices whose position lies in a set \(U\) of points. We first describe the interface of such a set of vertices by what happens at the boundary of \(U\). A segment \(e\) of a curve (\cref{sec:passing}) \emph{crosses} \(U\) if
\begin{align*}
e\cap U\ne\emptyset\ne e\setminus U.
\end{align*}
The set \(U\) \emph{splits} the cluster of a representative \(w\) (\cref{sec:witness-choice}) if some member \(v\) of the cluster satisfies
\begin{align*}
v\in U,
\qquad
\hat w_l\notin U.
\end{align*}

\begin{lemma}[Across a cut]\label{lem:across}
Let \(U\) be a set of points, and let \(S\) be the set of vertices whose position lies in \(U\). Every vertex of \(\partial S\) is an end of a segment that crosses \(U\), or one of the two witnesses \(z_w,z_w'\) of a cluster that \(U\) splits.
\end{lemma}

\begin{proof}
Let \(v\in S\) have a neighbour \(u\notin S\), and let \(e\) be the segment between them. By the definition of \(S\), the positions of the two ends satisfy
\begin{align*}
v\in e\cap U,
\qquad
u\in e\setminus U.
\end{align*}
So \(e\) crosses \(U\), and \(v\) and \(u\) are its ends. This covers the vertices of \(S\) with a neighbour outside \(S\). It also covers the witnesses that a vertex of \(S\) has outside \(S\) on its own chain.

It remains to treat a witness \(u\) of \(v\) on a chain \(X_l\) that encloses \(v\), with
\begin{align*}
u\in\wit_l(v)\setminus S,
\qquad
v\in S.
\end{align*}
We first show that \(v\) is not tied with a vertex of \(X_l\). Suppose that a vertex \(x\) of \(X_l\) is tied with \(v\). Then \(x\) has the position of \(v\), so it lies in \(S\). It is both bounds of \(v\) on \(X_l\) (see after \eqref{eq:parts}). So by Part~1 of \cref{lem:witness}, it is the only witness of \(v\) on \(X_l\). Then \(u=x\in S\), contrary to \(u\notin S\). So Part~2 of \cref{lem:witness} applies, and there remain the two cases of \cref{def:witness}.

\emph{The chain \(X_l\) touches \(v\).} Then \(u\) is an end of a segment \(e\) of \(X_l\) that contains the position of \(v\), by Part~2 of \cref{lem:witness}. So
\begin{align*}
v\in e\cap U,
\qquad
u\in e\setminus U,
\end{align*}
and \(e\) crosses \(U\).

\emph{The chain \(X_l\) misses \(v\).} Let \(w\) be the representative of the cluster of \(v\). Then \(u\) is an end of the segment \(e\) from \(z_w\) to \(z_w'\), which contains \(\hat w_l\):
\begin{align*}
u\in\wit_l(v)=\{z_w,z_w'\},
\qquad
\hat w_l\in e.
\end{align*}
If \(\hat w_l\) lies in \(U\), then \(\hat w_l\in e\cap U\) and \(u\in e\setminus U\), so \(e\) crosses \(U\). Otherwise
\begin{align*}
v\in U,
\qquad
\hat w_l\notin U,
\end{align*}
so \(U\) splits the cluster of \(w\), and \(u\) is one of its two witnesses.
\end{proof}

So the interface of a cut has at most twice as many vertices as there are segments that cross it and clusters that it splits. We count both for the level sets of a function. Let \(\rho\) be a real function of points that \emph{changes by at most the distance}, that is,
\begin{align*}
\lvert\rho(y)-\rho(x)\rvert\le\lVert y-x\rVert\qquad\text{for all points } x,y.
\end{align*}
For a sphere, \(\rho\) will be the distance from the centre. The cut at the level \(c\) is \(U=\{x : \rho(x)<c\}\). A segment that crosses this set contains a point with \(\rho=c\), because \(\rho\) is continuous. A cluster that this set splits has a member \(v\) with \(\rho(v)<c\le\rho(\hat w_l)\). Accordingly, let \(n_\rho(c)\) be the number of segments that contain a point with \(\rho=c\). For \(i\ne l\), let \(m^{il}_\rho(c)\) be the number of clusters of \(X_i\) for \(X_l\) that have a member \(v\) with
\begin{align*}
\rho(v)\le c\le\rho(\hat w_l),
\end{align*}
where \(w\) is the representative of the cluster. The count allows equality on the left, which only enlarges it. The proof of \cref{lem:grid} needs this equality for \(\rho=-\pi_{ak}\). There a cluster with \(\pi_{ak}(\hat w_l)<c\le\pi_{ak}(v)\) gives
\begin{align*}
\rho(v)\le-c<\rho(\hat w_l).
\end{align*}

We bound the two counts on average over the levels \(c\) in an interval \((a,b)\) (\cref{lem:levels}). A cluster is counted only at levels within \(\tfrac32d_l(w)\) of \(\rho(\hat w_l)\) (\cref{lem:groups}). So it contributes at most \(\min\{\tfrac32d_l(w),b-a\}\) to the integral of \(m^{il}_\rho\). For the clusters with a small passing distance, we therefore need a bound on the sum of the passing distances. For those with a large one, we need a bound on their number. The next lemma gives both, for clusters with members within distance \(D\) of one another.

\begin{lemma}[Room for clusters]\label{lem:room}
Fix \(i\ne l\), and consider a set of clusters of \(X_i\) for \(X_l\). For the representative \(w\) of every cluster of the set, choose a member \(v_w\) of its cluster, and write \(d_w\coloneqq d_l(w)\). Suppose that any two of the chosen members have distance at most \(D\), where \(D>0\). Then
\begin{itemize}
\item {\bf Part 1.} the clusters with \(d_w\le kD\) satisfy \(\sum_wd_w\le(2k^2+k)D\), and
\item {\bf Part 2.} at most \(2k+10\) clusters have \(d_w>kD\).
\end{itemize}
\end{lemma}

\begin{proof}
The interval \(I(w)\) has the centre \(\varphi_i(w)\) and the radius \(\varepsilon d_w\). We first locate these intervals. By \eqref{eq:progress-lipschitz} and by the first estimate of \cref{lem:groups}, any two representatives \(w,w'\) of the set satisfy
\begin{align*}
\lvert\varphi_i(v_w)-\varphi_i(v_{w'})\rvert\le(k-1)D,
\qquad
\lvert\varphi_i(w)-\varphi_i(v_w)\rvert\le2\varepsilon d_w.
\end{align*}
Let \(\tau>0\), and consider the clusters with \(d_w\le\tau\). The first bound confines the numbers \(\varphi_i(v_w)\) to an interval of length \((k-1)D\). The second moves each centre \(\varphi_i(w)\) by at most \(2\varepsilon\tau\) from \(\varphi_i(v_w)\). The radius of \(I(w)\) is at most \(\varepsilon\tau\). So the intervals \(I(w)\) of these clusters all lie in one interval of length
\begin{align}
(k-1)D+2\cdot2\varepsilon\tau+2\cdot\varepsilon\tau=(k-1)D+6\varepsilon\tau.
\label{eq:room}
\end{align}
The intervals are pairwise disjoint by the room requirement of \cref{sec:witness-choice}, and \(1/\varepsilon=4k\).

For Part~1, apply \eqref{eq:room} with \(\tau=kD\). The lengths \(2\varepsilon d_w\) of the disjoint intervals add up to at most the length of the interval that contains them:
\begin{align*}
2\varepsilon\sum_wd_w\le(k-1)D+6\varepsilon kD.
\end{align*}
Dividing by \(2\varepsilon=1/(2k)\) gives
\begin{align*}
\sum_wd_w\le\frac{(k-1)D+6\varepsilon kD}{2\varepsilon}=2k(k-1)D+3kD=(2k^2+k)D.
\end{align*}

For Part~2, let \(d_{\max}\) and \(d_{\min}>kD\) be the largest and the smallest \(d_w\) among the clusters with \(d_w>kD\). We first show that their sizes are comparable. By the fourth estimate of \cref{lem:groups} and by \eqref{eq:passing-lipschitz}, respectively, any two representatives \(w,w'\) of the set satisfy
\begin{align*}
d_w/2\le d_l(v_w)\le d_w,
\qquad
\lvert d_l(v_w)-d_l(v_{w'})\rvert\le kD.
\end{align*}
Take \(w\) and \(w'\) with \(d_w=d_{\max}\) and \(d_{w'}=d_{\min}\). Then
\begin{align*}
d_{\max}=d_w\le2d_l(v_w)\le2\bigl(d_l(v_{w'})+kD\bigr)\le2\bigl(d_{\min}+kD\bigr)<4d_{\min}.
\end{align*}
The last step uses \(d_{\min}>kD\).

Next we count the clusters with \(d_w>kD\). By \eqref{eq:room} with \(\tau=d_{\max}\), their intervals lie in an interval of length \((k-1)D+6\varepsilon d_{\max}\). These intervals are disjoint, and each of them has length at least \(2\varepsilon d_{\min}\). So their number is at most
\begin{align*}
\frac{(k-1)D+6\varepsilon d_{\max}}{2\varepsilon d_{\min}}
=\frac{2k(k-1)D}{d_{\min}}+\frac{3d_{\max}}{d_{\min}}
<2(k-1)+12
=2k+10.
\end{align*}
The inequality uses \(d_{\min}>kD\) and \(d_{\max}<4d_{\min}\).
\end{proof}

The next lemma carries out the two counts. \Cref{lem:sphere,lem:grid} use it to choose a radius and a shift.

\begin{lemma}[Levels on average]\label{lem:levels}
Let \(\rho\) be a function of points with \(\lvert\rho(y)-\rho(x)\rvert\le\lVert y-x\rVert\) for all points \(x,y\). Let \(a<b\) and \(D\ge b-a\), and suppose that any two points of the curves \(\gamma_1,\dots,\gamma_k\) at which \(\rho<b\) have distance at most \(D\). Then for all \(i\ne l\),
\begin{align}
\int_a^bn_\rho(c)\mathrm{d}c&\le k(k-1)D,
\label{eq:level-segments}\\
\int_a^bm^{il}_\rho(c)\mathrm{d}c&\le(3k^2+4k+10)D.
\label{eq:level-groups}
\end{align}
\end{lemma}

\begin{proof}
We first prove \eqref{eq:level-segments}, one segment at a time. Write \(\lambda\) for the Lebesgue measure on \(\R\), and \(\rho(e)\coloneqq\{\rho(x) : x\in e\}\) for the set of values that \(\rho\) attains on a segment \(e\). A number \(c\) is counted by \(n_\rho(c)\) once for every segment \(e\) with \(c\in\rho(e)\). So the integral is a sum over all segments:
\begin{align*}
\int_a^bn_\rho(c)\mathrm{d}c=\sum_e\lambda\bigl(\rho(e)\cap(a,b)\bigr).
\end{align*}

We bound one term. Let \(e\) be a segment of \(\gamma_l\), between consecutive vertices \(z,z'\) of \(X_l\). It consists of the points \(\gamma_l(s)\) with \(s\) in the interval
\begin{align*}
I_e\coloneqq[\varphi_l(z),\varphi_l(z')].
\end{align*}
By the hypothesis on \(\rho\) and by \eqref{eq:path-lipschitz},
\begin{align*}
\bigl\lvert\rho(\gamma_l(s'))-\rho(\gamma_l(s))\bigr\rvert\le\lVert\gamma_l(s')-\gamma_l(s)\rVert\le\lvert s'-s\rvert.
\end{align*}
A real function of \(s\) that changes by at most \(\lvert s'-s\rvert\) maps every interval onto a set of at most the same length. It follows, by covering a set of measure \(\mu\) with intervals of total length close to \(\mu\), that the function maps this set onto a set of measure at most \(\mu\). We apply this to the function \(s\mapsto\rho(\gamma_l(s))\) on \(I_e\). Every value in \(\rho(e)\cap(a,b)\) is attained at some \(s\in I_e\), and there \(\rho(\gamma_l(s))<b\). So this set of values is contained in the image of the set on the right below, and
\begin{align*}
\lambda\bigl(\rho(e)\cap(a,b)\bigr)\le\lambda\bigl(\{s\in I_e : \rho(\gamma_l(s))<b\}\bigr).
\end{align*}

Next we add up these bounds over the segments \(e\) of \(\gamma_l\). The intervals \(I_e\) meet only at their ends. So the right-hand sides add up to at most the measure of the set
\begin{align*}
M\coloneqq\{s : \rho(\gamma_l(s))<b\}.
\end{align*}
Let \(s,s'\in M\). The points \(\gamma_l(s)\) and \(\gamma_l(s')\) have distance at most \(D\) by hypothesis. So \eqref{eq:progress-lipschitz} gives
\begin{align*}
\lvert s'-s\rvert=\bigl\lvert\varphi_l(\gamma_l(s'))-\varphi_l(\gamma_l(s))\bigr\rvert\le(k-1)D.
\end{align*}
Hence \(M\) lies in an interval of length \((k-1)D\), and
\begin{align*}
\sum_{e}\lambda\bigl(\rho(e)\cap(a,b)\bigr)
\le\lambda(M)
\le(k-1)D,
\end{align*}
where \(e\) runs over the segments of \(\gamma_l\). Summing over the \(k\) chains gives \eqref{eq:level-segments}.

We now prove \eqref{eq:level-groups}. Fix \(i\ne l\). For the cluster of a representative \(w\), let \(T_w\) be the set of numbers \(c\in(a,b)\) at which this cluster is counted by \(m^{il}_\rho(c)\). As for the segments,
\begin{align*}
\int_a^bm^{il}_\rho(c)\mathrm{d}c=\sum_w\lambda(T_w).
\end{align*}
We first bound one term. A member \(v\) of the cluster of \(w\) satisfies \(\lVert\hat w_l-v\rVert\le\tfrac32d_l(w)\) by \cref{lem:groups}. Since \(\rho\) changes by at most the distance,
\begin{align*}
\rho(\hat w_l)-\tfrac32d_l(w)\le\rho(v).
\end{align*}
So \(T_w\) lies in the interval of length \(\tfrac32d_l(w)\) that ends at \(\rho(\hat w_l)\), and in \((a,b)\):
\begin{align*}
\lambda(T_w)\le\min\bigl\{\tfrac32d_l(w),\ b-a\bigr\}\le\min\bigl\{\tfrac32d_l(w),\ D\bigr\}.
\end{align*}

Only the clusters with \(T_w\ne\emptyset\) contribute. Such a cluster has a member \(v\) with \(\rho(v)\le c\) for some \(c<b\). Choose such a member \(v_w\) in every such cluster. The position of a vertex is a point of the curve of its chain. So any two of the chosen members have distance at most \(D\), and \cref{lem:room} applies with \(d_w=d_l(w)\). For the clusters with \(d_w\le kD\), we use \(\lambda(T_w)\le\tfrac32d_w\). For those with \(d_w>kD\), we use \(\lambda(T_w)\le D\). So each of the latter is charged the whole range of levels, and this needs the hypothesis \(D\ge b-a\). The two parts of \cref{lem:room} give
\begin{align*}
\sum_w\lambda(T_w)
\le\underbrace{\tfrac32(2k^2+k)D}_{d_w\le kD}
+\underbrace{(2k+10)D}_{d_w>kD}
\le(3k^2+4k+10)D.
\end{align*}
\end{proof}

\subsection{Spheres and grids}

We first cut by a sphere around a point \(o\), with a radius in an interval \((R,2R)\). \Cref{sec:general-proof} uses such spheres to cut a configuration into shells.

\begin{lemma}[Spheres]\label{lem:sphere}
Let \(o\) be a point and \(R>0\). There is a radius \(c\in(R,2R)\) such that the set \(S\coloneqq\{v\in X : \lVert v-o\rVert<c\}\) satisfies
\begin{align*}
\lvert\partial S\rvert\le s_k\coloneqq32(k+1)^4.
\end{align*}
\end{lemma}

\begin{proof}
Let \(\rho(x)\coloneqq\lVert x-o\rVert\). By the triangle inequality, \(\rho\) changes by at most the distance. For \(c>0\) let
\begin{align*}
U_c\coloneqq\{x : \rho(x)<c\},
\qquad
S_c\coloneqq\{v\in X : \rho(v)<c\}.
\end{align*}

We first bound the interface of \(S_c\) by the two counts of \cref{lem:levels}. A segment that crosses \(U_c\) contains a point with \(\rho<c\) and a point with \(\rho\ge c\). Since \(\rho\) is continuous, it also contains a point with \(\rho=c\). A cluster that \(U_c\) splits, with representative \(w\), has a member \(v\) with
\begin{align*}
\rho(v)<c\le\rho(\hat w_l).
\end{align*}
So \cref{lem:across} gives, for every \(c>0\),
\begin{align}
\lvert\partial S_c\rvert\le2n_\rho(c)+2\sum_{i\ne l}m^{il}_\rho(c),
\label{eq:sphere-count}
\end{align}
where the sum runs over all ordered pairs \((i,l)\) of distinct indices.

It remains to choose the radius. Any two points with \(\rho<2R\) have distance less than \(4R\). So \cref{lem:levels} applies with \(a=R\), \(b=2R\) and \(D=4R\). The sum in \eqref{eq:sphere-count} has \(k(k-1)\) terms. So the integral of the right-hand side of \eqref{eq:sphere-count} over \(c\in(R,2R)\) is at most
\begin{align*}
2\cdot k(k-1)\cdot4R+2k(k-1)\cdot(3k^2+4k+10)\cdot4R=8k(k-1)(3k^2+4k+11)R\le32(k+1)^4R.
\end{align*}
The last step uses
\begin{align*}
k(k-1)\le(k+1)^2,
\qquad
3k^2+4k+11\le4(k+1)^2
\qquad(k\ge2).
\end{align*}
The range of radii has length \(R\). So some \(c\in(R,2R)\) has a right-hand side of at most \(32(k+1)^4\).
\end{proof}

For the grid, fix a spacing \(h>0\) and a shift \(\vartheta\in[0,h)^{k-1}\). Two points \(x,y\) lie in the same \emph{cell} if
\begin{align*}
\Bigl\lfloor\frac{\pi_{ak}(x)-\vartheta_a}{h}\Bigr\rfloor=\Bigl\lfloor\frac{\pi_{ak}(y)-\vartheta_a}{h}\Bigr\rfloor\qquad(1\le a<k).
\end{align*}
Cells are convex. Two points \(x,y\) of one cell have \(\lVert y-x\rVert<2h\), because \(\pi_{ab}=\pi_{ak}-\pi_{bk}\).

The cells will be the parts of a partition that is processed part by part by \cref{lem:serial}. Whatever the order of the cells, the union of the first \(j\) parts is the set of vertices in a union of cells. The next lemma provides one set \(W\) that contains the interfaces of all these unions. In \cref{sec:general-proof} it is applied to one block of a configuration, with \(\Delta/h<32\).

\begin{lemma}[Grids]\label{lem:grid}
Suppose that any two vertices of \(X\) have distance at most \(\Delta\), and let \(h>0\). There are a shift \(\vartheta\) and a set \(W\) of at most
\begin{align*}
q_k\Delta/h,\qquad q_k\coloneqq12k^5,
\end{align*}
vertices such that \(\partial S\subseteq W\) for every set \(S\) that consists of the vertices in a union of cells.
\end{lemma}

\begin{proof}
For a shift \(\vartheta\), let \(W(\vartheta)\) consist of two kinds of vertices. The first are the ends of the segments whose two ends lie in different cells. The second are the witnesses \(z_w,z_w'\) of the clusters that have a member in a cell different from the cell of their passing point \(\hat w_l\). We first show that, for the set \(S\) of vertices in a union of cells,
\begin{align*}
\partial S\subseteq W(\vartheta).
\end{align*}
Apply \cref{lem:across} to the set \(U\) of points of these cells. A segment that crosses \(U\) contains points of two different cells. Since cells are convex, its two ends lie in different cells. A cluster that \(U\) splits has a member in a cell of \(U\), while its passing point lies in another cell.

Next we bound the size of \(W(\vartheta)\) by the counts of \cref{lem:levels}. We use them for the functions \(\pi_{ak}\) and \(-\pi_{ak}\), \(a<k\), which change by at most the distance. Two points lie in different cells only if, for some \(a<k\), a number \(c\in\vartheta_a+h\mathbb Z\) separates their values of \(\pi_{ak}\). That is, one value is smaller than \(c\) and the other is at least \(c\). If these are the two ends of a segment, the segment contains a point with \(\pi_{ak}=c\). If they are a member \(v\) and the passing point \(\hat w_l\) of a cluster, then
\begin{align*}
\pi_{ak}(v)<c\le\pi_{ak}(\hat w_l)
\qquad\text{or}\qquad
-\pi_{ak}(v)\le-c<-\pi_{ak}(\hat w_l).
\end{align*}
So, with \(c\) running over \(\vartheta_a+h\mathbb Z\),
\begin{align}
\lvert W(\vartheta)\rvert\le2\sum_{a<k}\sum_c\Bigl(n_{\pi_{ak}}(c)+\sum_{i\ne l}\bigl(m^{il}_{\pi_{ak}}(c)+m^{il}_{-\pi_{ak}}(-c)\bigr)\Bigr).
\label{eq:grid-count}
\end{align}

It remains to average over the shift. If \(\vartheta_a\) is uniform in \([0,h)\), then for every function \(F\ge0\)
\begin{align*}
\mathbb E\Bigl[\sum_{c\in\vartheta_a+h\mathbb Z}F(c)\Bigr]=\frac1h\int_{\R}F(c)\mathrm{d}c,
\end{align*}
because every real number is \(\vartheta_a+nh\) for exactly one \(\vartheta_a\in[0,h)\) and one integer \(n\).

We bound these integrals by \cref{lem:levels}. Fix \(a<k\), and let \([c^-,c^+]\) be the smallest interval that contains the values of \(\pi_{ak}\) at all vertices. A pair coordinate changes by at most the distance, so
\begin{align*}
c^+-c^-\le\Delta.
\end{align*}
Every point of a curve lies on a segment between two vertices. So \(\pi_{ak}\) takes values in \([c^-,c^+]\) on the curves. Moreover, any two points \(x,y\) of the curves have distance at most \(\Delta\). Indeed, let \(x\) lie between the vertices \(u,u'\), and \(y\) between \(v,v'\). Then \(y-x\) is a convex combination of the four differences of a vertex of the second pair and a vertex of the first. The distance is a norm, so
\begin{align*}
\lVert y-x\rVert\le\max\bigl\{\lVert v-u\rVert,\lVert v-u'\rVert,\lVert v'-u\rVert,\lVert v'-u'\rVert\bigr\}\le\Delta.
\end{align*}
The three counts in \eqref{eq:grid-count} vanish for \(c\notin[c^-,c^+]\). Indeed, the first is positive only if \(\pi_{ak}=c\) at a point of a curve. The other two are positive only if \(c\) lies between the values of \(\pi_{ak}\) at a vertex and at a passing point. A passing point is a point of a curve. So \cref{lem:levels} applies to \(\pi_{ak}\) on \((c^-,c^+)\) and to \(-\pi_{ak}\) on \((-c^+,-c^-)\), with \(D=\Delta\). It gives
\begin{align*}
\int_{\R}n_{\pi_{ak}}(c)\mathrm{d}c\le k(k-1)\Delta,
\qquad
\int_{\R}m^{il}_{\pm\pi_{ak}}(c)\mathrm{d}c\le(3k^2+4k+10)\Delta.
\end{align*}
If \(c^-=c^+\), the integrals are zero. There are \(k-1\) coordinates \(a\) and \(k(k-1)\) ordered pairs \((i,l)\). With the factor \(2\) of \eqref{eq:grid-count} and the factor \(1/h\) of the average, a shift chosen uniformly at random has
\begin{align*}
\mathbb E\bigl[\lvert W(\vartheta)\rvert\bigr]
&\le\frac{2(k-1)}{h}\Bigl(k(k-1)\Delta+2k(k-1)(3k^2+4k+10)\Delta\Bigr)\\
&=2k(k-1)^2(6k^2+8k+21)\Delta/h
\le12k^5\Delta/h.
\end{align*}
The last step holds because
\begin{align*}
(k-1)^2(6k^2+8k+21)=6k^4-4k^3+11k^2-34k+21\le6k^4\qquad(k\ge2).
\end{align*}
Some shift attains at most the expected size.
\end{proof}

\section{Proof of the width theorem}
\label{sec:general-proof}

This section proves \cref{thm:width}. We prove a bound in terms of the packing number (\cref{def:packing}), by induction on it. The configuration is cut by spheres and grids into cells around a smallest ball that is not sortable (\cref{fig:shells}). A cell is either so small that it is sortable, or far from the ball compared with its own diameter. In the second case, the next lemma separates it from the ball.

\begin{lemma}[Far sets are separated]\label{lem:far-separated}
Let \(o\) be a vertex of a configuration \(X\), let \(r>0\) and \(j\ge1\), and let \(Y^*\) and \(Y\) be sub-configurations of \(X\). Suppose that, for all \(u\in Y^*\) and \(y,y'\in Y\),
\begin{align*}
\lVert u-o\rVert\le r,
\qquad
\lVert y-o\rVert>2^{j+1}r,
\qquad
\lVert y'-y\rVert<2^jr.
\end{align*}
Then \(Y\) and \(Y^*\) are separated.
\end{lemma}

\begin{proof}
Any two vertices of \(Y^*\) have distance at most \(2r\), since both have distance at most \(r\) from \(o\).

We first show that no vertex \(y\in Y\) lies on a chain between two vertices \(u,u'\) of \(Y^*\). Suppose otherwise. Then \eqref{eq:dist-monotone} gives
\begin{align*}
\lVert y-u\rVert\le\lVert u'-u\rVert\le2r.
\end{align*}
As \(j\ge1\), it follows that
\begin{align*}
\lVert y-o\rVert\le\lVert y-u\rVert+\lVert u-o\rVert\le2r+r=3r<2^{j+1}r,
\end{align*}
a contradiction.

Next we show that no vertex \(u\in Y^*\) lies on a chain between two vertices \(y,y'\) of \(Y\). Suppose otherwise. Then \eqref{eq:dist-monotone} gives again
\begin{align*}
\lVert u-y\rVert\le\lVert y'-y\rVert<2^jr.
\end{align*}
Hence
\begin{align*}
\lVert y-o\rVert\le\lVert y-u\rVert+\lVert u-o\rVert<2^jr+r\le2^{j+1}r,
\end{align*}
a contradiction again.

It follows that every \(i\) with \(Y_i\ne\emptyset\ne Y^*_i\) has
\begin{align*}
Y_i<_iY^*_i
\qquad\text{or}\qquad
Y^*_i<_iY_i.
\end{align*}
This is what separated means (\cref{sec:matching}). Suppose that some chain \(X_i\) violates this. Then it has vertices \(u_1,u_2\) of \(Y^*\) and \(y_1,y_2\) of \(Y\) such that \(u_1\) comes before \(y_1\) and \(y_2\) comes before \(u_2\). If \(y_1\) comes before \(u_2\), then \(y_1\) lies between the two vertices \(u_1,u_2\) of \(Y^*\). Otherwise \(u_2\) comes before \(y_1\) and after \(y_2\), so it lies between two vertices of \(Y\). Both were excluded.
\end{proof}

\begin{theorem}[Width and packing number]\label{thm:width-packing}
Every configuration \(X\) with \(k\ge2\) chains has width at most \(4k+c_k\nu(X)\), where \(c_k\coloneqq1000(k+1)^5\). In particular, its width is at most \(g(k)\), as \cref{thm:width} states.
\end{theorem}

\begin{proof}
The second statement follows from the first. Indeed, \(\nu(X)<N_k\) by \cref{lem:packing}, and hence
\begin{align*}
4k+c_k\nu(X)\le4k+1000(k+1)^5N_k=g(k).
\end{align*}
We prove the first statement by induction on \(\nu(X)\). If \(\nu(X)=0\), then \(X\) is sortable (\cref{sec:matching}) and has width at most \(4k\) by \cref{lem:sortable}.

Let \(\nu(X)\ge1\), so that \(X\) is not sortable. Put
\begin{align*}
\omega\coloneqq4k+c_k(\nu(X)-1).
\end{align*}
We cut \(X\) into cells of width at most \(\omega\).

\emph{A smallest ball that is not sortable.} For a vertex \(o\) and \(r\ge0\) let \(X(o,r)\coloneqq\{v\in X : \lVert v-o\rVert\le r\}\). Choose a vertex \(o\) and a number \(r\) such that the sub-configuration
\begin{align*}
Y^*\coloneqq X(o,r)
\end{align*}
is not sortable and \(r\) is as small as possible.

Such a choice exists, because \(X(o,r)=X\) for large \(r\) and only the finitely many distances between vertices matter. Moreover \(r>0\), since all vertices of \(X(o,0)\) have the same position. Such a configuration is sortable by Part~3 of \cref{lem:sortable-simple}.

\emph{Shells and cells.} We first cut \(X\) into shells. The radii of the spheres and the spacings of the grids are fixed multiples of \(r\) that double from one shell to the next. For \(j\ge0\), put
\begin{align*}
R_j\coloneqq2^{j+2}r,
\qquad
h_j\coloneqq
\begin{cases}
r/2 & (j=0),\\
2^{j-1}r & (j\ge1).
\end{cases}
\end{align*}
The spacings are chosen for the cells. Two vertices in one cell of a grid of spacing \(h\) have distance less than \(2h\). In the innermost block this is \(2h_0=r\). So a cell lies in a ball of radius less than \(r\), which is sortable by the choice of \(r\). The vertices of the shell \(V_j\) lie outside the sphere of radius \(r_{j-1}>R_{j-1}\) chosen below. Hence a cell \(Y\) of \(V_j\) and a vertex \(y\in Y\) satisfy
\begin{align*}
\max_{y'\in Y}\lVert y'-y\rVert<2h_j=2^jr,
\qquad
\lVert y-o\rVert>R_{j-1}=2^{j+1}r.
\end{align*}
Then \cref{lem:far-separated} separates the cell from \(Y^*\). Let \(J\) be so large that \(R_J\) exceeds all distances from \(o\). For \(0\le j\le J\) take the radius \(r_j\in(R_j,2R_j)\) of \cref{lem:sphere} for the point \(o\), so that
\begin{align*}
S_j\coloneqq\{v : \lVert v-o\rVert<r_j\},
\qquad
\lvert\partial S_j\rvert\le s_k.
\end{align*}
The blocks are \(V_0\) and the shells \(V_1,\dots,V_J\):
\begin{align*}
V_0\coloneqq S_0,
\qquad
V_j\coloneqq S_j\setminus S_{j-1}\quad(1\le j\le J).
\end{align*}
They partition \(X\).

Next we cut every block by a grid. Every block \(V_j\) is a configuration of its own. We give it its own witnesses, chosen as in \cref{sec:witnesses} inside \(V_j\). This is legitimate, because the width of a configuration is defined without witnesses. The witnesses only serve to construct an order of small width. So \cref{lem:serial} is applied below with two systems of witnesses: those of \(X\) for the blocks, and those of \(V_j\) for the cells of \(V_j\). Any two vertices of \(V_j\) have distance less than
\begin{align*}
2r_j<4R_j=32h_j.
\end{align*}
So \cref{lem:grid}, applied to \(V_j\), gives a grid of spacing \(h_j\). For this grid, the interface of every union of cells of \(V_j\) has at most \(32q_k\) vertices.

\emph{Every cell has a smaller packing number.} Let \(Y\) be the set of vertices of \(V_j\) in one cell. Any two of them have distance less than \(2h_j\). We show \(\nu(Y)\le\nu(X)-1\).

Let \(j=0\), and fix a vertex \(y\in Y\). Then \(2h_0=r\), so
\begin{align*}
r'\coloneqq\max_{v\in Y}\lVert v-y\rVert<r,
\qquad
Y\subseteq X(y,r').
\end{align*}
By the choice of \(r\), the configuration \(X(y,r')\) is sortable. Hence so is its sub-configuration \(Y\) (Part~2 of \cref{lem:sortable-simple}), and \(\nu(Y)=0\).

Let \(j\ge1\). Then \(Y\) and \(Y^*\) satisfy the hypotheses of \cref{lem:far-separated}: for all \(u\in Y^*\) and \(y,y'\in Y\),
\begin{align*}
\lVert u-o\rVert\le r,
\qquad
\lVert y-o\rVert>2^{j+1}r,
\qquad
\lVert y'-y\rVert<2^jr.
\end{align*}
The first bound is the definition of \(Y^*\). The second holds because every vertex of \(Y\) lies outside \(S_{j-1}\). So its distance from \(o\) is at least \(r_{j-1}\), and
\begin{align*}
r_{j-1}>R_{j-1}=2^{j+1}r.
\end{align*}
The third holds because \(2h_j=2^jr\). So \(Y\) and \(Y^*\) are separated. Every member of a packing in \(Y\) (\cref{def:packing}) is a subset of \(Y\). It is therefore separated from \(Y^*\). Moreover, \(Y^*\) is not sortable. So a largest packing in \(Y\), together with \(Y^*\), is a packing in \(X\), and
\begin{align*}
\nu(Y)+1\le\nu(X).
\end{align*}

\emph{The count.} By the induction hypothesis and since \(\nu(Y)\le\nu(X)-1\), every cell \(Y\) has width at most
\begin{align*}
4k+c_k\nu(Y)\le4k+c_k(\nu(X)-1)=\omega.
\end{align*}
We apply \cref{lem:serial} twice. First apply it to a block \(V_j\), partitioned into its cells, with \(s=32q_k\). It shows that \(V_j\) has width at most \(\omega+64q_k\).

Next apply it to \(X\), partitioned into the blocks \(V_0,\dots,V_J\), with \(s=s_k\). It shows that \(X\) has width at most
\begin{align*}
\omega+\underbrace{64q_k}_{\text{cells of a block}}+\underbrace{2s_k}_{\text{blocks}}
=\omega+768k^5+64(k+1)^4
\le\omega+c_k,
\end{align*}
which is \(4k+c_k\nu(X)\).
\end{proof}

\section{Open problems}
\label{sec:four}

Our algorithm runs in \(n^{g(k)+O(1)}\) time with \(g(k)=2^{O(2^kk\log k)}\). For acyclic and for planar digraphs, \(n^{O(k)}\)-time algorithms are known~\cite{bk17}, and since the problem is W[1]-hard when parameterized by \(k\)~\cite{ctw24}, an exponent that does not depend on \(k\) is not to be expected.

\begin{question}\label{q:time}
Can directed \(k\)-DSP with positive weights be solved in \(n^{k^{O(1)}}\) time, or even in \(n^{O(k)}\) time?
\end{question}

By \cref{lem:transfer} and Part~3 of \cref{thm:sweep}, the first bound would follow from a bound polynomial in \(k\) on the width of configurations, a question about chains of points alone (\cref{sec:config}).

\begin{question}\label{q:memory}
Is the largest width of a configuration with \(k\) chains bounded by a polynomial in \(k\)?
\end{question}

In our bound (\cref{thm:width-packing}), the cuts cost polynomially many stored vertices per level of the induction, and the doubly exponential size of \(g\) comes from the number of levels, which is bounded by the packing number (\cref{lem:packing}).

\PaperPrintBibliography

\section*{AI methodology}

We used GPT 6.2 Astra (OpenAI) and Claude Opus 5.5 and Claude Fable 5.2 (Anthropic) throughout this work, including the development of the algorithm and its proofs and the writing of the manuscript. The authors are responsible for all content.

\PaperEndMatter

\end{document}